\documentclass[useAMS,usenatbib]{mn2e}

\usepackage{graphicx}

\newcommand{\be}{\begin{equation}}
\newcommand{\ee}{\end{equation}}
\newcommand{\bea}{\begin{eqnarray}}
\newcommand{\nn}{\nonumber}
\newcommand{\eea}{\end{eqnarray}}

\newcommand{\p}{\partial}

\title[An analytic spacetime exterior to neutron stars]{An all-purpose metric for the exterior of any
kind of rotating neutron star}
\author[G.~Pappas and T.~A.~Apostolatos]{George Pappas$^{1,2}$\thanks{E-mail:
georgios.pappas@guest.uni-tuebingen.de} and Theocharis A.
Apostolatos$^{1}$\thanks{E-mail: thapostol@phys.uoa.gr}\\
$^{1}$Section of Astrophysics, Astronomy, and
Mechanics, Department of Physics, University of Athens,
\\Panepistimiopolis Zografos GR15783, Athens, Greece\\
$^{2}$Theoretical Astrophysics, IAAT, Eberhard Karls University of T$\ddot{u}$bingen, T$\ddot{u}$bingen 72076, Germany}

\begin{document}

\date{}

\pagerange{\pageref{firstpage}--\pageref{lastpage}} \pubyear{2002}

\maketitle

\label{firstpage}

\begin{abstract}
We have tested the appropriateness of two-soliton analytic metric to describe
the exterior of all types of neutron stars, no matter what their equation of state or
rotation rate is. The particular analytic solution of the vaccuum Einstein equations
proved quite adjustable to mimic the metric functions of all numerically constructed
neutron-star models that we used as a testbed. The neutron-star models covered
a wide range of stiffness, with regard to the equation of state of
their interior, and all rotation rates up to the maximum possible rotation rate allowed for
each such star. Apart of the metric functions themselves,
we have compared the radius of the innermost stable circular orbit
$R_{\rm{ISCO}}$, the orbital frequency $\Omega\equiv\frac{d\phi}{dt}$ of
circular geodesics, and their epicyclic frequencies $\Omega_{\rho}, \Omega_z$, as well as the
change of the energy of circular orbits per logarithmic change of orbital frequency $\Delta\tilde{E}$.
All these quantities, calculated by means of the two-soliton analytic metric, fitted with good accuracy
the corresponding numerical ones as in previous analogous comparisons
(although previous attempts were restricted to neutron star models with either high or low rotation rates).
We believe that this particular analytic solution could be considered as an analytic faithful representation of the
gravitation field of any rotating neutron star with such accuracy,
that one could explore the interior structure of a neutron star by using this space-time to interpret observations
of astrophysical processes that take place around it.
\end{abstract}


\section{Introduction}

The amount and accuracy of modern observations in various parts of
the electromagnetic spectrum has increased dramatically. In order
to give astrophysically plausible explanations of the various
problems related to the observations we have to rely on
theoretical assumptions that are at least as accurate as the data
we are trying to analyze. There is a large class of observations
(see e.g., \cite{derKlis}) that are related to the astrophysical environment
of compact relativistic objects (AGNs, LMXBs, etc).
Furthermore, the anticipated successful gravitational wave detection will open a new window to
observe such objects. In order to understand these phenomena, one has to have a sufficiently
accurate analytic description of the spacetime around such compact objects.
If the central object is a black hole, there is a unique
choice in the framework of general relativity: the Kerr spacetime.
On the other hand the geometry around a rotating neutron star is
much more complicated, since it depends on many parameters related
to the internal structure of the neutron star and the way it rotates.

The assumption that the geometry around such an object is approximately that of
a Schwarzschild, or a Kerr metric (see for example \cite{derKlis})
is very simplistic and it may lead to erroneous conclusions about the actual
astrophysical processes that take place in the close neighborhood
of the star itself (cf.~\cite{PappasQPOs,PachonNew}).

One can alternatively rely on numerical codes that are able to
describe the geometry around a realistic neutron star in a tabular
form on a given grid with sufficiently high accuracy. There are various groups
(see \cite{Sterg}, and for an extended list of numerical schemes
see \cite{Lrr}), that have acquired expertise in building relativistic
models of astrophysical objects with adjustable physical
characteristics and constructing the metric inside and outside
such objects by solving numerically the full Einstein equations in stationary, axisymmetric cases.

Although studying astrophysical phenomena in a geometric
background that has been constructed numerically is plausible,
there are certain drawbacks in using such metrics: (i)
Computing various physical quantities of a system, like the
orbital frequencies, or the innermost circular orbit, from a
metric that is given in a tabular form is not very practical and
is often plagued by numerical errors.
(ii) Astrophysical observations from the environment of a compact
object could be used to read the physical parameters that are
related to the structure of the compact object such as its mass,
its equation of state, its rotation, or to obtain the law of its differential
rotation, etc. This would be very difficult to achieve with a
metric that is numerically constructed. Solving the inverse
problem by means of  a numerical metric is a blind process that can not
be easily led by physical insight. Instead an analytic expression for the
corresponding spacetime would be much more preferable.

There are various analytic metrics that have been used in the past
to describe the exterior geometry of a neutron star. As mentioned
above, the Schwarzschild metric is not accurate enough for
rotating neutron stars, while the Kerr metric is good only for a
collapsed object (a black hole) but it fails to describe the
exterior of a neutron star, as comparisons of Kerr with numerical
geometries of rotating neutron stars by \cite{BertSter} have shown.

The Hartle-Thorne metric of \cite{HT}, which has been constructed as
an approximate solution of the vacuum Einstein equations (VEE)
for the exterior of a slowly rotating star, has been extensively used by
various authors to describe neutron stars of low rotation rate (see for example
\cite{Bertietal}). Finally, various other analytic solutions of VEE have been constructed and some of them have been
used, especially during the last decade, to describe the exterior geometries
of neutron stars (see \cite{Stuart,BertSter,Pachon,Pappas2,Teich}).
Such solutions are based on the formalism developed by \cite{ernst1,ernst2} which reformulates
Einstein equations in the case of axisymmetric, stationary
space-times. Manko et al. and Sibgatullin (see the articles of
\cite{sib1,SibManko,twosoliton,manko2,Mankoetal})
have used various analytic methods to produce such space-times
parameterized by various parameters that have different physical
context depending on the type of each solution.

Such an analytic solution, with its parameters appropriately adjusted to
match numerical models of neutron stars, could then be used to describe the
stationary properties of the space-time around the neutron star itself;
that is, study the geodesics in the exterior of the neutron star.
More specifically, from the analytic solution we could obtain
bounds of motion for test particles orbiting the neutron star,
find the location of the innermost stable circular orbit (ISCO), compute the orbital frequency of the
circular orbits on the equatorial plane as well as the epicyclic
frequencies around it and perform any sort of dynamical analysis on the
geodesics (for example see \cite{Lukes}). These properties of the space-time could be used
to study quantitatively astrophysical phenomena that take place in the vicinity of
neutron stars, such as accretion discs. Inversely one could use the astrophysical observations
related to such phenomena to determine the parameters describing the analytic space-time and from
that acquire information for the central object.

The central issue with analytic metrics is whether one can find
solutions that are able to describe with sufficient faithfulness all kinds
of rotating neutron stars; either slowly or rapidly rotating ones, or even
differentially rotating ones.

One solution that has been recently used by \cite{Stuart} and later by
\cite{BertSter} to describe the exterior space-time of rotating
neutron stars is the three-parameter solution of \cite{Mankoetal} (also mentioned as Manko et al.).
Although this solution was shown to match quite well
the space-time of highly rotating neutron stars, it failed to match
the slowly rotating ones. The reason for failing to describe slow rotation is that in the zero angular momentum
limit, this particular solution has a non vanishing quadrupole
moment, while one would expect slowly rotating neutron stars
to be approximately spherically symmetric. This problem of the Manko et al.
solution was not considered disappointing by Berti and Stergioulas, since
the space-time around slowly rotating stars could be described equally well by the Hartle-Thorne approximation.

The three-parameter solution of Manko et al. is a special case of the
so called two-soliton solution, which was constructed by
\cite{twosoliton}. The two-soliton is a four-parameter analytic metric which,
contrary to the previous one, can be continuously reduced to a
Scwarzschild or a Kerr metric while it does not suffer from the
problematic constraints of the Manko et al. solution with respect to the
anomalous behavior of its quadrupole moment. Actually the first four multiple moments of the
two-soliton solution can be freely chosen. Of course the analytic form of the two-soliton solution is not as
compact as the Manko et al. solution, but this is the price one has to pay in
order to cover the whole range of the physical parameters of a neutron star with a single analytic metric.

In the present work, that constitutes the extension and completion
of preliminary results presented by \cite{Pappas2}, that were suitably
corrected with respect to the right extraction of the  multipole moments of the
numerical space-time as was recently demonstrated by \cite{PappasMoments},
we are using this two-soliton solution to describe the space-time
around a wide range of numerically constructed rotating neutron
stars. We use the numerical multipole moments to set the multipole
moments of the analytic space-time. Then we examine how well the
two metrics match each other. Moreover we have performed comparisons between
astrophysicaly relevant geometric quantities produced from the
numerical and the analytic space-times, like the position of the
innermost stable circular orbit (ISCO), the orbital frequencies,
the epicyclic frequencies that are related to slightly
non-circular and slightly non-equatorial orbits, and the change of
energy of the circular orbits per logarithmic change of the
orbital frequency, $\Delta\tilde{E}$. The overall picture is that the new metric matches the numerical one
with excellent accuracy for all rotation rates and all equations of state.

The rest of the paper is organized as follows: In Sec.~2
the proposed analytic solution (two-soliton) is briefly presented
and some of its properties are thoroughly analyzed. The parameter space of
the two-soliton is investigated and it is shown how to obtain the
limiting cases of Schwarzschild, Kerr and Manko et al.
A brief discussion of the physical properties of the space-time such as
the presence of singularities, horizons, ergoregions and regions
of closed timelike curves (CTCs) is also given. In Sec.~3 we
show how we match the analytic two-soliton solution to a specific numerical one
by matching  the first four multipole moments, and show why this is generally the best choice.
In Sec.~4 we discuss various criteria that could be used to compare the two metrics.
Finally, in Sec.~5 the final comparison criteria and the results of the corresponding
comparisons are presented. In Sec.~6 we give an overview of the conclusions obtained by our study.

\section{The two-soliton solution}

The vacuum region of a stationary and axially symmetric space-time can be described by the Papapetrou line element,
which was first used by \cite{Papapetrou},
\be
\label{Pap} ds^2=-f\left(dt-\omega d\phi\right)^2+
      f^{-1}\left[ e^{2\gamma} \left( d\rho^2+dz^2 \right)+
      \rho^2 d\phi^2 \right],
\ee
where $f,\;\omega,$ and $\gamma$ are functions of the
Weyl-Papapetrou coordinates ($\rho,z$). By introducing the complex
potential $\mathcal{E}(\rho,z)=f(\rho,z)+\imath\psi(\rho,z)$,
\cite{ernst1} reformulated the Einstein field equations for this type of
space-times in a concise complex equation
\be \label{ErnstE}
(Re(\mathcal{E}))\nabla^2\mathcal{E}=\nabla\mathcal{E}\cdot\nabla\mathcal{E}.
\ee

The real part of the Ernst potential $\mathcal{E}$ is the metric
function $f$, which is also the norm of the timelike Killing
vector related with stationarity, while $\psi$ is a scalar
potential related with the twist of the same vector.

A general procedure for generating solutions of the Ernst
equations was developed by \cite{sib1,SibManko,manko2,twosoliton}. Each
solution of the Ernst equation is produced from a choice of the Ernst potential along
the axis of symmetry of the metric in the form of a rational function
\be
\mathcal{E}(\rho=0,z)=e(z)=\frac{P(z)}{R(z)},
\label{Eernst}
\ee
where $P(z), R(z)$ are polynomials of $z$ of order $n$
with complex coefficients in general. The algorithm developed by
\cite{manko2} works as follows: First the Ernst potential along the axis is expressed in the form
\be
e(z)=1+\sum^n_{k=1}\frac{e_k}{z-\beta_k},
\label{eofz}
\ee
where $\beta_k$ are the roots of the polynomial $R(z)$ and $e_k$
are complex coefficients appropriately chosen so that the latter
form of $e(z)$ (eq.~\ref{eofz}) is equal to the former one (eq.~\ref{Eernst}).
Subsequently one determines the $2n$ roots of  equation
\be
\label{charpoly}
e(z)+e^*(z)=0,
\ee
where $^*$ denotes  complex conjugation. These roots are denoted as $\xi_k$, with $k=1,2,\ldots,2n$
and from these one defines the $2n$ complex functions $R_k=\sqrt{\rho^2+(z-\xi_k)^2}$. All these
functions and roots are then used as building blocks for the following determinants
\be E_\pm = \left| \begin{array}{cccc}
  1 & 1 & \cdots & 1\\
  \pm1 & &  &\\
   \vdots &  & &\\
   \pm1 & & &\\
     & & \bf{C} &  \\
  0 & & &\\
  \vdots &   & &\\
  0 & & &\\
\end{array}\right|\\
\ee
\be G = \left| \begin{array}{cccc}
  0 & R_1+\xi_1-z & \cdots & R_{2n}+\xi_{2n}-z \\
  -1 & & &  \\
   \vdots & & & \\
  -1 & &  &  \\
    & & \bf{C} &  \\
   0 & & &  \\
  \vdots & & & \\
  0 & & &
\end{array}\right|\ee
\be H = \left| \begin{array}{cccc}
  z & 1 & \cdots & 1 \\
  -\beta_1 & & &  \\
 \vdots & & & \\
 -\beta_n & &  &  \\
   & & \bf{C} &  \\
  e^*_1 &  &  & \\
   \vdots & & & \\
  e^*_n & & &
\end{array}\right|\ee
\be K_0 = \left| {{\begin{array}{ccc}
 {\frac{1 }{\xi _1 - \beta _1 }}  & \cdots  &
{\frac{1}{\xi _{2n} - \beta _1 }}  \\
  \vdots  & \ddots  & \vdots  \\
  {\frac{1}{\xi _1 - \beta _n }}  & \cdots  &
{\frac{1}{\xi _{2n} - \beta _n }}  \\
{\frac{e^*_1}{\xi _1 - \beta _1^\ast }}  & \cdots  &
{\frac{e^*_1}{\xi _{2n} - \beta _1^\ast }}  \\
 \vdots  & \ddots  & \vdots  \\
  {\frac{e^*_n}{\xi _1 - \beta _n^\ast }}  & \cdots  &
{\frac{e^*_n}{\xi _{2n} - \beta _n^\ast }}  \\
\end{array} }} \right|\ee
where $\bf{C}$ is the $2n\times2n$ matrix
\be
\bf{C} = \left(
{{\begin{array}{ccc}
 {\frac{R_1 }{\xi _1 - \beta _1 }}  & \cdots  &
{\frac{R_{2n} }{\xi _{2n} - \beta _1 }}   \\
  \vdots  & \ddots  & \vdots  \\
  {\frac{R_1 }{\xi _1 - \beta _n }}  & \cdots  &
{\frac{R_{2n} }{\xi _{2n} - \beta _n }}  \\
{\frac{e^*_1}{\xi _1 - \beta _1^\ast }}  & \cdots  &
{\frac{e^*_1}{\xi _{2n} - \beta _1^\ast }}  \\
 \vdots  & \ddots  & \vdots  \\
  {\frac{e^*_n}{\xi _1 - \beta _n^\ast }}  & \cdots  &
{\frac{e^*_n}{\xi _{2n} - \beta _n^\ast }}  \\
\end{array} }} \right).
\ee
The Ernst potential and the metric functions are finally expressed in terms
of the determinants given above as:
\bea
\label{metricfunc1}
{\mathcal E}(\rho ,z) &=& \frac{E_ + }{E_ - },\\
f(\rho ,z) &=& \frac{E_ + E_- ^\ast + E_+^\ast E_- }{2E_ - E_-^\ast },\\
e^{2\gamma (\rho ,z)}&=& \frac{E_ + E_-^\ast + E_+^\ast E_ -
}{2K_0 K_0^\ast
\prod\nolimits_{k = 1}^{2n} {R_k } }, \\
\omega (\rho ,z) &=& \frac{2 \; \Im\left[ {E_- H^\ast - E_-^\ast G}
\right]}{E_ + E_-^\ast + E_+^\ast E_ - } .
\label{metricfunc2}
\eea

We should note that due to the form of the metric functions, the
parameters $e_k$ and their complex conjugates $e^*_k$ that appear in the
determinants cancel out (the $\prod_{k=1}^n e_k e^*_k$ is a common
factor of all products of determinants that show up in the metric
functions), so they do not affect the final expressions.

The vacuum two-soliton solution (proposed by \cite{twosoliton}) is
a special case of the previous general axisymmetric solution that
is obtained from the ansatz (see also \cite{Sotiriou})
\be
\label{2soliton}
e(z)=\frac{(z-M-ia)(z+ib)-k}{(z+M-ia)(z+ib)-k},
\ee
where all the parameters $M, a, k, b$ are real. From the
Ernst potential along the axis one can compute the mass and
mass-current moments of this space-time. Particularly, for the
two-soliton space-time the first five non-vanishing moments are:
\bea
\label{moments2}
M_0&=&M,\quad  M_2=-(a^2-k)M,\nn\\
 M_4&=&\left[ a^4 - (3a^2 -2ab + b^2)k + k^2 - \frac{1}{7}kM^2\right] M\nn\\
J_1&=& aM ,\quad J_3=- [a^3 -(2a - b)k]M.
\eea
The mass moments of odd order and the mass-current moments of even
order are zero due to reflection symmetry with respect to the
equatorial plane $(z=0)$ of the space-time (this is actually
ensured by restricting all parameters of eq.~(\ref{2soliton}) to assume real values).
From the moments we see that the parameter $M$ corresponds to the mass monopole of the
space-time, the parameter $a$ is the reduced angular momentum, $k$
is the deviation of the reduced quadrupole from the corresponding
Kerr quadrupole (the one that has the same $M$ and $a$),
and $b$ is associated with the deviation of the current octupole moment from the current octupole of the
corresponding Kerr.

\begin{table*}
 \centering
 \begin{minipage}{0.7\textwidth}
  \caption{Classification of the various two-soliton solutions depending on the values of the parameters
  $d,~\kappa_{\pm},~\xi_{\pm},~R_{\pm}$ and $r_{\pm}$. The table also shows the various conjugation relations
  between the parameters. The types of solutions that have degeneracies are indicated with an asterisk ($^*$).
  $\Re$ means real, $\Im$ means imaginary, and ${\mathbf C}$
  means complex.}
  \centering
\begin{tabular}{||c||c|c|c|c|c|c|c|c|c||}\hline\hline
  Case   & $d^2$ & $\kappa_+^2$ & $\kappa_-^2$ & $\xi_+$    & $\xi_-$    &  $R_+$     & $R_-$      & $r_+$      & $r_-$ \\
  \hline\hline
  $Ia$   & $>0$  & $>0$         & $>0$         &  $\Re$     & $\Re$      &  $\Re$     & $\Re$      &  $\Re$     & $\Re$  \\ \hline
  $Ib^*$ & $>0$  & $>0$         & $=0$         &  $\Re$     & $\Re$      &  $\Re$     & $\Re$      &  $\Re$     & $\Re$  \\
         &       &              &              &            &$=\xi_+$    &$\quad\quad\quad$&       & $=R_+$     &$=R_-$  \\
         \hline\hline
  $IIa$  & $>0$  & $>0$         & $<0$         &$\mathbf{C}$&$\mathbf{C}$&$\mathbf{C}$&$\mathbf{C}$&$\mathbf{C}$&$\mathbf{C}$  \\
         &       &              &              &$\quad\quad\quad$&$=(\xi_+)^*$&       &            &$=(R_+)^*$  & $=(R_-)^*$  \\ \hline
  $IIb^*$& $>0$  & $=0$         & $<0$         &$\Im$       &$\Im$       &$\mathbf{C}$&$\mathbf{C}$&$\mathbf{C}$&$\mathbf{C}$  \\
         &       &              &              &            &$=(\xi_+)^*$&            & $=(R_+)^*$ & $=R_-$     & $=R_+$  \\\hline
  $IIc$  & $>0$  & $<0$         & $<0$         &$\Im$       &$\Im$        &$\mathbf{C}$&$\mathbf{C}$&$\mathbf{C}$&$\mathbf{C}$  \\
         &       &              &              &            &            &            & $=(R_+)^*$ &            & $=(r_+)^*$
         \\\hline\hline
  $III$  & $<0$  & $\mathbf{C}$ & $\mathbf{C}$ &$\Re$       &$\Im$       &$\Re$       &$\Re$       &$\mathbf{C}$&$\mathbf{C}$  \\
         &       &           &$=(\kappa_+^2)^*$&            &            &            &            &            & $=(r_+)^*$
         \\\hline\hline
  $IVa^*$& $=0$  & $>0$         &$=\kappa_+^2$ &  $\Re$     & $=0$       &  $\Re$     & $\Re$      &  $\Re$     & $\Re$  \\
         &       &              &              &$=\kappa_+$ &            &            &            &            & $=r_+$  \\\hline
  $IVb^*$& $=0$  & $=0$         &$=0$          &  $=0$      & $=0$       &            &      -     &    -       &  -   \\\hline
  $IVc^*$& $=0$  & $<0$         &$=\kappa_+^2$ &  $\Im$     & $=0$       &$\mathbf{C}$&$\mathbf{C}$&  $\Re$     & $\Re$  \\
         &       &              &              &$=\kappa_+$ &            &            &            &            & $=r_+$
         \\\hline\hline
\end{tabular}
  \end{minipage}
\end{table*}

For the two-soliton ansatz (\ref{2soliton}), the characteristic equation (\ref{charpoly}) takes the form,
\be
\label{poly}
z^4-(M^2-a^2-b^2+2k)z^2+(k-ab)^2-b^2M^2=0.
\ee
Since the coefficients of the polynomial are real, the roots can be either real, or conjugate pairs. The four roots of
(\ref{poly}) can be written as,
\be
\xi_1=-\xi_3=\xi_+,\;\;\xi_2=-\xi_4=\xi_-,
\ee
where,
\be
\xi_{\pm}=\frac{1}{2}\left(\kappa_+\pm\kappa_-\right),
\ee
with
\be
\kappa_{\pm}=\sqrt{M^2-a^2-b^2+2k\pm2d},
\ee
and
\be
d=\sqrt{(k-ab)^2-b^2M^2}.
\ee
Using these symbols for the four roots we redefine the four corresponding functions $R_k$ as
\be
R_{\pm}=\sqrt{\rho^2+(z\pm
\xi_+)^2},\;r_{\pm}=\sqrt{\rho^2+(z\pm \xi_-)^2}.
\ee

Next we proceed to classify the various types of solutions depending on whether the four roots
have real, purely imaginary or complex values. This classification is outlined in Table 1.

\begin{enumerate}
\renewcommand{\theenumi}{$\bullet$ Case \Roman{enumi}:}
 \item This case is characterized by two real roots $\xi_{\pm}$.
 The Kerr family of solutions, which corresponds to
$k=0$ is definitely not included in this family of solutions.
\begin{enumerate}
\renewcommand{\theenumii}{\Roman{enumi}\alph{enumii}:}
\item This subfamily of case I is the simplest to compute, since all
functions $R_{\pm},r_{\pm}$ are real.
\item This is a degenerate case where the roots $\xi_{\pm}$ coincide.
The degeneracy is due to  $\kappa_-$ being zero
 which corresponds to the parameter constraint $M^2-a^2-b^2+2k-2d=0$.
 In such degenerate cases, the computation of the metric function is not straightforward since
 the expressions for the metric become
 indeterminate, and a limiting procedure should then be applied. In the
reduced-parameter space $({a}/{M},{b}/{M},{k}/{M^2})$, the previous
constraint corresponds to a two-dimensional surface. 
 \end{enumerate}
 \item In this case, the roots $\xi_{\pm}$ are either complex or
 imaginary, since $\kappa_-^2<0$. Furthermore this means that there
 are nonvanishing values of $(\rho,z)$ where the functions $R_{\pm},r_{\pm}$
 assume zero value,  which then leads to singularities at the corresponding points.
\begin{enumerate}
\renewcommand{\theenumii}{\Roman{enumi}\alph{enumii}:}
\item  This sub-case, as with Case I, belongs to a class of solutions that
cannot have a vanishing parameter $k$.
\item Here, a degeneracy shows up again as in Case Ib, which admits
the same treatment (limiting procedure) as in the former situation. Contrary to all previous cases, Case IIb admits
 a Kerr solution that belongs to the hyper-extreme branch  ($|a|>M$).
 Similarly to Case  Ib this solution is also represented
 by a two dimensional surface in the reduced-parameter space.
\item This case is similar to the previous one, without the degeneracy
in the roots $\xi_{\pm}$. It also includes hyper-extreme Kerr solutions.
\end{enumerate}
 \item In this case one of the $\xi_+, \xi_-$ is real while the other one is imaginary.
 Thus the same type of singularity issues,
 as in Case II arise. In particular such problematic behavior shows up on the $z=0$ plane.
 The Kerr and the  Schwarzschild solutions lie entirely within this family of solutions.

 \item All types of solutions belonging to this case are degenerate
 (there is a special constraint between the parameters) and
 as such are probably of no interest to realistic neutron stars. Subcases IVa and IVc have one double root
 ($\xi_-=0$), while subcase IVb has a quadruple root ($\xi_+=\xi_-=0$)
 and the computation of the metric functions needs special treatment. We should also note that
 cases IVb and IVc include the extreme Kerr solution ($|a|=M$) as a special case.
\end{enumerate}

As we can see from the classification, the two-soliton solution
can produce a very rich family of analytic solutions with the
classical solutions of Schwarzschild and Kerr being special cases
of the general solution. Also the Manko et al. solution of \cite{Mankoetal} that has been
used previously by \cite{BertSter,Stuart} to match the exterior
space-time of rotating neutron stars is a special case of the two-soliton solution
as we will see next.

\begin{figure*}
\centering
\includegraphics[width=.32\textwidth]{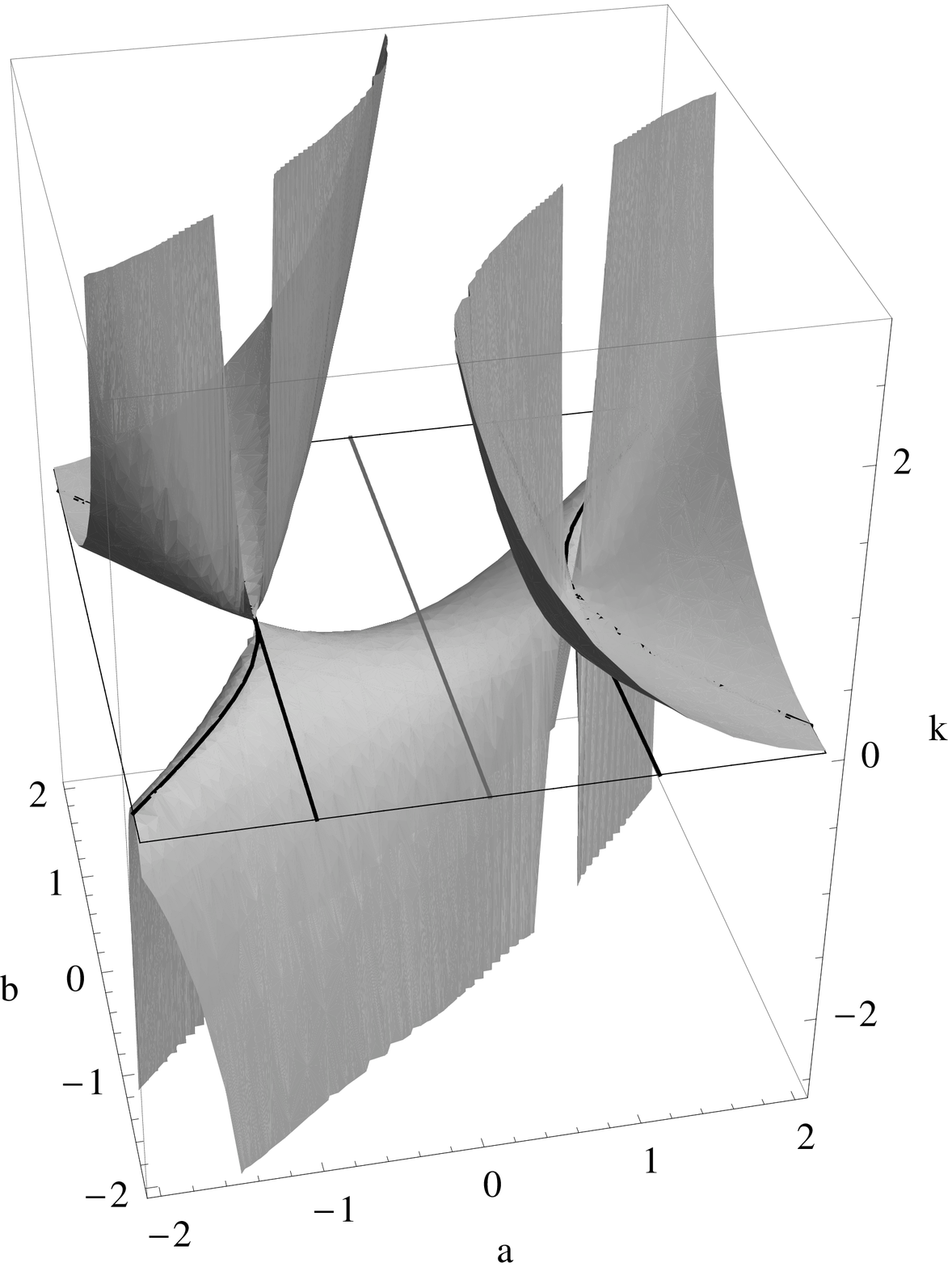}
\includegraphics[width=.32\textwidth]{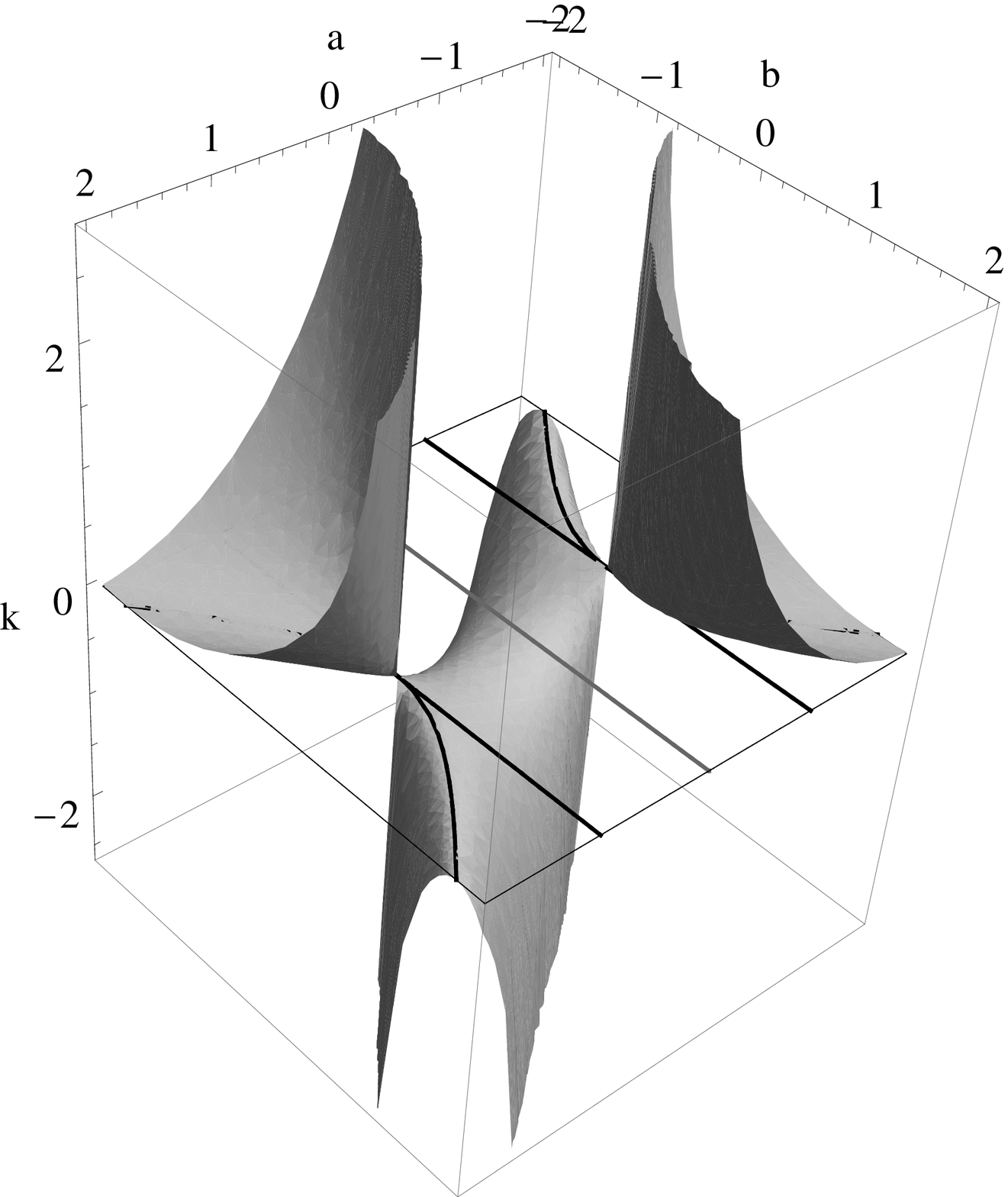}
\includegraphics[width=.32\textwidth]{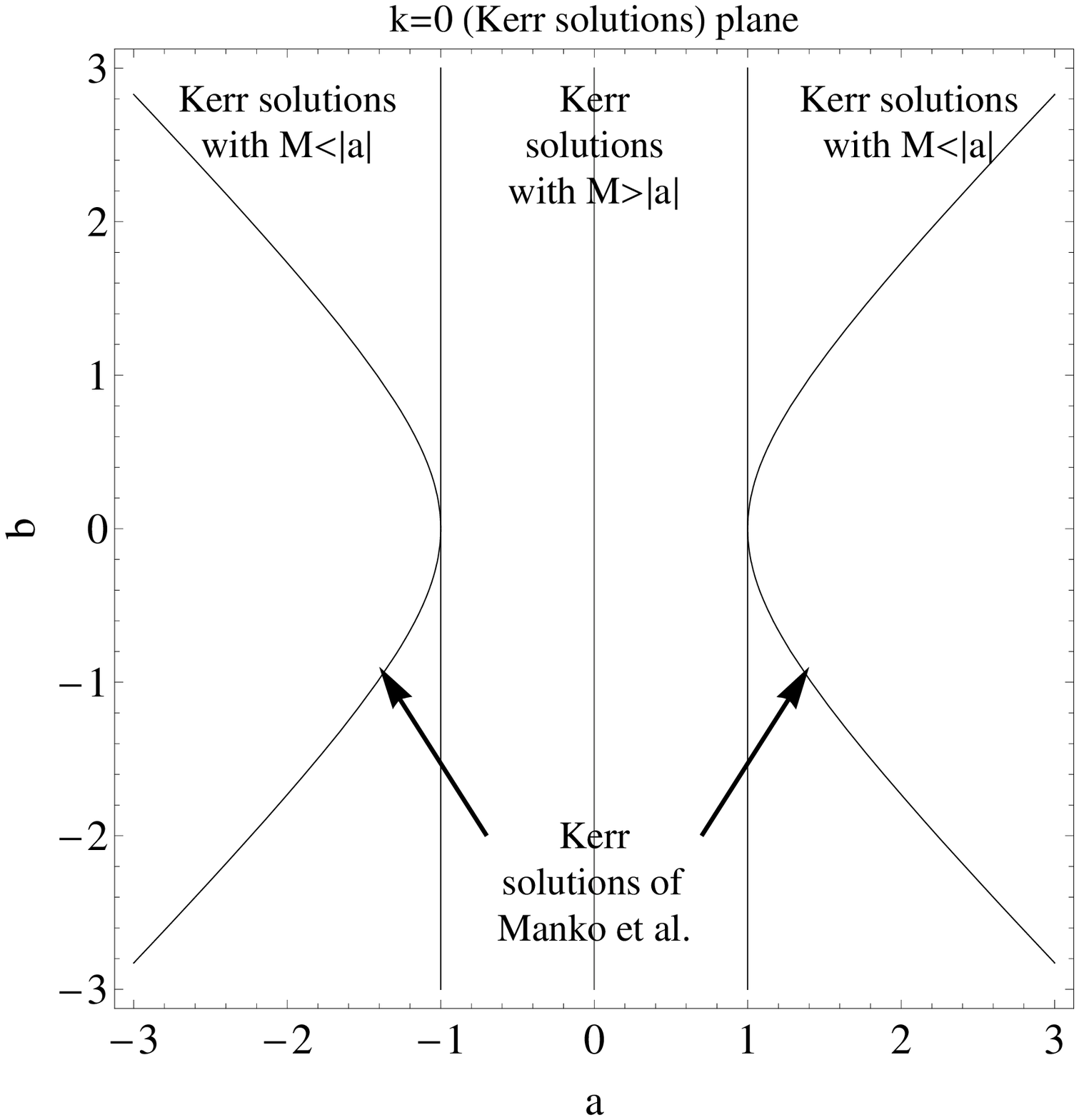}
\caption{The first two plots show the parameter space $(a,b,k)$ of
the two soliton solution for a particular mass ($M=1$) from two
different view points so that the intricate foldings of the surface are better
understood. The 2D surface plotted is the constraint of
$k$ which corresponds to the solution of Manko et al. used by
Berti~\&~Stergioulas (2004). The third plot is the $k=0$ plane of the
parameter space which corresponds to all Kerr solutions.
The plots clearly show that the Manko et al. solution has no set of parameters to describe the case  $k=0$
which corresponds to the Kerr and the Schwarzschild (for $a=0$) solutions, since there is no intersection of
the constraint of $k$ and the Kerr plane in the appropriate range of parameters.
The two hyperbolas plotted on the plane $k=0$ are the only points were the constraint of $k$ touches the plane
tangentially. As we can see these hyperbolas correspond to $|a| > M$, i.e., to hyper-extreme Kerr
space-times.}
\label{parspacefig}
\end{figure*}

All types of solutions discussed above can be represented in a
three dimensional parameter space, the reduced-parameter space that was
mentioned in Case IIb. Although the two-soliton solution is
characterized by four parameters, one of them, the monopole mass
$M$, is simply a scaling parameter which can be used to reduce the
rest of the parameters to dimensionless ones. The three
dimensionless parameters thus formed,
$({a}/{M},{b}/{M},{k}/{M^2})$, are related to the
multipole moments (see eq.~(\ref{moments2})) of the corresponding
space-time in the following way: The first parameter ${a}/{M}$
is the spin parameter (where $a$ is the reduced angular momentum) which is the only
parameter, besides the mass, that uniquely characterizes a Kerr
space-time. The second parameter ${k}/{M^2}$ expresses the
deviation of the quadrupole moment of the solution from the
quadrupole moment of the corresponding Kerr (the one with the same
${a}/{M}$ value); an increase of the value of ${k}/{M^2}$
produces solutions that are less oblate than Kerr. The final
parameter ${b}/{M}$ controls in a linear fashion the current
octupole moment. The actual deviation of the two soliton octupole
moment from the Kerr octupole moment depends on all three
parameters ${a}/{M}$, ${k}/{M^2}$ and ${b}/{M}$.
Of course the higher moments are also affected by these parameters.

In this three dimensional parameter space, the plane
$k/M^2=0$ corresponds to all types of Kerr solutions. This
is clear from the form of the Ernst potential along the axis,
where if one sets $k=0$ it reduces to the Ernst potential of the
Kerr solution,
\be
e(z)=\frac{z-M-ia}{z+M-ia}.
\ee
Obviously in this case the parameter $b/M$ is
redundant; thus each line ${a}/{M}=\textrm{const}$, which is
parallel to the ${b}/{M}$ axis on the plane ${k}/{M^2}=0$,
corresponds to a single Kerr (modulo the mass of the black hole).

As mentioned in the introduction, the solution of \cite{Mankoetal}
has been used to describe the exterior of rotating neutron stars.
As it was briefly discussed above this solution is included in the two-soliton solution and
can be obtained by imposing a specific constraint on the
two-soliton parameters. The Manko et al. solution is obtained by
setting
\be
\label{thek}
k=-\frac{1}{4}\left[M^2-(a-b)^2\right]-\frac{M^2b^2}{M^2-(a-b)^2}+ab.
\ee
This constraint defines a surface in the three parameter space
$({a}/{M}, {b}/{M}, {k}/{M^2})$ (see
Figure \ref{parspacefig}). The particular solution, depending on the
values of $a, b$, falls under either Case Ib or Case IIb, where either
$\kappa_-$ or $\kappa_+$ is equal to zero, respectively. We should note
that the Manko et al. solution is the union of these two
cases. By substituting the above expression for $k$ (eq.~(\ref{thek}))
in the formula for the quadrupole moment (\ref{moments2}), the
quadrupole moment takes the following value when $a=0$,
\be
M_2=-\frac{M}{4}\frac{(M^2+b^2)^2}{M^2-b^2}.
\ee
This is why the quadrupole moment of the Manko et al. solution does
not vanish in the limit of zero rotation. From the above
expression one can see that the metric is not spherically
symmetric as one would expect for a non rotating object.
Especially for $|b|<M$ the metric is oblate while for $|b|>M$ the
metric is prolate.

\begin{figure*}
\centering
\includegraphics[width=.32\textwidth]{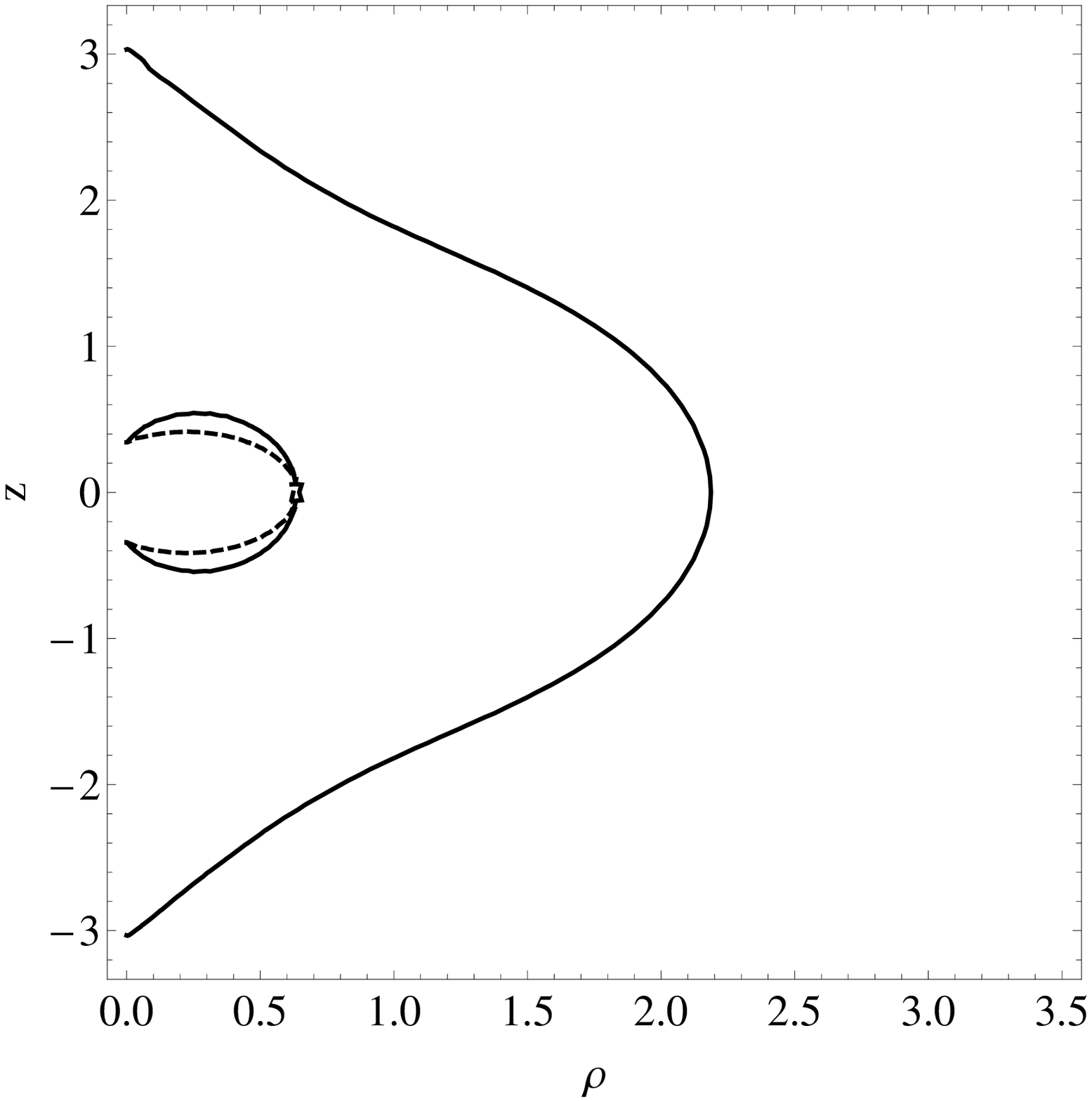}
\includegraphics[width=.32\textwidth]{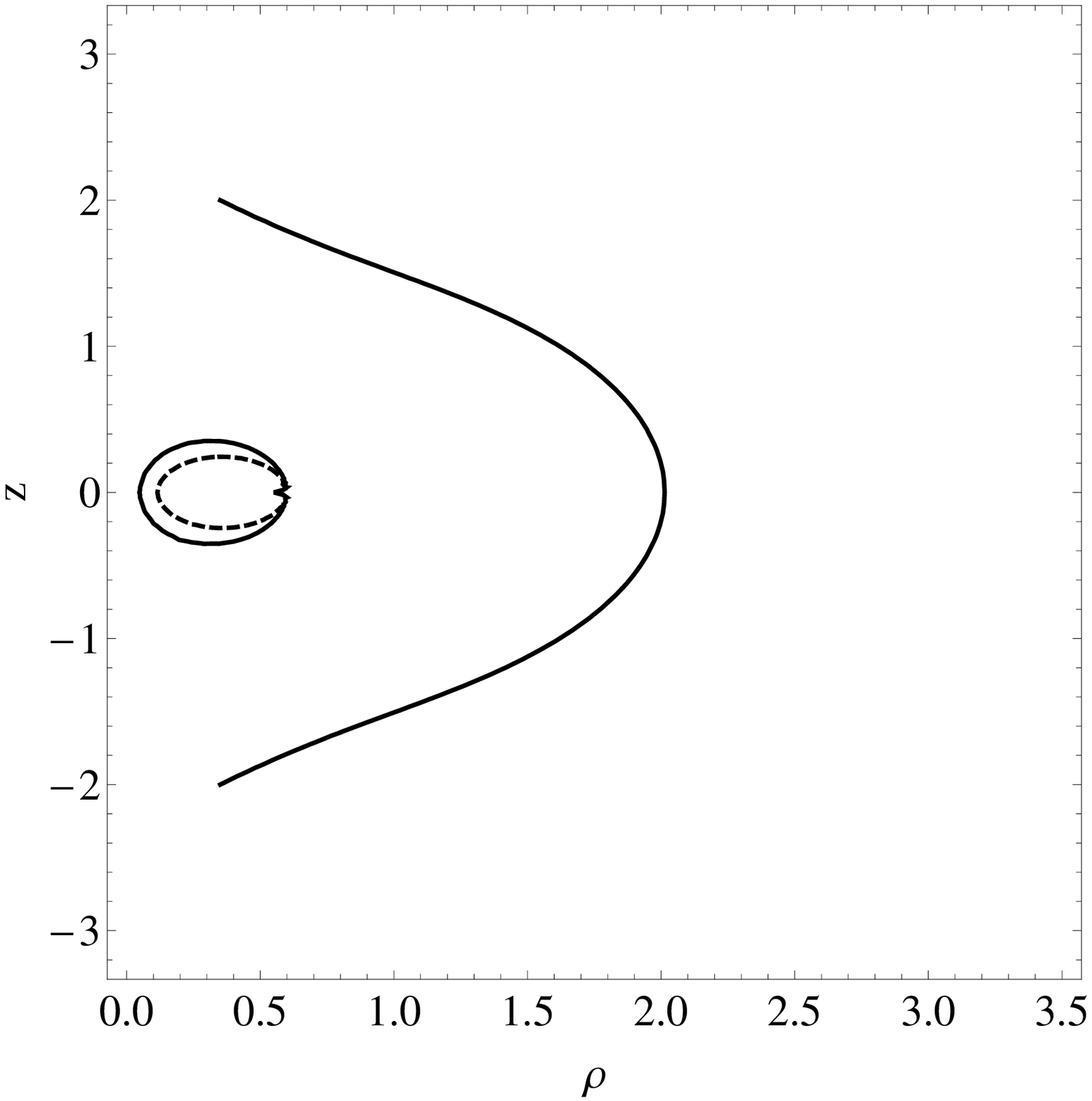}
\includegraphics[width=.32\textwidth]{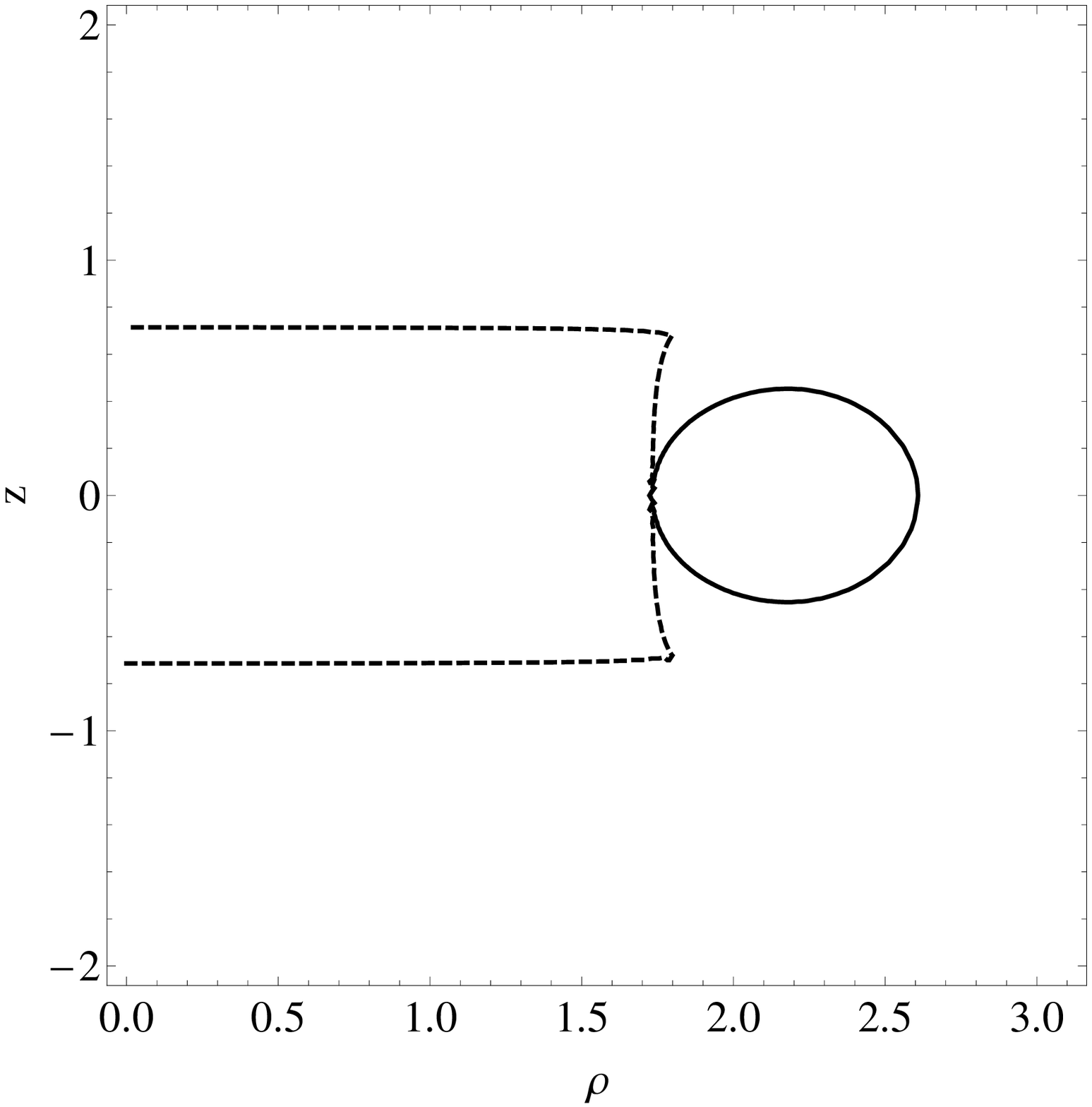}
\includegraphics[width=.32\textwidth]{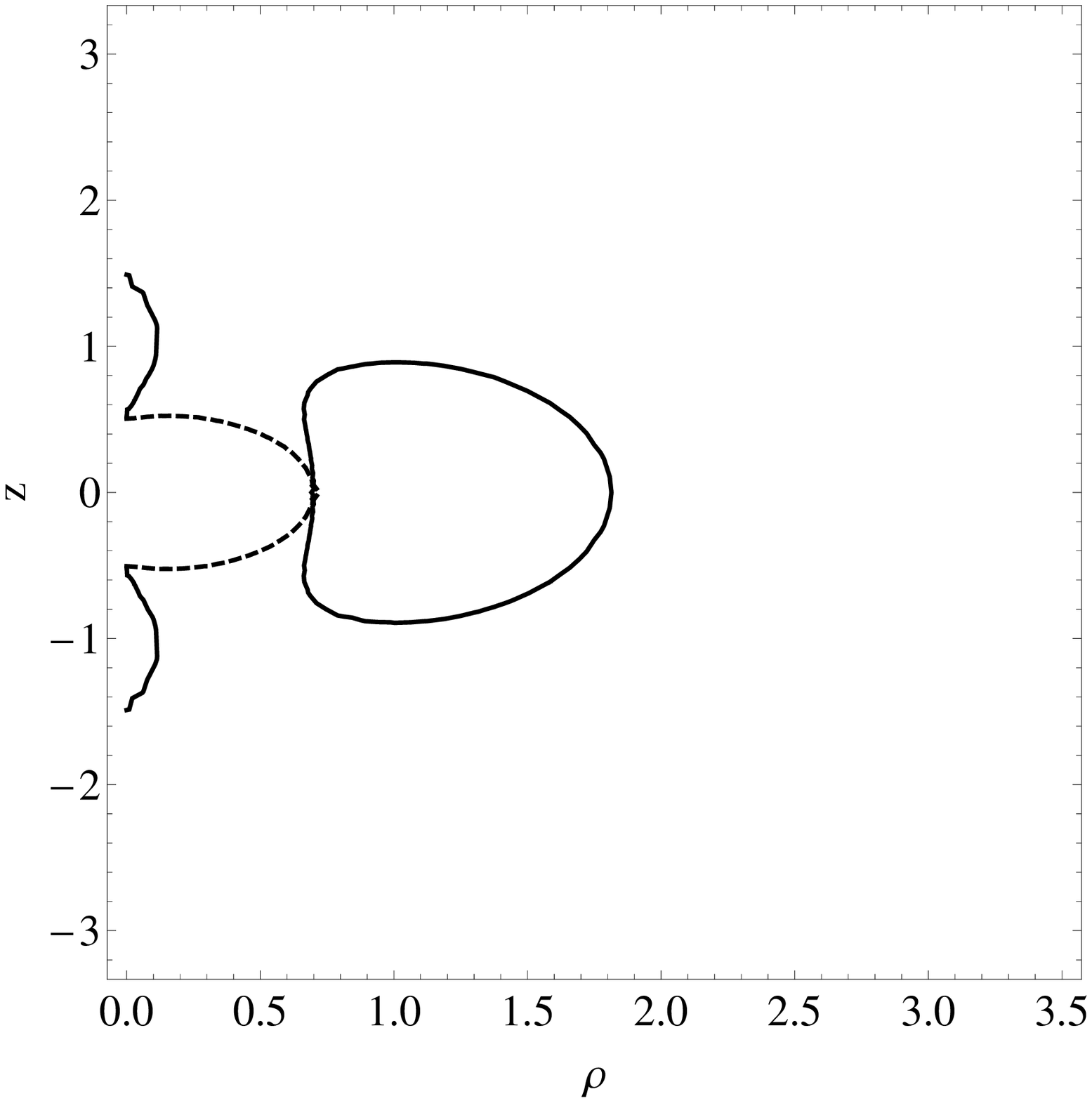}
\includegraphics[width=.32\textwidth]{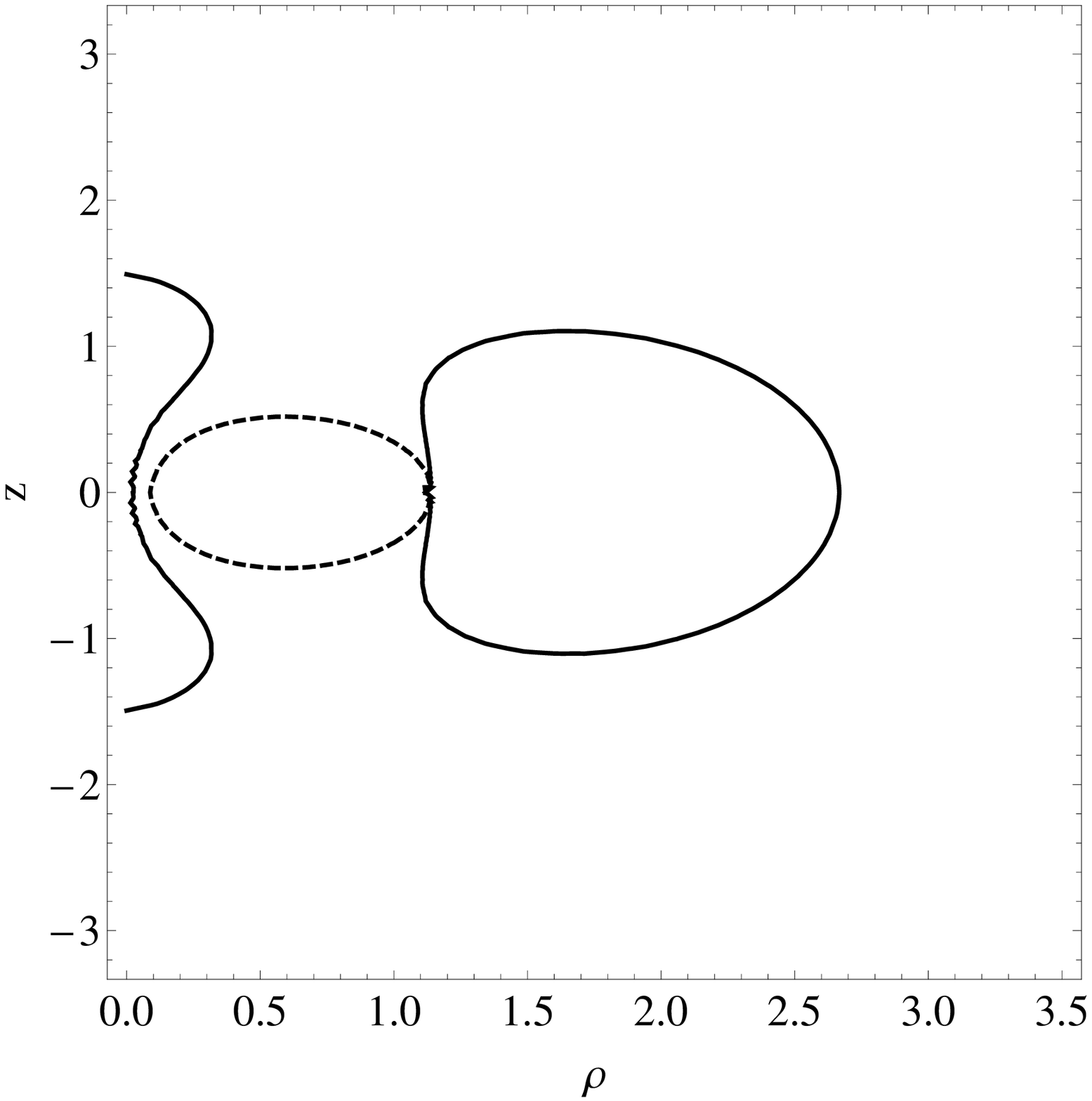}
\includegraphics[width=.32\textwidth]{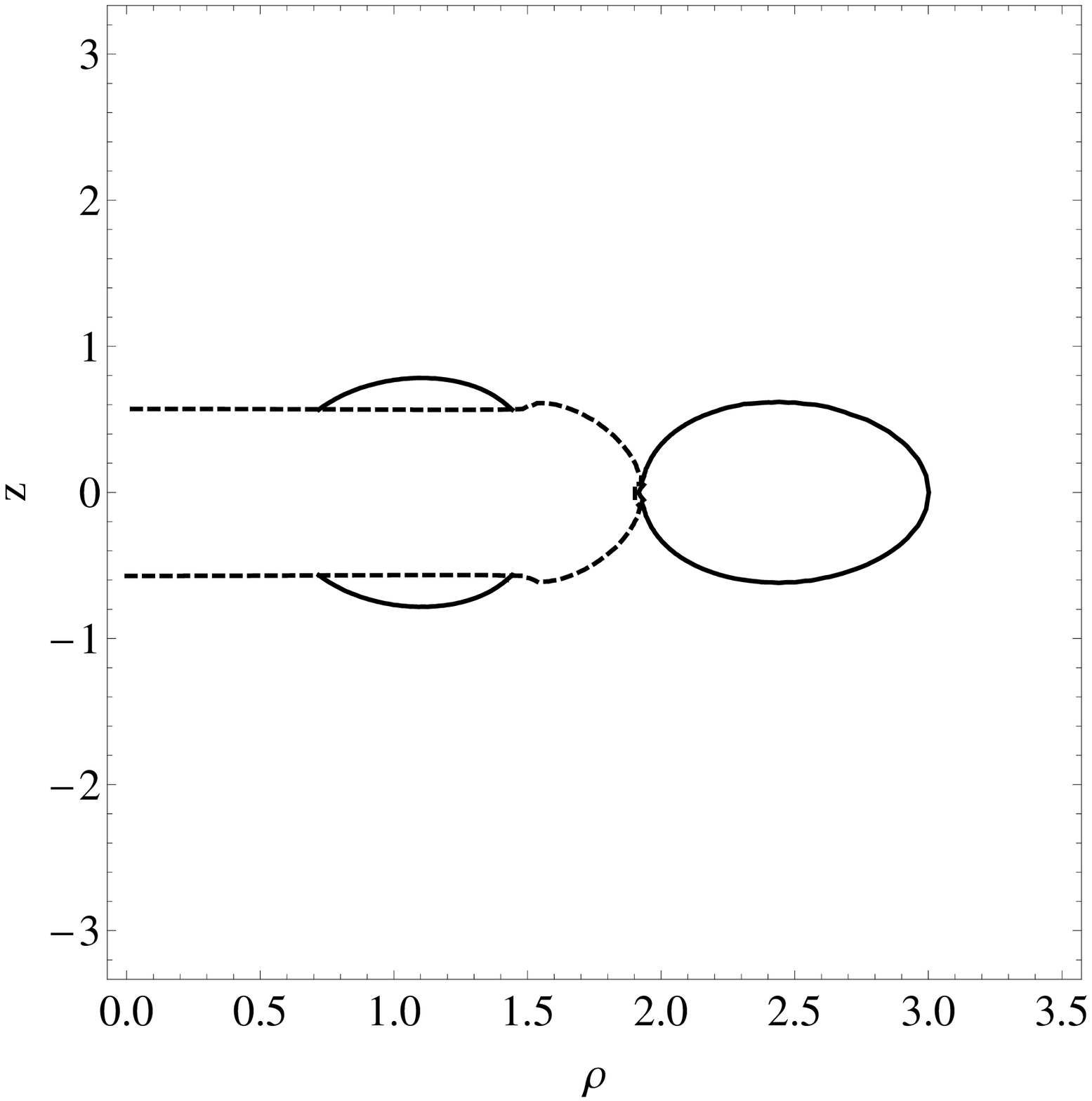}
\caption{Typical types of the surfaces that define the static
limit and the regions of CTCs for the various cases of the
two-soliton. On the left panels we have two typical figures that
correspond to Case Ia, in the middle panels the figures correspond to
Case IIa and on the right panels the figures correspond to Case III. The
solid curves correspond to the static limit and the dotted curves
correspond to the boundary of the regions with CTCs. For the Case
Ia, the region of CTCs is in contact with the axis of symmetry while
the same also applies to the surface of the static limit. For the Case
IIa the static limit is in contact with the axis, but the region
of CTCs has been detached. Finally for the Case III the inverse is true, that is
the region of CTCs is in contact with the axis of symmetry while the
static limit is detached from the axis. In all cases the upper plots correspond to slow rotation
while the lower plots correspond to faster rotation. As the rotation rate
increases the upper configurations are continuously deformed to the lower ones.}
\label{SLSsCTCs}
\end{figure*}
This anomalous behavior of the quadrupole moment is an important drawback for using the Manko et al. solution
to describe every rotating neutron star and it was pointed out by
\cite{BertSter}. In fact this analytic metric is good to match only rapidly rotating
neutron stars.

As shown by \cite{Mankoetal} this particular metric turns into a
Kerr metric if $b^2=a^2-M^2$. Since all the parameters are
assumed real, then this corresponds to a hyperextreme Kerr metric ($a
\geq M$). In Figure \ref{parspacefig} this is represented by the two
hyperbolas that lay outside the strip $|a|<M$ on the Kerr plane (the plane
$k=0$) along which the intricate surface of the Manko et al. solution
tangentially touches the corresponding plane.

The two-soliton solution, which we will thoroughly study later on,
is a much better metric to describe the
exterior of an arbitrary rotating neutron star than the Manko et al. solution because
(i) the former has 4 independent parameters
(compared to the 3 independent parameters of the latter one) that
offer more flexibility to adjust the metric, and (ii) these 4
parameters are able to cover the whole space of the first 4
moments of the space-time, while the first 4 moments of the latter
metric are actually correlated with each other through the
dependence of $k$ on the 3 independent parameters $M, a, b$,
that was mentioned previously.
Exactly this restriction renders the Manko et al. solution
inappropriate to describe slowly rotating neutron stars.

Before closing this section we will give a brief description
of the space-time characteristics of the two-soliton solution for the
range of parameter that we are going to use.

A horizon of a space-time is the boundary between the region where
stationary observers can exist and the region where such
observers cannot exist. For a stationary and axially symmetric
space-time, the stationary observers are those that have a
four-velocity that is a linear combination of the timelike and the
spacelike Killing vectors that the space-time possesses, i.e.
\be
u^\mu=\gamma (\xi^\mu+\Omega \eta^\mu)
\ee
where $\xi^\mu, \eta^\mu$ are the timelike and spacelike Killing fields
respectively and $\Omega$ is the observer's angular velocity. The factor
$\gamma$ is meant to normalize the four-velocity so that
$g_{\mu \nu}u^\mu u^\nu=-1$. In order for the four-velocity to be
timelike, $\gamma$ should satisfy the equation
\be
\gamma^{-2}=-g_{tt}-2\Omega g_{t\phi}-\Omega^2
g_{\phi\phi},
\ee
and it should be $\gamma^{-2}>0$, which corresponds to an $\Omega$
taking values between the two roots
\be
\Omega_{\pm}=\frac{-g_{t\phi}\pm\sqrt{(g_{t\phi})^2-g_{tt}g_{\phi\phi}}}{g_{\phi\phi}}.
\ee

This condition can not be satisfied when
$(g_{t\phi})^2-g_{tt}g_{\phi\phi}\leq 0$. Thus the condition
$(g_{t\phi})^2-g_{tt}g_{\phi\phi}= 0$ defines the horizon. In the
case of the two-soliton,  expressed in the Weyl-Papapetrou
coordinates, this condition corresponds to having $\rho=0$, since
$\rho^2=(g_{t\phi})^2-g_{tt}g_{\phi\phi}$. Thus the issue of
horizons is something that we will not have to face; in these
coordinates the whole space described corresponds to the exterior of any possible horizon.

Another issue is the existence of singularities.
Singularities might arise where the metric functions have
infinities. From equations (\ref{metricfunc1}-\ref{metricfunc2})
one can see that singularities might exist where the functions
$R_{\pm}=\sqrt{\rho^2+(z\pm \xi_+)^2},\,r_{\pm}=\sqrt{\rho^2+(z\pm
\xi_-)^2}$ go to zero, or where the determinant $E_-$ goes to zero,
or where $E_ + E_-^\ast + E_+^\ast E_ -$ goes to zero. Whether or
not some of these quantities vanish depends on which Case
the solution belongs to. A thorough investigation of the
singularities of the two-soliton is out of the scope of this
analysis; so we should only point out that for all the neutron star
models that we have studied and the corresponding parameters
of the two-soliton solution, any such singularities, when present, are always confined in the
region covered by the interior of the neutron star and thus
they do not pose any computational problems in our analysis.

The final issue is with regard to the existence of ergoregions
(i.e., regions where $g_{\mu \nu}\xi^\mu\xi^\nu=g_{tt}<0$) and regions with
closed timelike curves (CTCs)  (i.e., regions where $g_{ab}\eta^a\eta^b=g_{\phi\phi}<0$). In
Figure \ref{SLSsCTCs} we have plotted the boundary surfaces of
such regions for the two-soliton metric. One can see that there are three distinct topologies
for these surfaces that are observed for the different two-soliton Cases.
In any case though, for all  the models used here,
these surfaces are  again confined at regions where the interior of the neutron star lays.

In the following sections we analyze the method that we are going to use to
obtain the right values for the parameters of the two-soliton
solution for each neutron star model and compare its properties
with the corresponding numerical metric. In order to do that, we
have constructed several sequences of numerical neutron star
models with the aid of the RNS numerical code of \cite{Sterg}. The
numerical neutron star models used are the same models used by
\cite{PappasMoments} for demonstrating how to correct the numerical multipole
moments. They are produced using three equations of state (EOSs),
i.e., AU, FPS and L. The scope of using these models is twofold. First we are using them
to provide the appropriate parameters describing realistic
neutron stars in order to build the corresponding analytic
metrics. Then we use them as a testbed against which to
compare the analytic metrics and thus test their accuracy. As we
have already mentioned, we will use as matching conditions between the analytic
and the numerical metrics the first four non-zero
multipole moments. For the neutron star models that we have
studied,  the corresponding  analytic space-times
that are produced belong to three of the Cases of the
aforementioned classification, i.e., to Cases Ia, IIa and III.

\section{Matching the analytic to the numerical solution}

\begin{figure*}
\centering
\includegraphics[width=.32\textwidth]{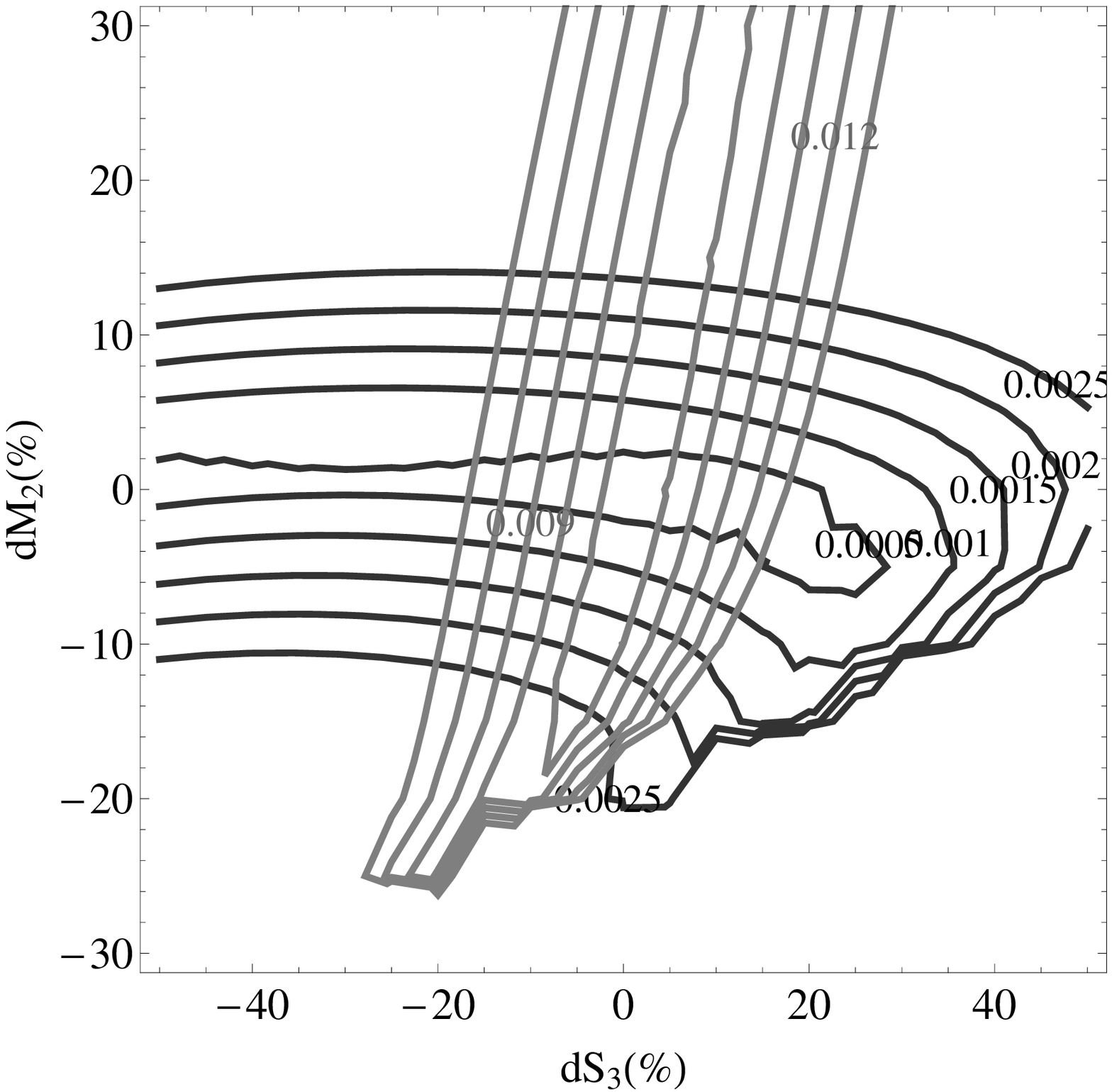}
\includegraphics[width=.32\textwidth]{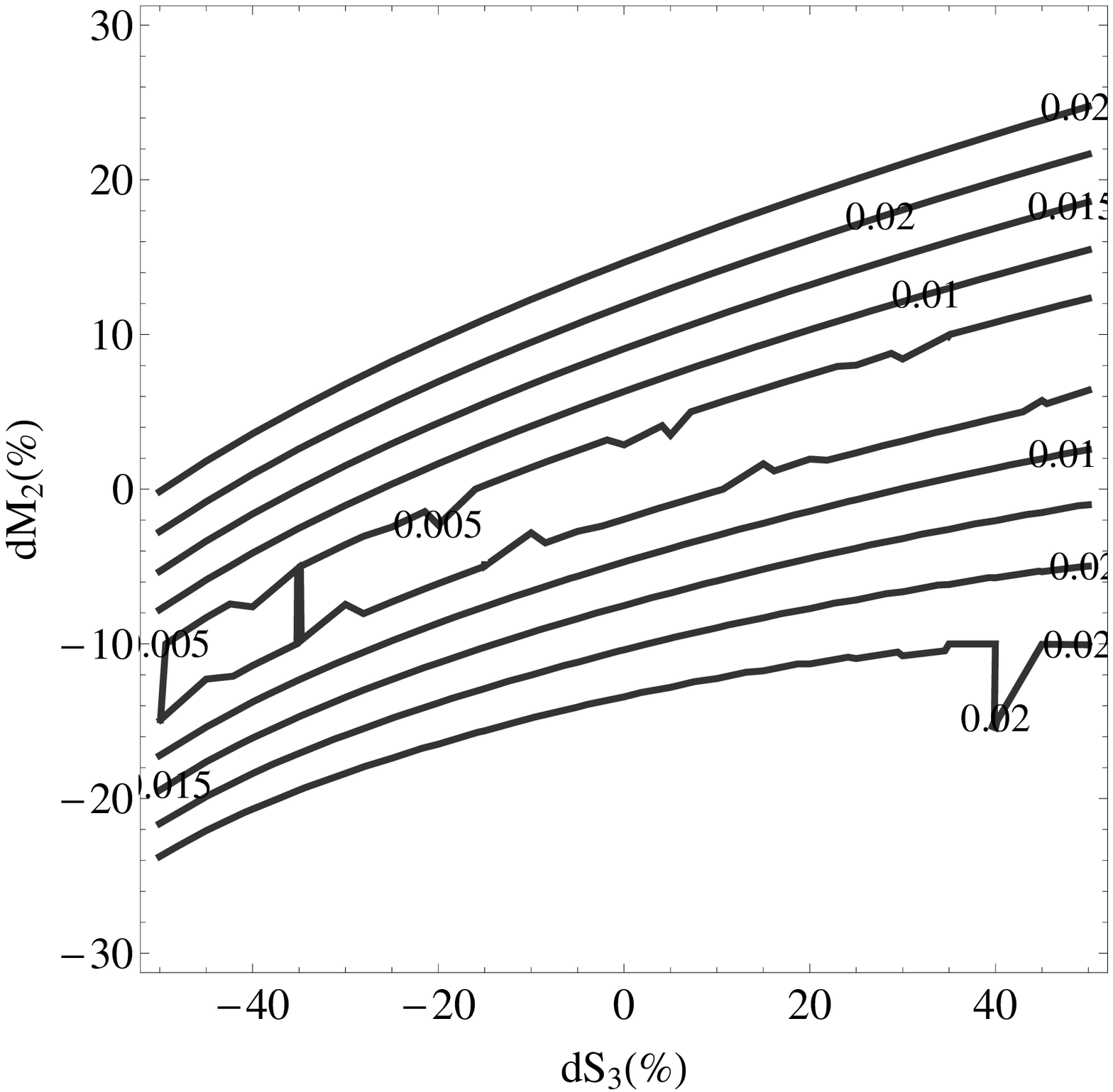}
\includegraphics[width=.32\textwidth]{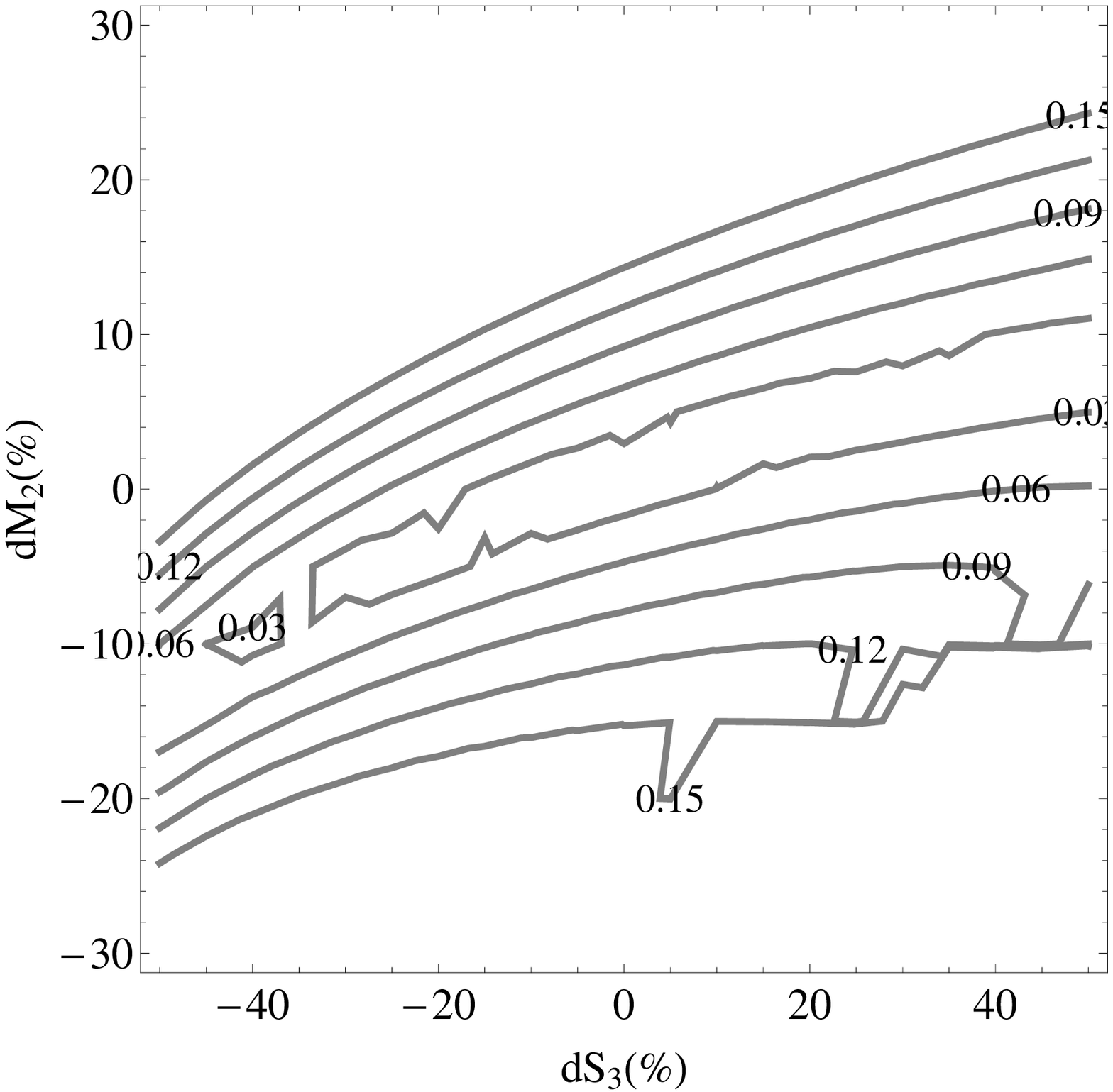}
\caption{Contour plots that point out what is the best choice for the
parameters of an analytic metric so that it matches well a numerical one. The
left plot shows the contour plots of the overall mismatch $\sigma_{ij}$
between the analytic and the numerical metric for the $tt$ (black curves) and
$t\phi$ components  (grey curves), respectively, as a function of the fractional deviation of
the quadrupole, $\delta M_2$, and the current octupole, $\delta S_3$, of the analytic metric
from those calculated directly from the
numerical metric (assuming the same mass and angular momentum though).
Since the contours of the $\sigma_{tt}$ (almost horizontal) are
orthogonal to the ones of $\sigma_{t\phi}$ (almost vertical), the
combination indicates an optimum choice for the multipole
moments of the analytic space-time. That choice is the moments that
have zero deviation from the moments of the numerical space-time.
The next two contour plots  are similar contour plots but the contours
correspond to different quantities. The  middle one is for the relative difference
between the analytic and numerical $R_{\rm{ISCO}}$, while the
right one is for the overall difference between the analytic
and numerical orbital frequency $\Omega$ (defined analogously to the
overall mismatch of the metric components). Both these latter plots are
consistent with the first one. All plots correspond to the model
$\# 15$ for the AU EOS of the models presented by Pappas~\&~Apostolatos (2012) and they give a representative
picture of what's happening with all numerical models that we have computed.}
\label{matching}
\end{figure*}

When one attempts to match an analytic solution to a numerical one, it is
desirable to find a suitable matching criterion
that would be characteristic of the whole structure of the particular
numerical space-time, instead of just a finite region of it.
That is, the matching  should be global and not local.
\cite{BertSter} have argued that a suitable
global condition should be the matching of the first few multipole moments.
Indeed, the full set of multipole moments (as defined relativistically
by \cite{geroch,hansen,fodor}) of a stationary and axially symmetric
space-time can fully specify the Ernst potential on the axis of symmetry. On the
other hand, when  the Ernst potential along the axis of symmetry is given,
there is a space-time which is unambiguously specified by that Ernst
potential as it was shown by \cite{XanthMoments,Xanth32,HauserV}. Thus, the
full set of multipole moments are uniquely characterizing a space-time and they can be used as a
global matching condition.

When the space-time of a neutron star model is constructed from a numerical algorithm,
one can evaluate its mass moments $M, Q,...$ and current moments $J, S_3,...$
with an accuracy depending on the grid used to present the
numerical metric (for further discussion see \cite{BertSter}, and
\cite{PappasMoments}). Practically, the first few numerically evaluated moments can be
used as matching conditions to the analytic
space-time. The first four nonzero multipole moments of the two-soliton solution
as a function of its parameters $M$, $a$, $b$, $k$ are shown in eq.~(\ref{moments2})
from which it is clear that once we specify the mass and the angular
momentum of the space-time, the parameter $k$ is uniquely
determined by the quadrupole moment $Q\equiv M_2$ ,while the parameter $b$
is uniquely determined by the current octupole $S_3\equiv J_3$. Thus,
having constrained the four parameters of the two-soliton, we have
completely specified  an analytic space-time that could be used to
describe the exterior of the particular neutron star model. What remains to be seen
is how well do the properties of the analytic space-time compare to those of the numerical one.

At this point there is an issue that should be addressed. Having
specified the first four non-zero moments of the two-soliton
metric, we have fixed all the higher moments of the space-time in a
specific manner related to the particular choice of the
analytic metric. These higher moments will probably
deviate from the ones of the numerical space-time. So the question
is, could one make a better choice when trying to match the analytic to
the numerical space-time than the one of setting  the first four
analytic moments exactly equal to the first four numerical moments? To answer
this question we have performed the following test. For several
numerical models of uniformly rotating neutron stars that we have constructed, we
formed a set of two-soliton space-times for each neutron star model
that have the same mass $M$ and angular momentum $J$ with the numerical model, while
the quadrupoles and the current octupoles of each single two-soliton space-time
take the values $M_2^{\rm (a) }=M_2^{\rm(n)}(1-\delta M_2 )$ and
 $S_3^{\rm (a) } =S_3^{\rm (n) }(1-\delta S_3 )$ respectively with various $\delta M_2$ and
$\delta S_3$ values. The quantities $\delta M_2$ and $\delta S_3$ denote the fractional differences of
the corresponding analytic moments of each two-soliton space-time from the numerical one. Then for
each one of these sets of moments we calculated the overall
mismatch between the analytic and the numerical metric functions,
which are defined (see \cite{PappasMoments}) as
\be
\sigma_{ij}=\left[ \int_{R_{\rm S}}^{\infty}(g_{ij}^n-g_{ij}^a)^2dr \right]^{1/2},
\label{sigmamis}
\ee
where $R_{\rm S}$ is the radius $r$ at the surface of the star,
and have thus constructed contour plots of $\sigma_{ij}$ on the plane
of $\delta S_3$ and $\delta M_2$, like the ones shown in Figure \ref{matching}.
The same type of contour plots were drawn for other quantities as
well, like the relative difference of the $R_{\rm ISCO}$ and
the overall difference between the analytic and numerical orbital
frequency $\Omega$ for circular, equatorial orbits (defined in the
same fashion as it was defined the overall metric mismatch in eq.~(\ref{sigmamis})).
The contour plots for the particular neutron star model that is shown in
Figure \ref{matching} are typical of the behavior that we observed in
all models. An important result is that the contours for the
overall mismatch $\sigma_{tt}$ combined with the contours for
$\sigma_{t\phi}$ give us the best choice for matching a numerical
to an analytic space-time; namely, the equation of the first four multipole
moments between the two space-times. That is because $g_{tt}$ seems to be
sensitive mainly in deviations from the numerical quadrupole and thus the
contours appear to be approximately horizontal and parallel to the axis of
$\delta S_3$, while $g_{t\phi}$ seems to be sensitive mainly in deviations
from the numerical current octupole and thus the contours appear
to be approximately vertical and parallel to the axis of $\delta M_2$
(the contours of $\sigma_{t \phi}$ are almost orthogonal to the contours of $\sigma_{tt}$).

We should note here that the exact position of the optimum point in the contour plots of
$\sigma_{tt},\sigma_{t\phi}$ did not deviate from $(0,0)$ by more than 3-4 per cent
in all cases studied. The largest deviations showed up in some of the fastest rotating models.

The conclusion is that what we expected to be true based on
theoretical considerations turns out to be exactly the case after the
implementation of the aforementioned test. Therefore in what follows, we
will set the first four non-zero multipole moments of the analytic space-time
equal to that of the corresponding numerical space-times.

\section{Criteria for the comparison of the analytic to the
numerical space-time}

Once we have constructed the analytic metric, appropriately matched to the
corresponding numerical one, we proceed to thoroughly compare the
two space-times. In order to do that, we should try again
to use criteria that are characteristic of the  geometric structure
of the whole space-time, and  if possible  coordinate independent.
It would be preferable if these criteria are also
related to quantities that are relevant to astrophysical observations. Thus, if two space-times are in good
agreement, with respect to these criteria, they could be considered
more or less equivalent.
On the other hand, such criteria, as well as possible
observations associated to them, could be used to distinguish
different space-times and consequently different compact
objects that are the sources of these space-times.

As a first criterion  of comparing the metrics we will use the
direct comparison between the analytic and the numerical metric
components themselves. Although the metric components are quantities that are not
coordinate independent, they have specific physical meanings and can
be related to observable quantities. Thus, the $g_{tt}$ component
is related to the gravitational redshift of a photon and the
injection energy of a particle. The $g_{t\phi}$ component is
related to the frame dragging effect, and the angular velocity
$\omega=-g_{t\phi}/g_{\phi\phi}$ of the zero angular momentum observers (ZAMOs).
Finally the $g_{\phi\phi}$ component is related to the circumference of a circle at a particular radial
distance and defines the circumferential radius
$R_{\rm circ}=C/2\pi=\sqrt{g_{\phi\phi}}$. Also $g_{\phi \phi}$ together with $g_{rr}$
are used to measure surface areas. So, if the relative difference
between the numerical and the analytic metric components
\bea
&&g_{tt}=-f,\quad g_{t\phi}=f\omega, \nn \\
&&g_{\phi\phi}=f^{-1}\rho^2-f\omega^2, \quad
g_{\rho\rho}=g_{zz}=f^{-1}e^{2\gamma}
\eea
is small, then one could consider the analytic metric as a good approximation of the numerical metric.

Another criterion for comparing an analytic to a numerical space-time is the location of the innermost stable
circular orbit (ISCO). Particles moving on the equatorial plane are governed by the equation of motion
(see for example \cite{Ryan})
\be
\label{Veff4}
-g_{\rho\rho}\left(\frac{d\rho}{d\tau}\right)^2=1-
\frac{\tilde{E}^2g_{\phi\phi}+2\tilde{E}\tilde{L}g_{t\phi}+\tilde{L}^2g_{tt}}{\rho^2}\equiv
V(\rho),
\ee
where $\tilde{E}$ and $\tilde{L}$ are the conserved energy and
angular momentum parallel to the axis of symmetry, per unit mass.
$V(\rho)$ is an effective potential for the radial motion and in
the case of orbits that are circular, we additionally have
the conditions $d\rho/d\tau=0$ and $d^2\rho/d\tau^2=0$,
which are equivalent to the conditions for a local
extremum of the potential, i.e., $V(\rho)=0$, and $dV(\rho)/d\rho=0$.
The radius of the ISCO is evaluated if we further demand the constraint
$d^2V(\rho)/d\rho^2=0$, the physical meaning of which is
that the position of the circular orbit is also a turning point
of the potential. From these three conditions we can evaluate a
specific $\rho_{\rm {ISCO}}$ and from that the
$R_{\rm {ISCO}}=\sqrt{g_{\phi\phi}(\rho_{\rm {ISCO}})}$,
which we then compare to the corresponding numerical one. The position
of the ISCO is of obvious astrophysical interest since it is the
inner radius of an accretion disc and recently it
has been used to evaluate the rotation parameter of black holes from
fitting the continuous spectrum of the accretion disc
around them (see work by \cite{Narayan}).

Another criterion for comparing the metrics can be the orbital
frequency of circular equatorial orbits $\Omega$. The orbital
frequency is given by the equation
\be \label{Omega4}
\Omega(\rho)=\frac{-g_{t\phi,\rho}+\sqrt{(g_{t\phi,\rho})^2-g_{tt,\rho}g_{\phi\phi,\rho}}}{g_{\phi\phi,\rho}}.
\ee
Apart from the orbital frequency one could also use the precession
frequencies of the almost circular and  almost equatorial
orbits, i.e., the precession of the periastron $\Omega_{\rho}$ and
the precession of the orbital plane $\Omega_z$. These frequencies
are derived from the perturbation of the equation of motion

\be
-g_{\rho\rho}\left(\frac{d\rho}{d\tau}\right)^2-g_{zz}\left(\frac{dz}{d\tau}\right)^2=
V(\rho,z) \ee
around the circular equatorial orbits. In this
expression, $V(\rho,z)$ is the same effective potential which was defined in
the second equation of (\ref{Veff4}) the $z$ dependence of which now
has not been omitted as in eq.~(\ref{Veff4}). The perturbation frequencies derived from the above
equation are then given with respect to the metric functions as
\bea
\kappa_a^2&=&-\frac{g^{aa}}{2}\left\{(g_{tt}+g_{t\phi}\Omega)^2\left(\frac{g_{\phi\phi}}{\rho^2}\right)_{,aa}\right.\nn\\
              &-&2(g_{tt}+g_{t\phi}\Omega)(g_{t\phi}+g_{\phi\phi}\Omega)\left(\frac{g_{t\phi}}{\rho^2}\right)_{,aa}\nn\\
              &+&\left.(g_{t\phi}+ g_{\phi\phi}\Omega)^2\left(\frac{g_{tt}}{\rho^2}\right)_{,aa}\right\},
\eea
where the index $a$ takes either the value $\rho$ or $z$ to
obtain the frequency of the radial or the vertical perturbation
respectively. These expressions are evaluated on the equatorial
plane (at $z=0$); thus they are functions of $\rho$ alone. The corresponding
precession frequencies are given by the difference between the orbital frequency and the perturbation
frequency:
\be
\Omega_a=\Omega-\kappa_a.
\ee
These quantities are quite interesting with respect to
astrophysical phenomena as well. More specifically, they can be
associated to the orbital motion of material accreting onto a
compact object through an accretion disc. These very frequencies have
been proposed to be connected to the observed quasi-periodic
modulation (QPOs) of the X-ray flux of accretion discs that are present in
X-ray binaries (see \cite{stella,derKlis,boutloukos,lamb}).

Finally, the last criterion that we will use to compare metrics is
the quantity $\Delta\tilde{E}$ of circular orbits, which expresses
the energy difference of the orbits per logarithmic orbital
frequency interval as one moves from one circular orbit to the
next towards the central object. This quantity is defined as
\be
\Delta\tilde{E}=-\Omega\frac{d\tilde{E}}{d\Omega},
\ee
where the energy per unit mass $\tilde{E}$ is given by the
expression
\be
\label{Etilde4}
\tilde{E}=\frac{-g_{tt}-g_{t\phi}\Omega}{\sqrt{-g_{tt}-2g_{t\phi}\Omega-g_{\phi\phi}\Omega^2}}.
\ee
The quantity $\Delta\tilde{E}$ is a measure of the energy that a
particle has to lose in order to move from one circular orbit to
another closer to the central object so that the frequency increases by one $e$-fold. The quantity,
$\Delta\tilde{E}$, is associated to the emission of gravitational
radiation and was used by \cite{Ryan} to measure the multipole
moments of the space-time from gravitational waves emitted by test particles orbiting
in that background. The same quantity can also be associated to accretion discs and in
particular, in the case of thin discs, it would correspond to the amount of
energy that the disc will radiate as a function of the radius
from the central object and thus it will be related to the
temperature profile of the disc and consequently to the total
luminosity of the disc (for a review on accretion discs see \cite{krolik}).

The last set of criteria, i.e., the frequencies
and $\Delta\tilde{E}$, are related to specific observable
properties of astrophysical systems, in particular of accretion discs around
compact objects; thus they are very useful and
relevant to astrophysics (for an application see the work by
\cite{PappasQPOs}).

\section{Results of the comparison}

\begin{figure*}
\centering
\includegraphics[width=0.32\textwidth]{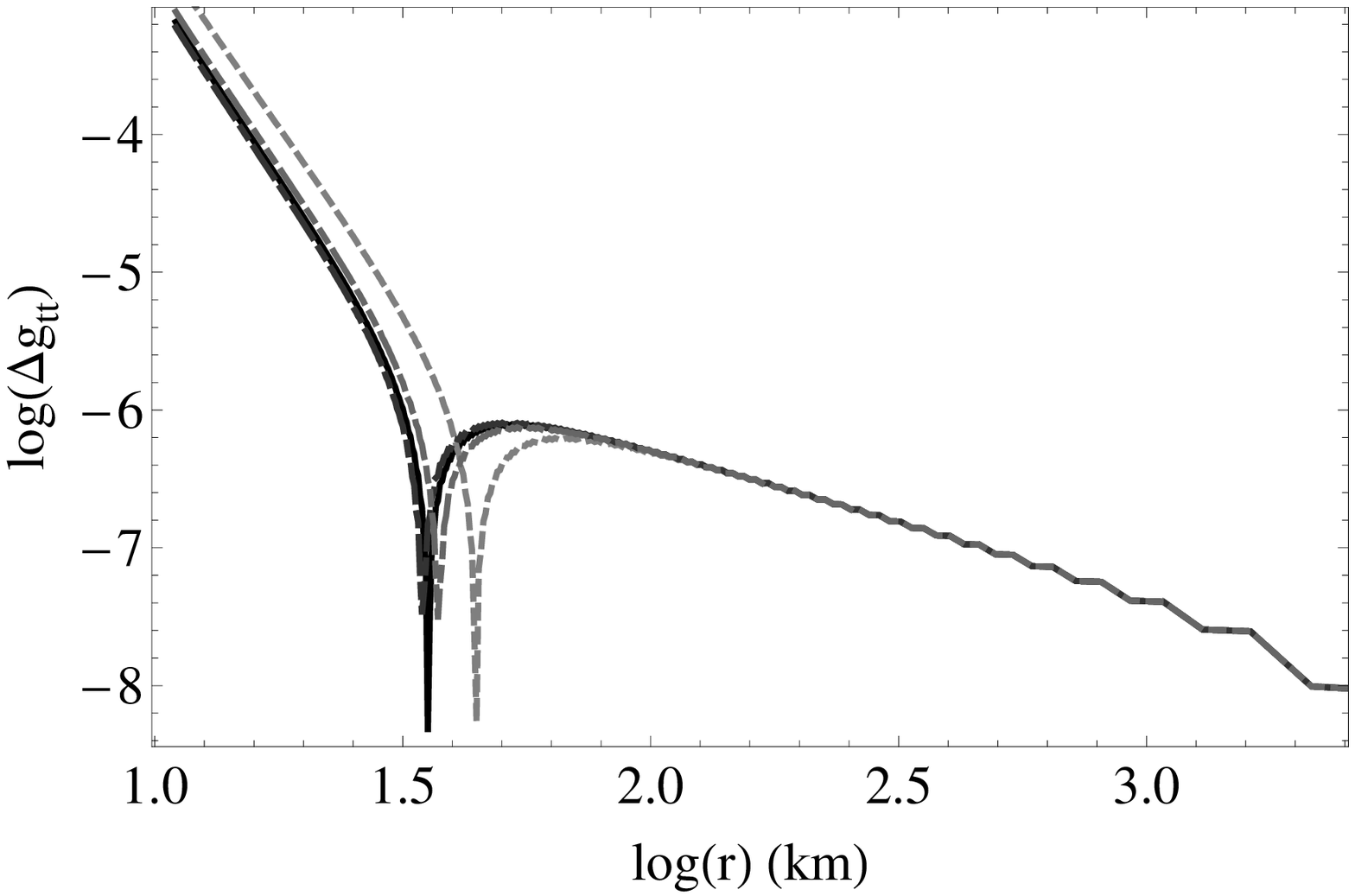}
\includegraphics[width=0.32\textwidth]{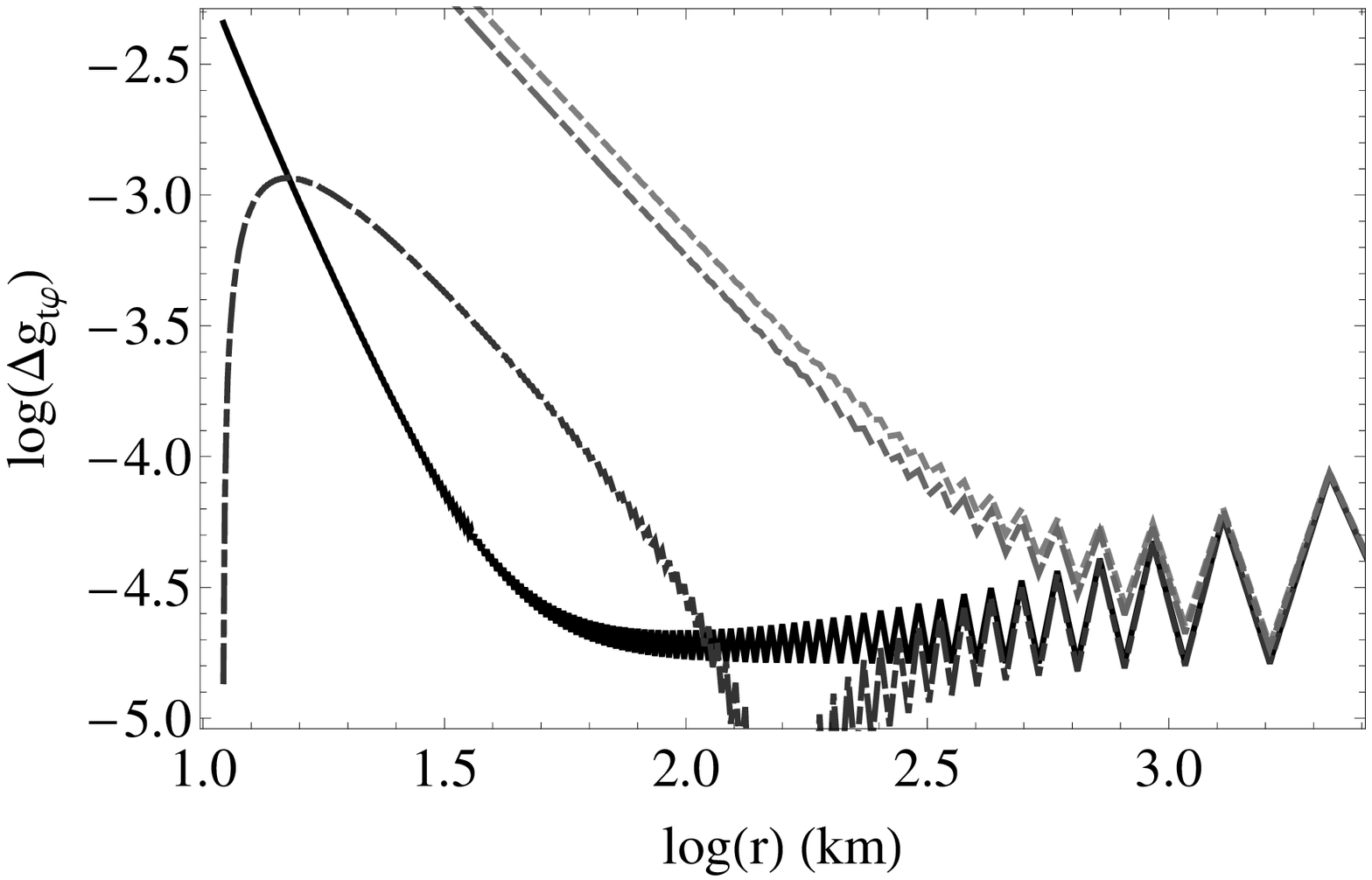}
\includegraphics[width=0.32\textwidth]{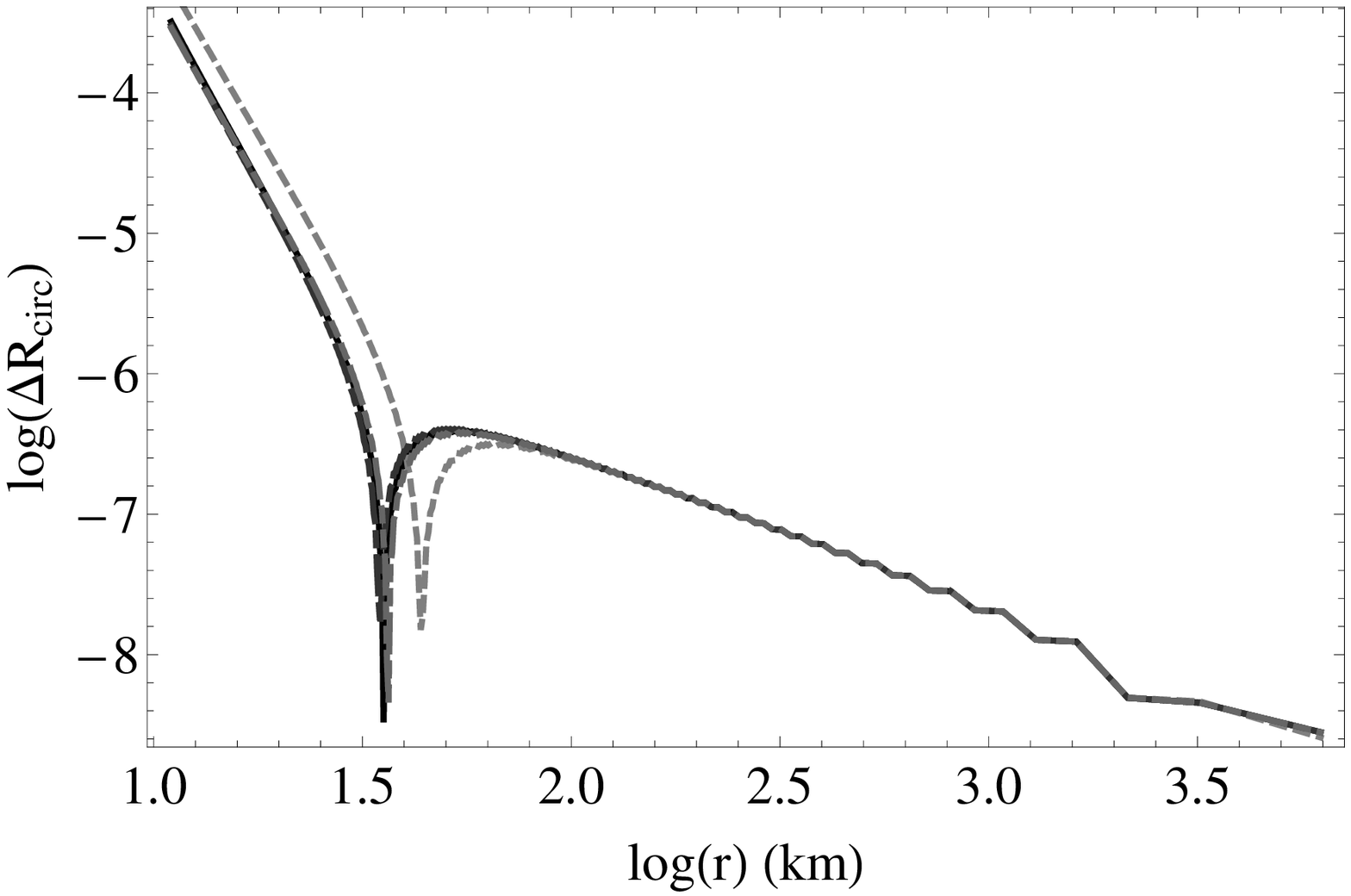}
\includegraphics[width=0.32\textwidth]{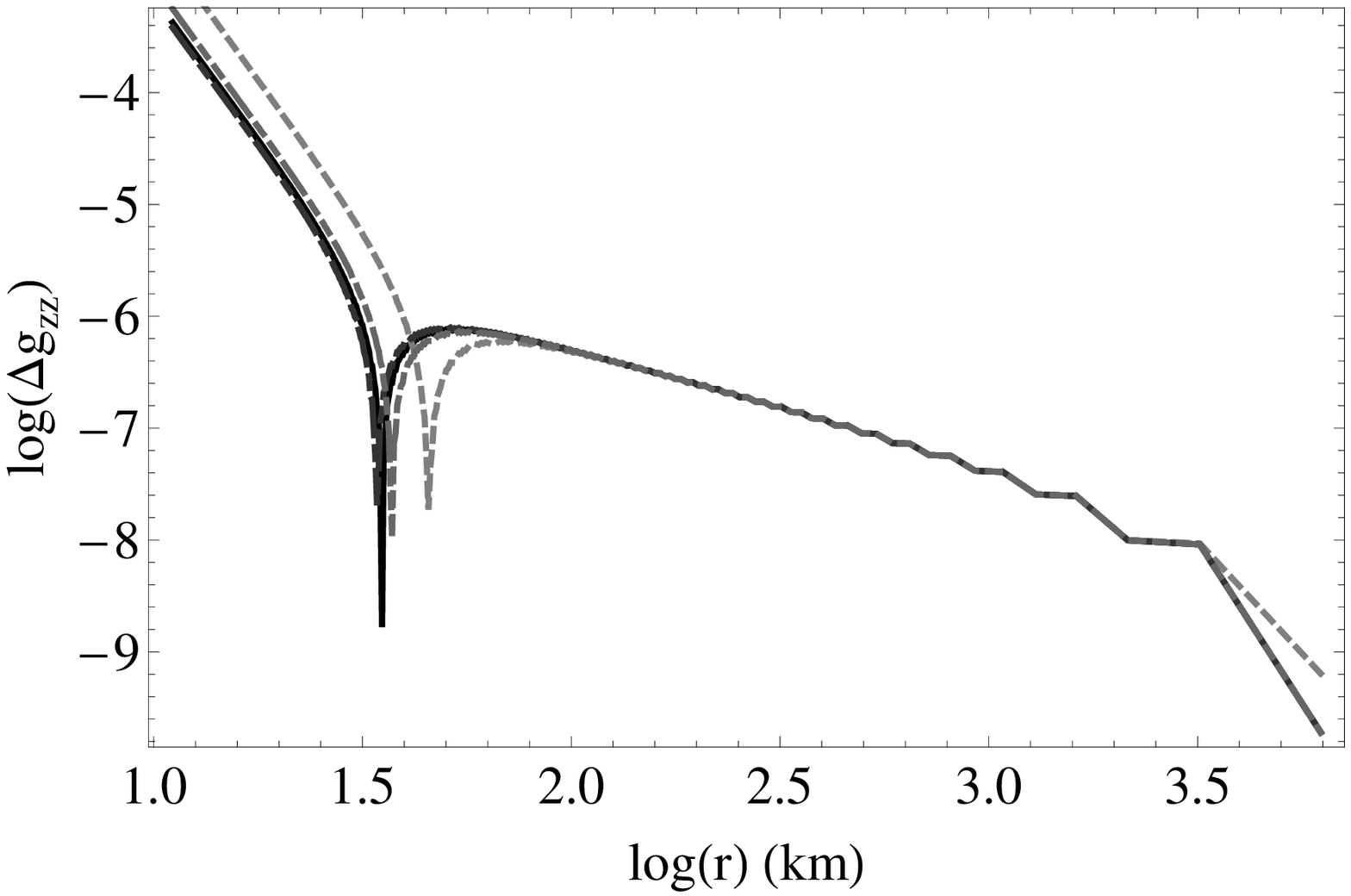}
\includegraphics[width=0.32\textwidth]{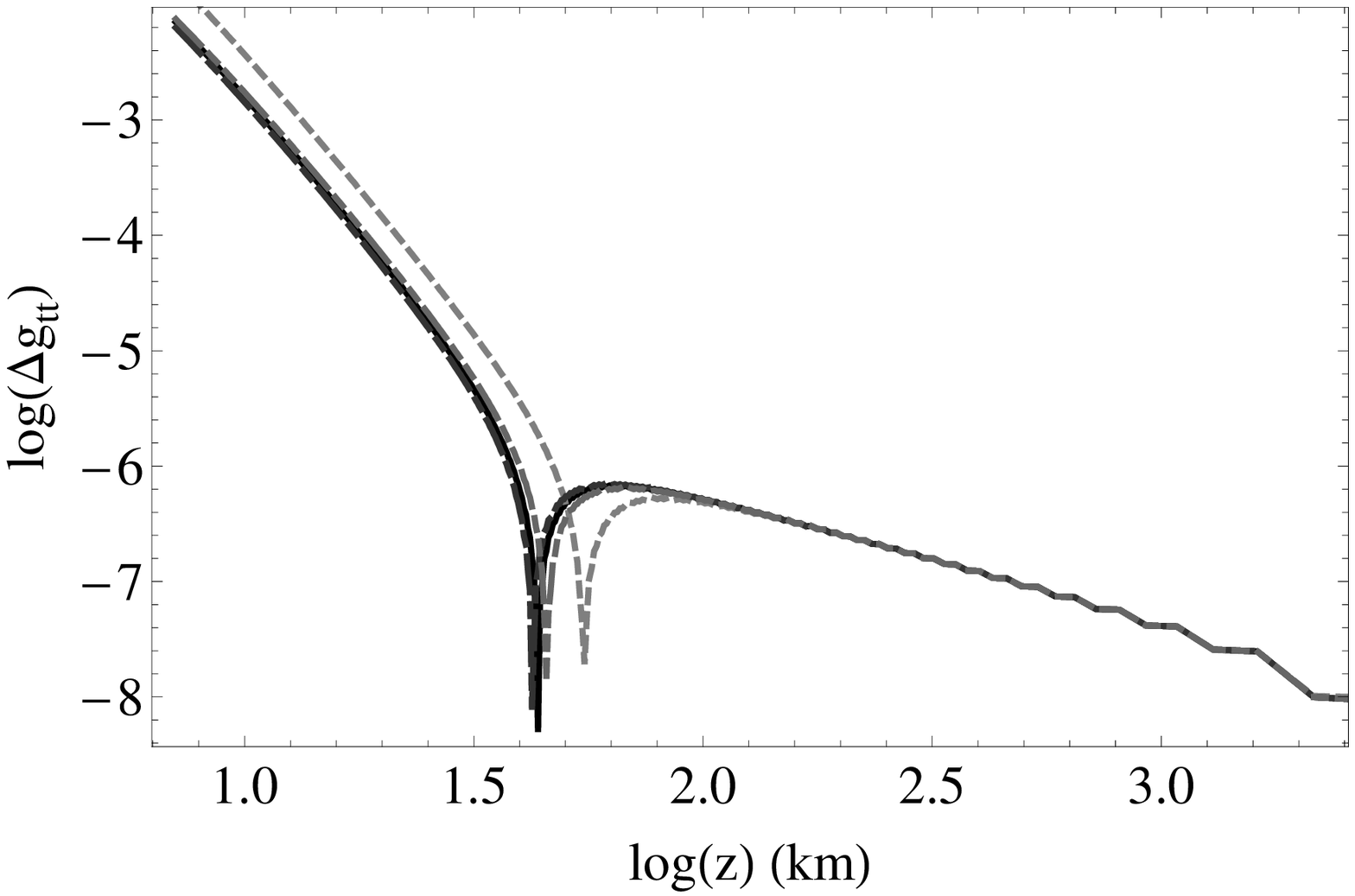}
\caption{These are the plots of the logarithm of the relative
difference of the analytic to the numerical metric components
$g_{tt}$, $g_{t\phi}$, $\sqrt{g_{\phi\phi}}=R_{\rm circ}$, and $g_{zz}$
on the equatorial plane, as well as  $g_{tt}$  on the axis of
symmetry. The plots shown here are indicative of all comparisons
performed for all the neutron star models that we have constructed. These
particular plots are drawn for the model $\#10$  of the AU EOS, the characteristics of which are
presented in Table II of the supplement of Pappas~\&~Apostolatos (2012). Different curves correspond to
different metrics, i.e., the two-soliton (solid curve), the Manko et al. solution
with the negative root for $b$ (dashed-dotted curve) and the one with
the positive root (dashed-double-dotted), and finally the Hartle-Thorne (dotted) metric.           
In most cases shown here, the Manko et al. curves are on top of the two-soliton curve.
The case of the $g_{t\phi}$ component, where the two-soliton
curve is almost an order of magnitude below the Manko et al. one
for a large interval of radii, is a notable exception.
The Hartle-Thorne's characteristic failure to describe $g_{t \phi}$ is also evident.}
\label{metricfunctions}
\end{figure*}

The results of the comparison between the analytic and the
numerical metrics, describing the exterior of realistic neutron
stars, that are presented here is indicative of all  comparisons
performed for every numerically constructed neutron star model.
The models that we have used as a testbed of comparison are briefly presented in the
Appendix and are the same models used in \cite{PappasMoments}.

For illustrative reasons
we have also plotted comparisons between the numerical and the Manko et al. solution (\cite{Mankoetal})
which was used by \cite{BertSter}, as well as
comparisons between the numerical and the Hartle-Thorne metric
(\cite{HT}). The reason for using these two metrics is that on the
one hand the Hartle-Thorne metric is considered to be a good
approximation of slowly rotating relativistic stars and on the
other hand the Manko et al. metric has been shown by \cite{BertSter} to be a good
approximation for relativistic stars with fast
rotation. In cases with slow rotation rates, for which corresponding models of the
Manko et al. metric cannot be constructed, we have used only the
Hartle-Thorne metric to compare, though. In the case of  models with fast rotation
besides the Manko et al. metric, we have used the
Hartle-Thorne metric as well. for these cases we treated the Hartle-Thorne metric as a three parameter exterior
metric, where the three parameters are the mass $M$, the angular
momentum $J$ and the reduced quadrupole $q=M_2/M^3$ of the neutron star. It should be noted that this is not
a consistent way to use the Hartle-Thorne metric, since the
quadrupole and the angular momentum in the Hartle-Thorne cannot
take arbitrary values, while the metric is essentially a two
parameter solution (parameterized by the central density of the
corresponding slowly rotating star and a small parameter
$\varepsilon$ that corresponds to the fraction of the angular velocity
of the star relative to the Keplerian angular velocity of the
surface of the star) that has to be properly matched to an
interior solution following the procedure described by
\cite{Bertietal}. Here we are taking some leeway in using
Hartle-Thorne for fast rotation since it is used simply for
illustrative purposes and not in order to draw any conclusions from it.

In Figure \ref{metricfunctions} we present the comparison of the
various analytic metric functions (using the two-soliton, the
Manko et al., and the Hartle-Thorne solutions) to the corresponding
numerical ones for a single model constructed using the EOS AU
(model $\#10$  of the AU EOS the characteristics of which are
presented in Table II of the supplement of \cite{PappasMoments}). The
figures display the relative difference between the various analytic and
the numerical metric functions $g_{tt},g_{t\phi},\sqrt{g_{\phi\phi}}$ and $g_{zz}=g_{\rho\rho}$
on the equatorial plane, as well as the function $g_{tt}$ on the axis of
symmetry.

The general picture we get from these figures is typical for all models
constructed using all three equations of state, that is AU, FPS and
L. The overall comparison of the two-soliton to the numerical
metrics shows that this analytic metric is an excellent substitute
of the numerical space-time both for slow and fast rotating models,
with an accuracy that is everywhere outside the neutron star
always better than about $1/1000$ for all the metric functions (there is an exception to that for the
comparison of the $g_{tt}$ metric component right at the pole where for
some models it's fractional difference is a bit smaller than
$1/100$). In comparison to the other two analytic metrics discussed above, we
see that for the models that a Manko et al. metric can be found,
that is for the rapidly rotating neutron stars, both this metric and the two-soliton metric perform
very well (actually the Manko et al. solution performs better than
\cite{BertSter} had initially found as was shown by
\cite{PappasMoments}) and there are only tiny
differences between the two-soliton and the Manko et al. analytic
metrics. The greatest difference between the two-soliton and the
Manko et al. appears to be in the $g_{t\phi}$ component of the
metric, where the two-soliton is usually clearly better. This was anticipated
because the $g_{t\phi}$ component of the metric is, as we have
shown in section 3, more sensitive to the value of  $S_3$, which can be
suitably adjusted in the two-soliton metric, but not in the Manko et al. solution. For
the rapidly rotating models the Hartle-Thorne is not such a good
representation of the numerical metric as the other two metrics. That also was
expected since the Hartle-Thorne metric is not suitable for
fast rotation. Hartle-Thorne's failure is more evident in the
$g_{t\phi}$ component of the metric which is consistent with
Hartle-Thorne's vanishing spin octupole $S_3$ (in the Appendix we show
that  $S_3=0$ for the Hartle-Thorne metric).

For the slowly rotating models, there are no Manko et al.
solutions to compare to the numerical metric, so in these cases
the only alternative is the Hartle-Thorne metric. We should say
again that the consistent way to calculate the Hartle-Thorne
parameters is the one described by \cite{Bertietal}, but as it is
discussed by \cite{PappasMoments} the parameters of the
Hartle-Thorne metric (specifically the parameter $q$ which is the
reduced quadrupole) consistently calculated are in very good
agrement with the numerical multipole moments, so these moments
were used straightforwardly for the construction of the
Hartle-Thorne metric.
Again we saw that the two-soliton performs better when compared
to the numerical space-time than the Hartle-Thorne
metric. We should  note though that the problem of
Hartle-Thorne's metric to acurately describe the $g_{t\phi}$ metric component is present
even at slow rotation.

Having demonstated the overall superiority of two-soliton,
compared to Manko et al. and Hartle-Thorne, to accurately
describe the metric functions of any numerical neutron star model,
in the following comparisons we will only compare the two-soliton
quantities to the corresponding numerical ones.
%
\begin{figure*}
\centering
\includegraphics[width=0.32\textwidth]{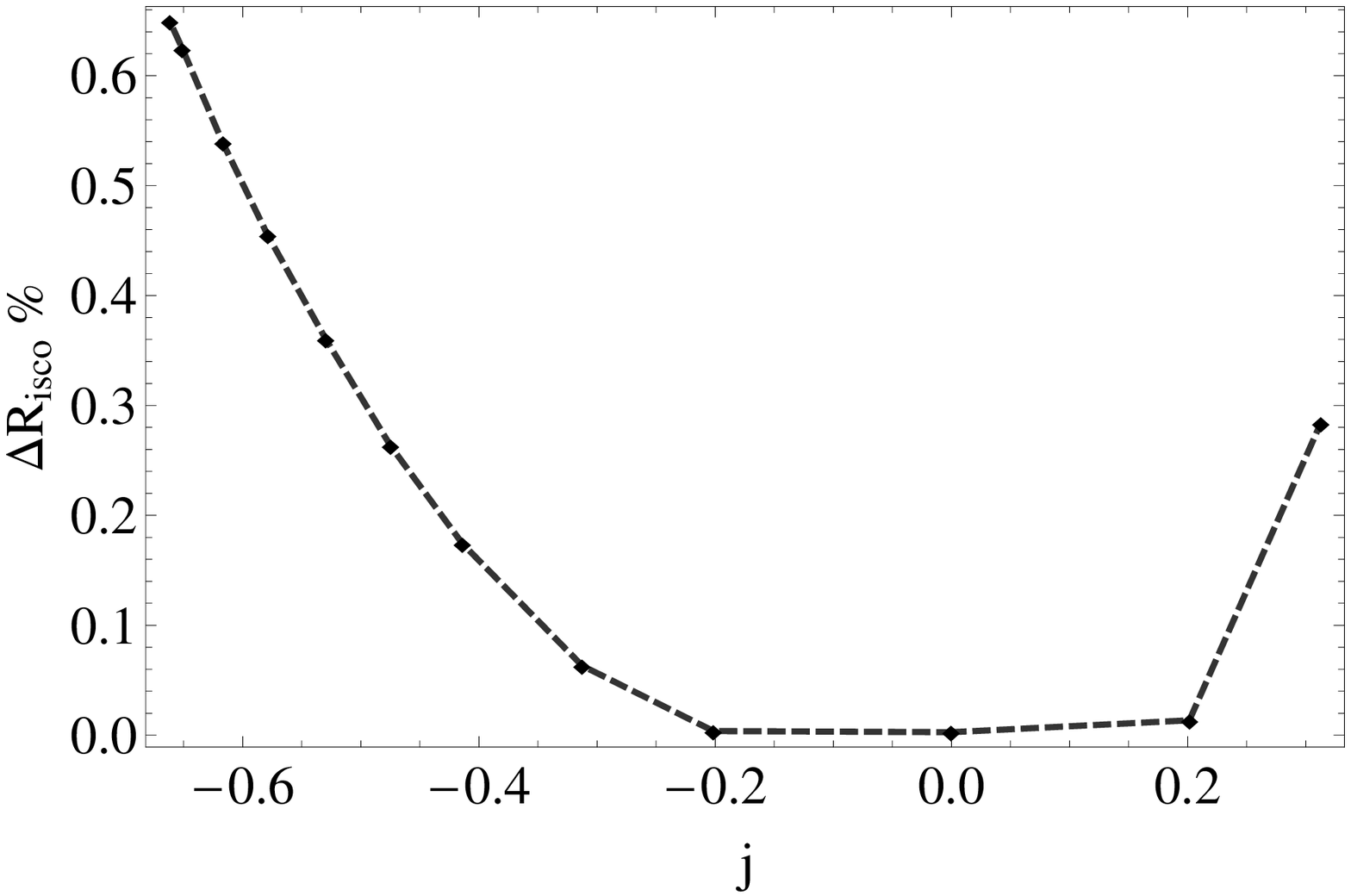}
\includegraphics[width=0.32\textwidth]{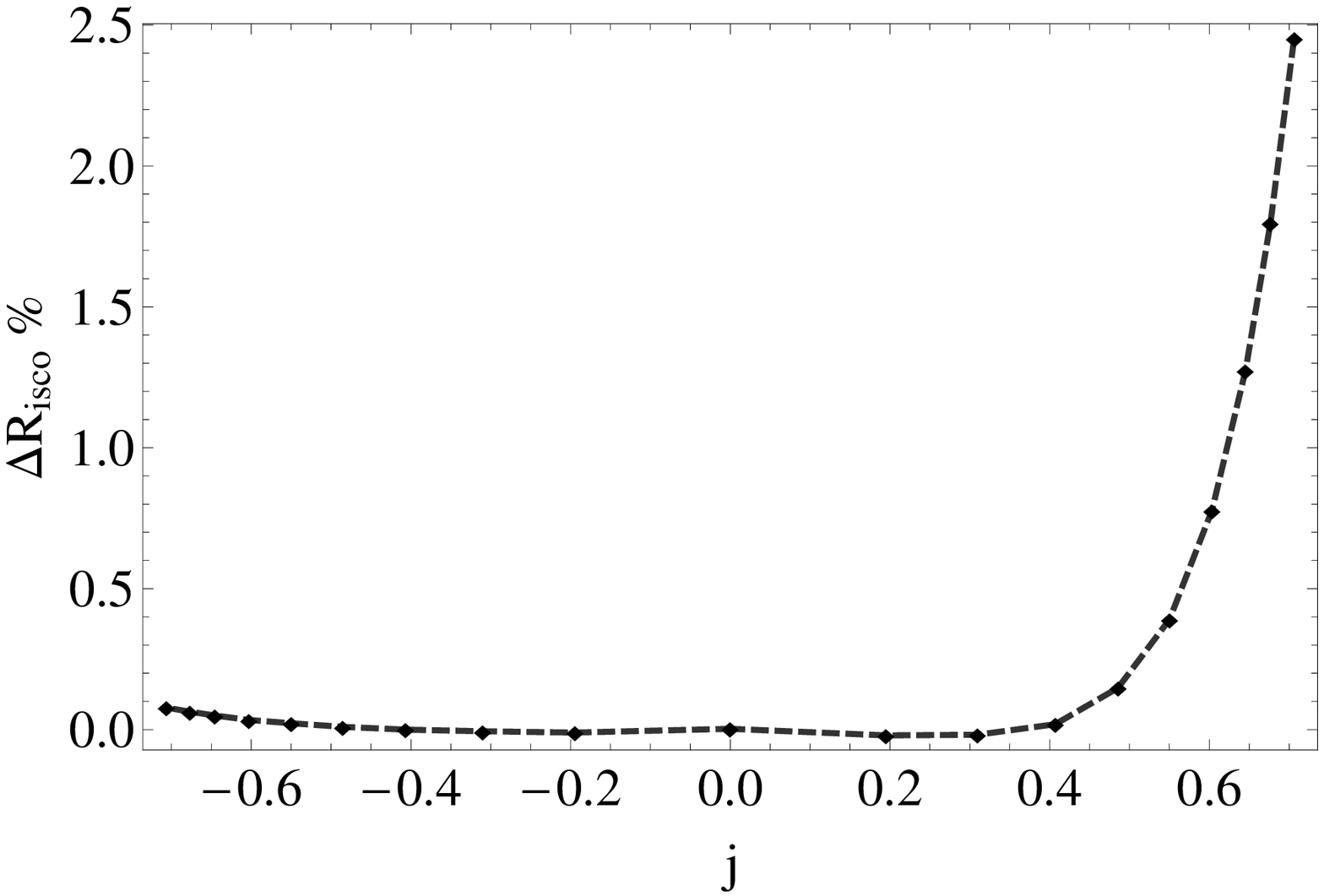}
\includegraphics[width=0.32\textwidth]{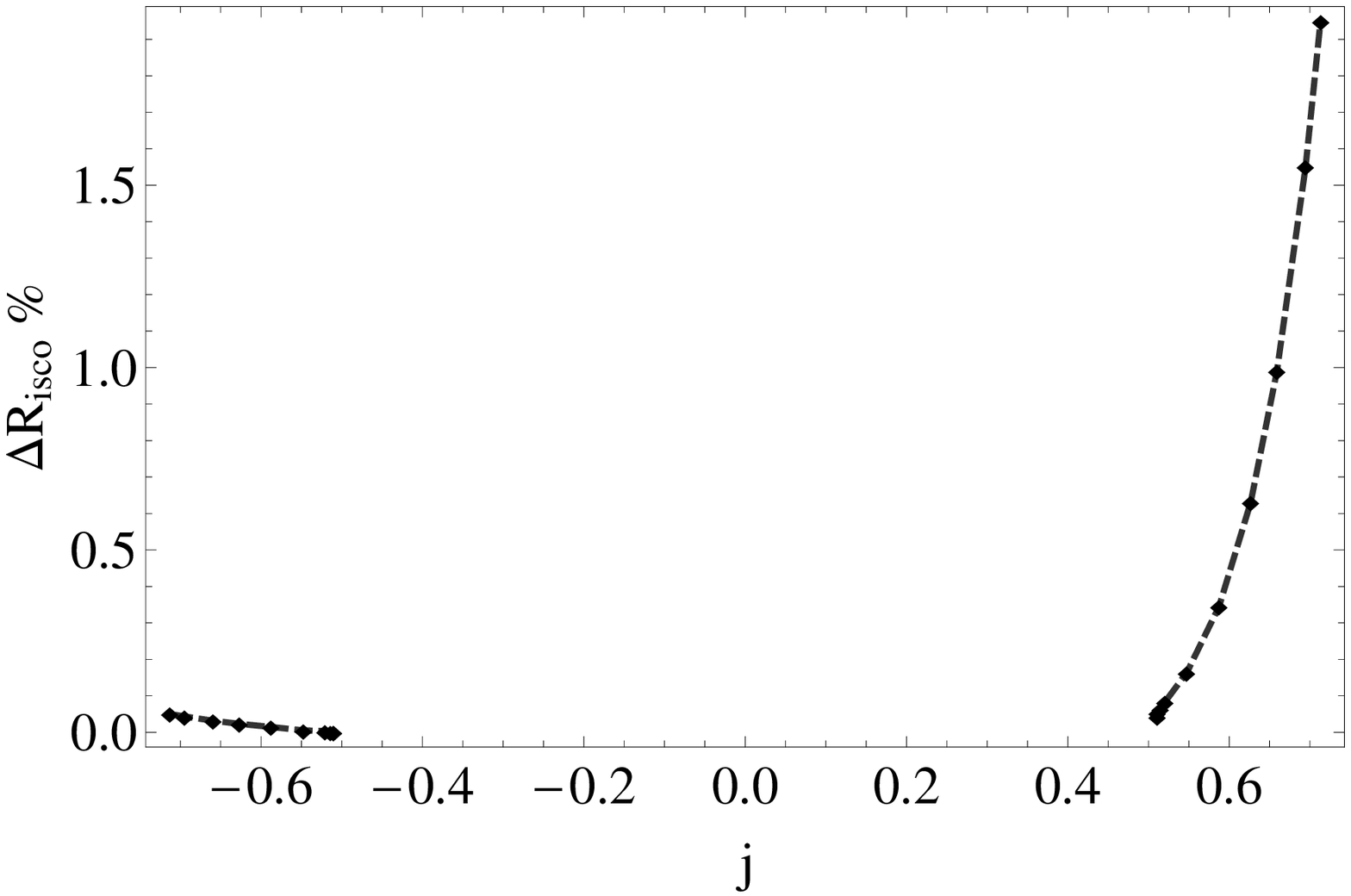}
\caption{These figures show the relative difference between the
numerical and the analytic $R_{\rm{ISCO}}$ for the
neutron star models calculated using the AU EOS. The relative
difference is shown only for the models that have their ISCO
outside the surface of the neutron star. The negative values of
$j$ correspond to counter-rotating orbits (retrograde). The
data for the co-rotating orbits can be found in Table 2.}
\label{figISCO}
\end{figure*}
%

The next quantity we have used for comparison is the position of the ISCO. In
Figure \ref{figISCO} we present the relative difference between
the numerical and the analytic ISCO for all  neutron star
models constructed with the AU EOS for both prograde and
retrograde orbits (the latter are indicated by negative parameter
$j\equiv J/M^2$). The general conclusion is that for all  models
constructed using all three EOSs the ISCO of the analytic metric
does not deviate by more than 4 per cent from the ISCO of the
corresponding numerical model and such deviations are observed for
the prograde orbits of the fastest rotating models. We should note
that we don't perform any comparison with numerical models
the radius of the surface of which on the equatorial plane is further than
the radius of the corresponding ISCO; therefore the points that would
correspond to these models are missing from the plots. At this
point we should mention that apart from the position of the
marginally stable circular orbit for the particles, there is also
the position of the unstable photon circular orbit that could also be
 used as a criterion for comparison. This orbit though
is usually, for the prograde case, below the surface of the
neutron star (while for the retrograde it is usually outside
the star) and it doesn't have an immediately measurable effect\footnote{
One could argue that it could be associated to the optics around
neutron stars and possibly to quasi-normal modes of the space-time
around the neutron star.}.

\begin{figure*}
\centering
\includegraphics[width=0.32\textwidth]{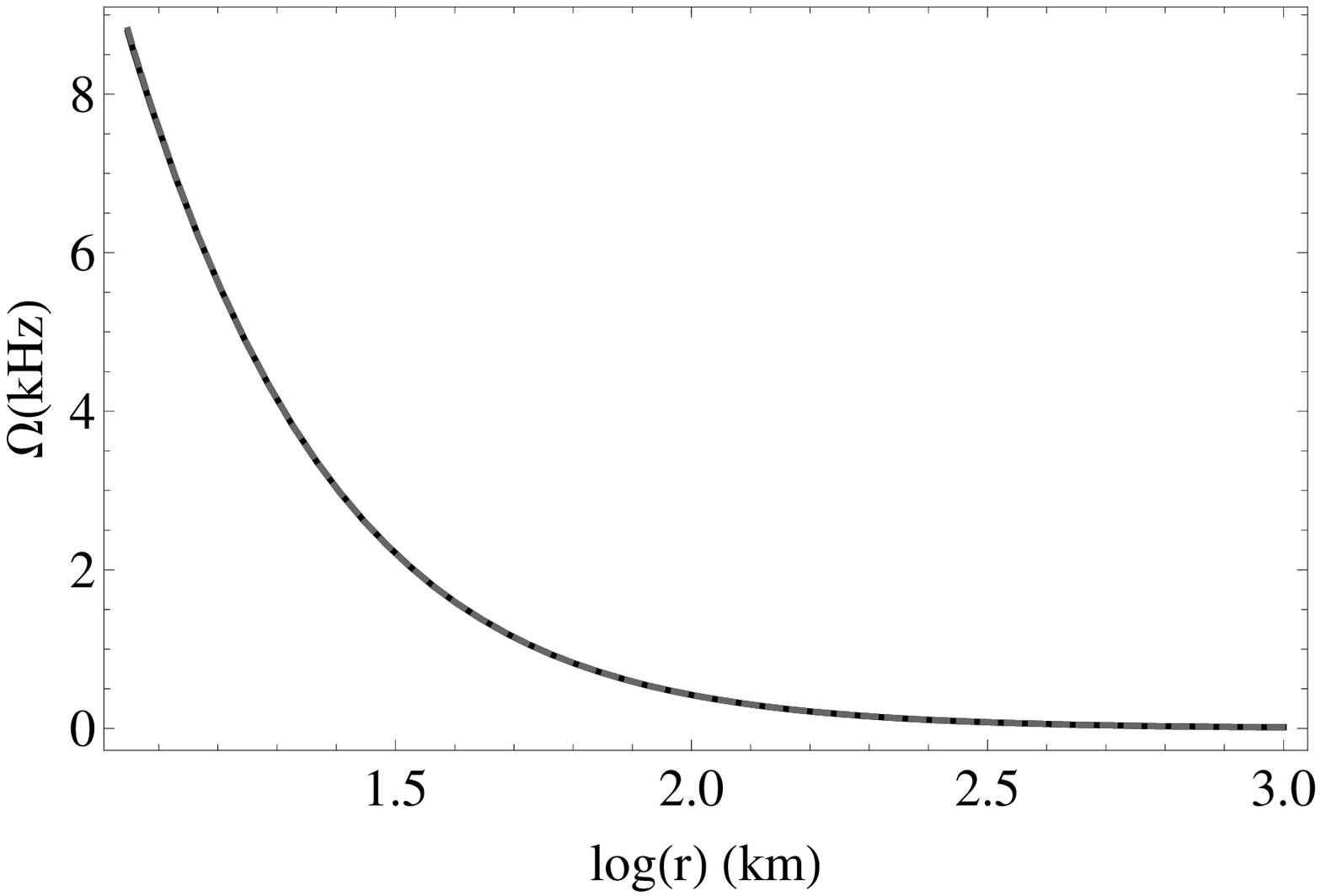}
\includegraphics[width=0.32\textwidth]{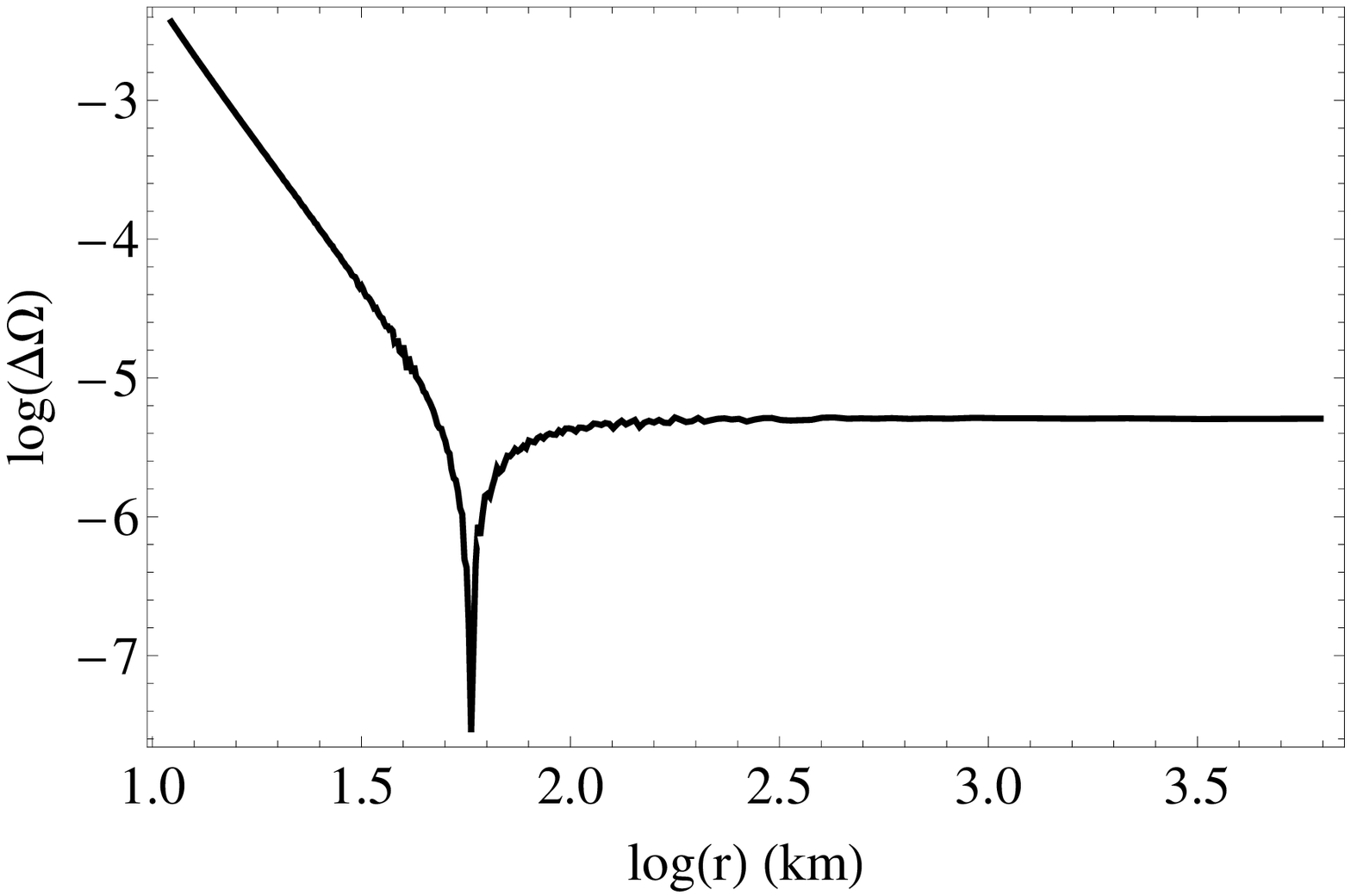}\\
\includegraphics[width=0.32\textwidth]{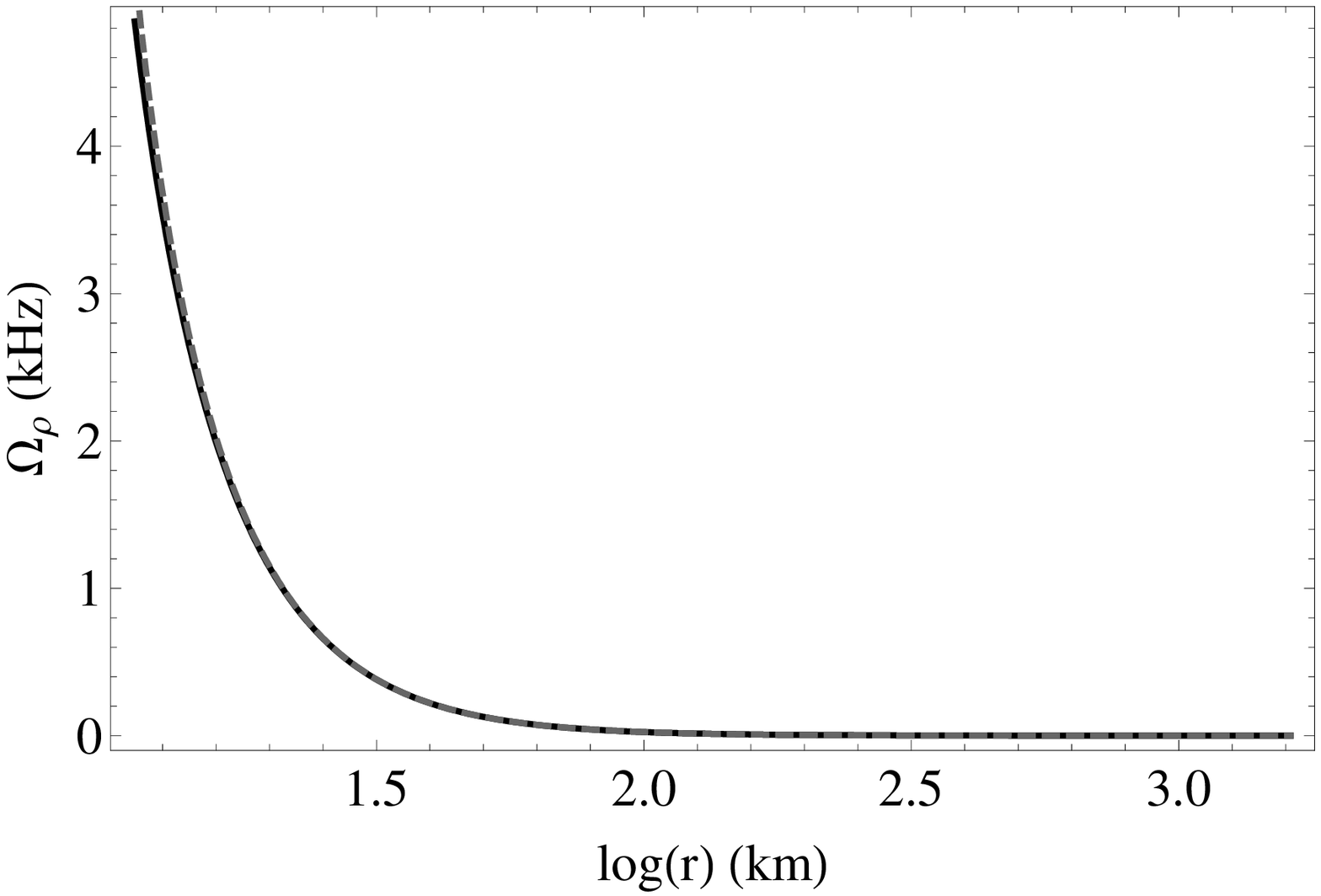}
\includegraphics[width=0.32\textwidth]{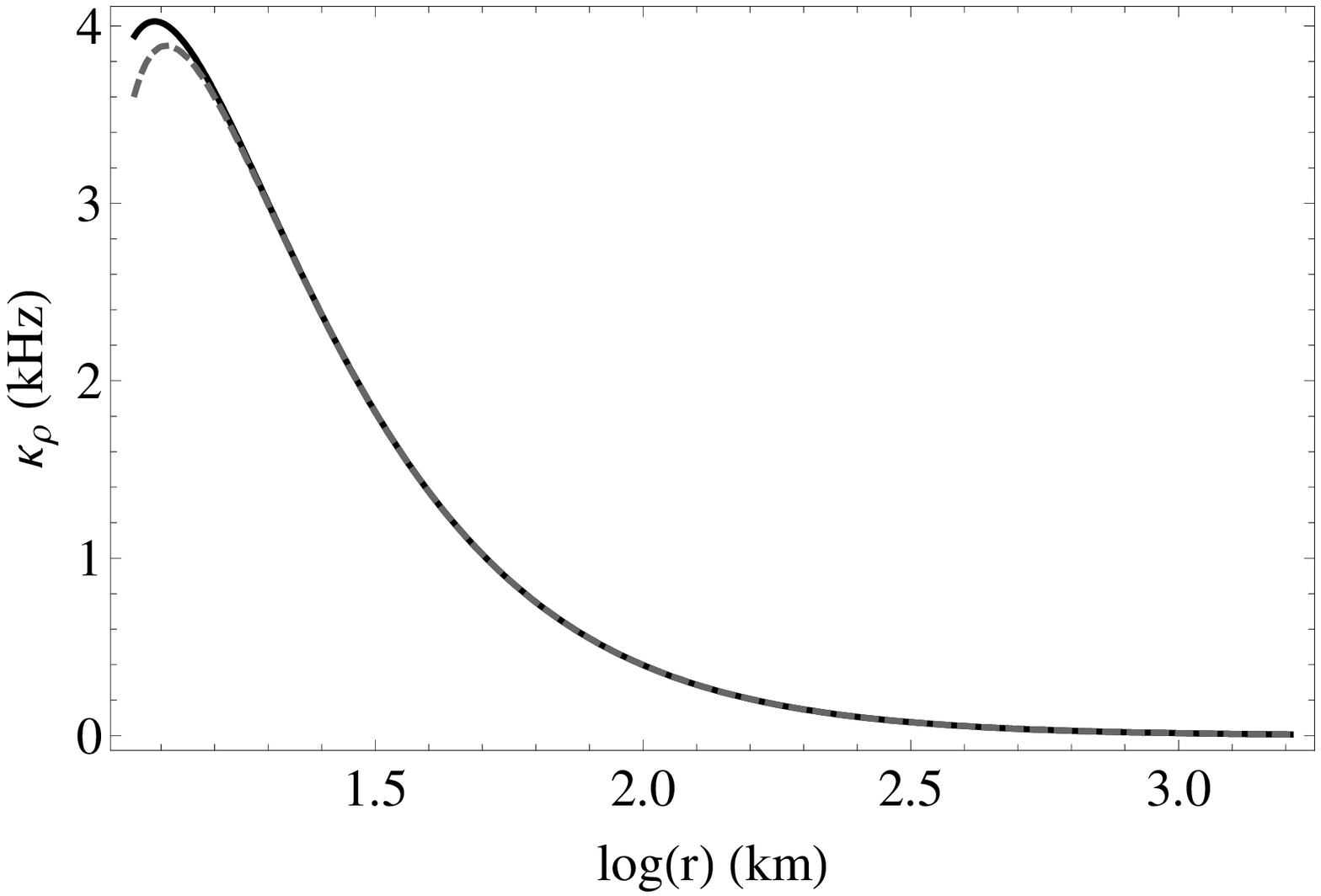}
\includegraphics[width=0.32\textwidth]{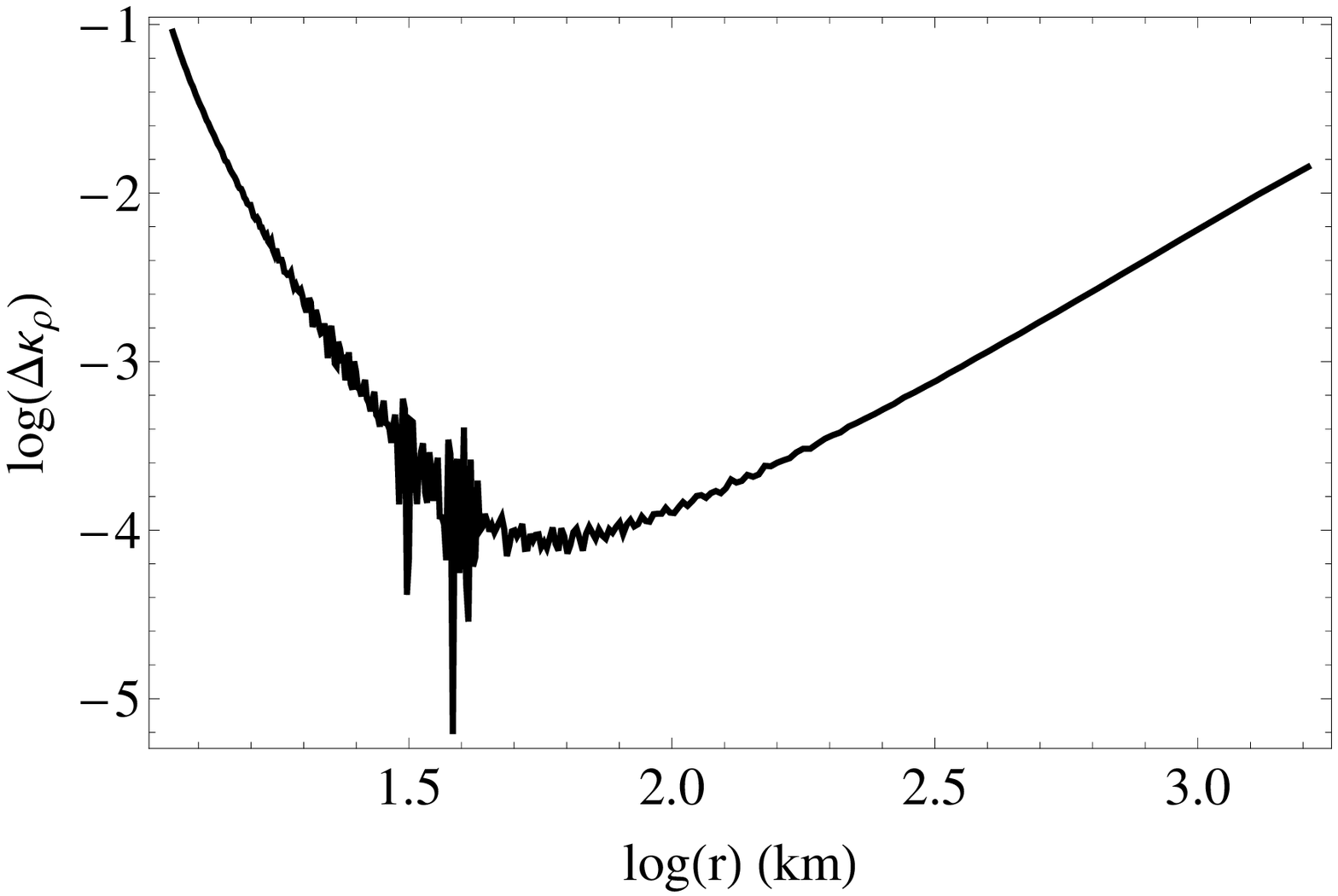}
\includegraphics[width=0.32\textwidth]{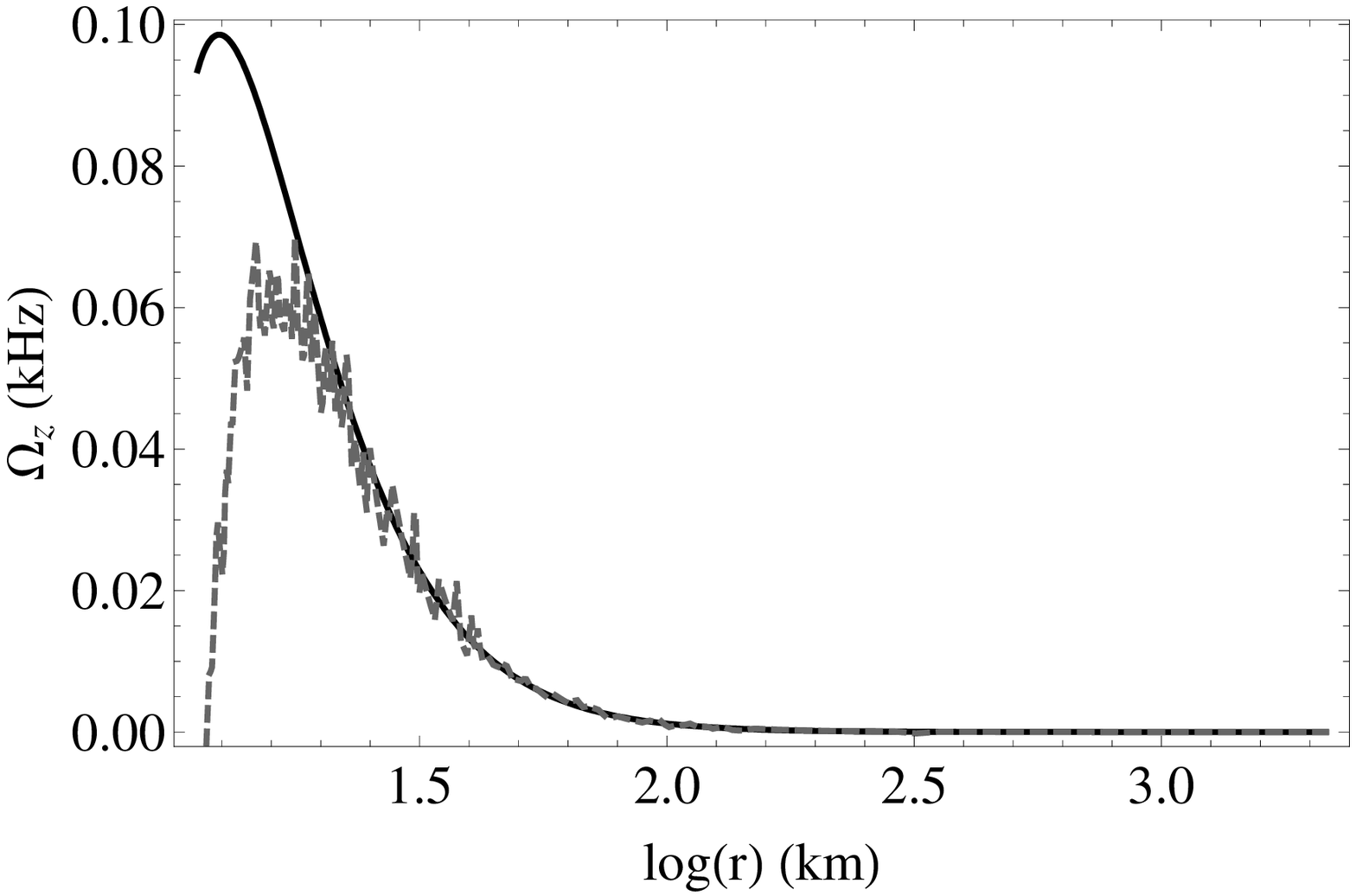}
\includegraphics[width=0.32\textwidth]{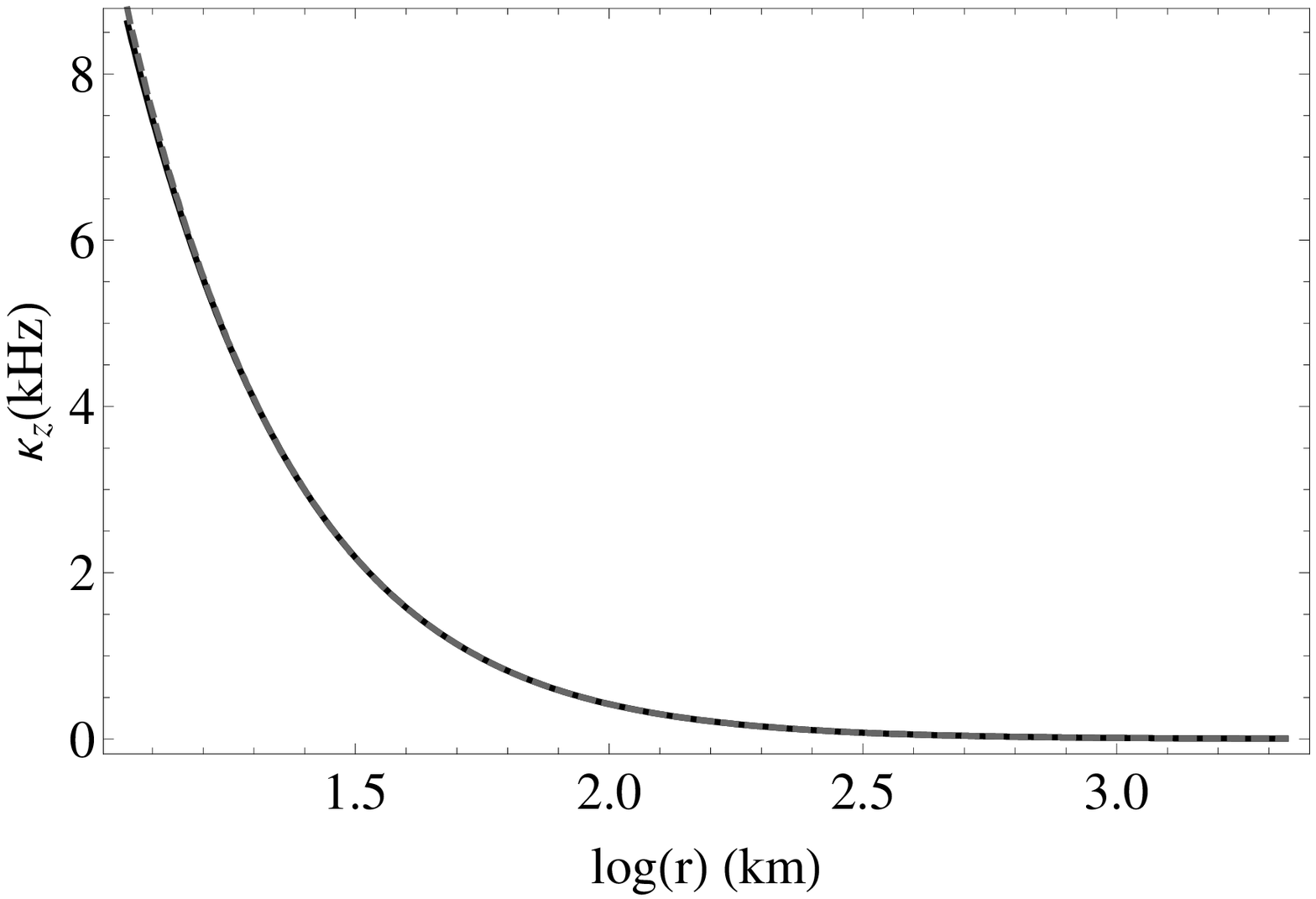}
\includegraphics[width=0.32\textwidth]{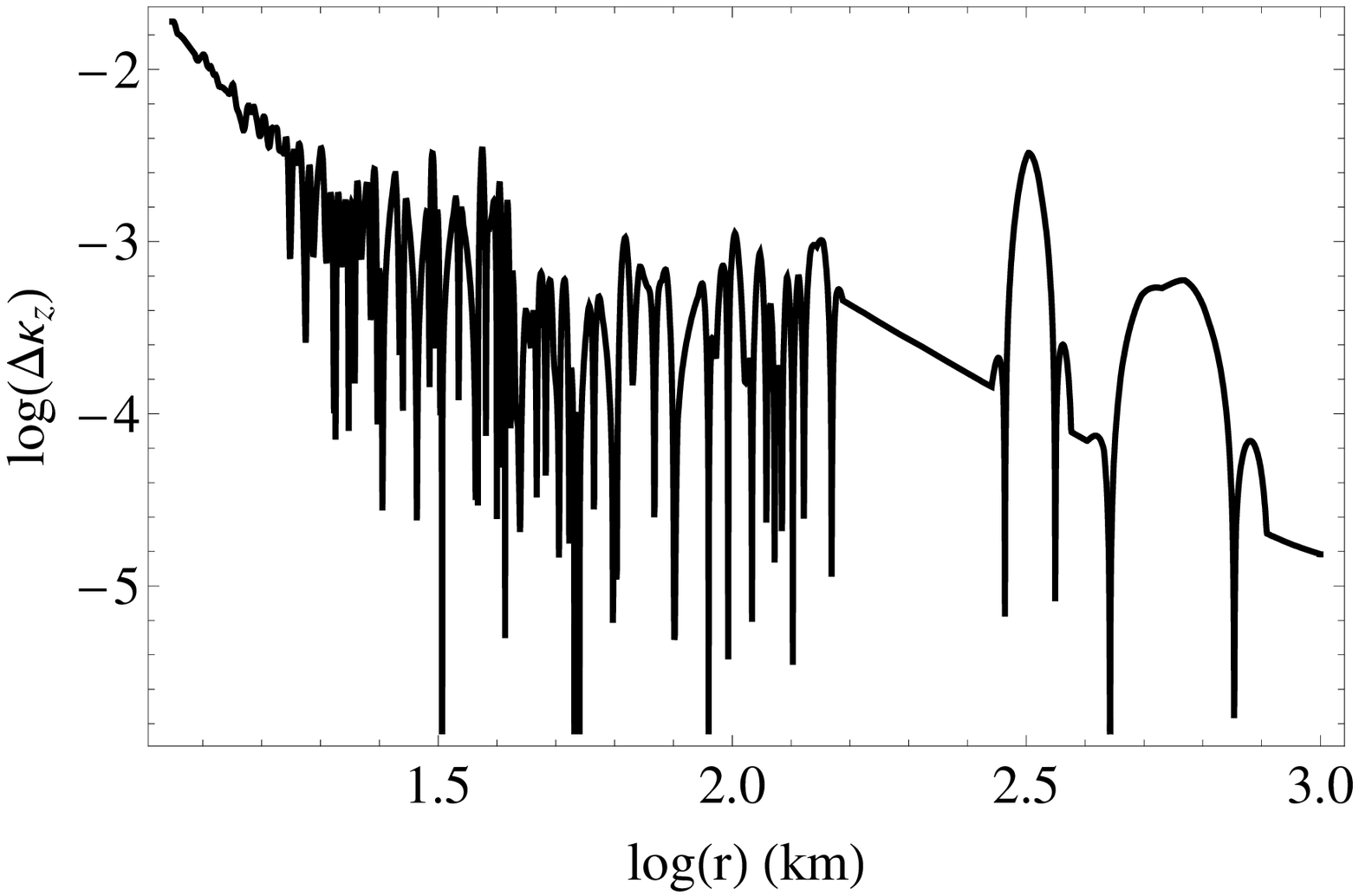}
\caption{\label{freqfig} The plots show the various frequencies
and the relative difference of the analytic to the corresponding
numerical ones. The top row of plots shows on the left the numerical (dotted)
and the analytic (solid) orbital frequency $\Omega$ for comparison reasons (they are hardly distinguishable),
while on the right it is plotted the logarithm of their relative difference. The
middle row  shows on the left the numerical (dotted) and analytic
(solid) precession frequencies $\Omega_{\rho}$, in the middle the
corresponding radial oscillations frequencies $\kappa_{\rho}$
(same correspondence of lines) and on the right the logarithm of the
relative difference of the latter oscillation frequencies. Finally the bottom row
 shows the corresponding plots for the plane precession (left) and the
vertical oscillation frequencies (middle), $\Omega_z$, and $\kappa_z$ respectively, as
well as the logarithm of the relative difference of the latter frequencies (right). All
plots are for the model $\#10$ of the AU EOS presented by Pappas~\&~Apostolatos (2012).}
\end{figure*}

We continue to the results of the comparison of the various
frequencies associated to the circular orbits on the equatorial
plane, i.e., $\Omega,\,\Omega_{\rho},\kappa_{\rho},\Omega_z$, and
$\kappa_z$. The analytic orbital frequency compares very well to
the numerical orbital frequency for all the models with the
relative difference being in all cases smaller than $\sim10^{-3}$.
This result is very important, since the orbital frequency
together with the $R_{\rm{ISCO}}$ are relevant to
observations from accretions discs. Typical plots of $\Omega$ and
the relative difference between the analytic and the numerical one
as functions of the logarithm of the distance from the central
object are shown in Figure \ref{freqfig}.

With regard to the comparison of the radial and vertical
perturbation frequencies $\kappa_{\rho}$, and $\kappa_z$,
respectively, and the corresponding precession frequencies
$\Omega_{\rho}$, and $\Omega_z$, things are more complicated. These
frequencies include second derivatives of the metric functions in
their calculation. Consequently the results of the corresponding numerical
calculations are plagued by accuracy problems. In particular
numerical calculation of the second derivatives induces
artificial oscillations in the results. These issues have been
discussed earlier by \cite{Bertietal}. Following the suggestions of
N. Stergioulas, we found two ways to mitigate the problem. The first
one is to calculate the frequencies directly in the coordinates
that the RNS code produces the metric functions so as to avoid any
numerical errors caused by the transformations of the coordinates
and the metric functions themselves. Then one would only have to
identify the coordinates of the points at which the frequencies
are calculated with the corresponding Weyl-Papapetrou coordinates
(the calculated frequencies themselves are the frequencies that
static observers at infinity measure so they don't depend on the
coordinate system used). The second one is to smooth out these 
artificial oscillations by taking a three point average of the
frequencies. The efficiency of this technique has been tested in the
case of the non-rotating models (the exterior of which are described by a
Schwarzschild metric) and has been verified to give trustworthy
results. Another thing that we should also consider is that in the
case of $\kappa_z$, the values are very close to the corresponding values of
$\Omega$; consequently there is  low accuracy in the
calculation of the precession frequency $\Omega_z=\Omega-\kappa_z$
in some cases. That is why we consider as a better indicator of
the actual ability of the analytic metric to capture the behavior
of the numerical metric, the deviations in the oscillation frequencies
$\kappa_{\rho}$ and $\kappa_z$, instead of $\Omega_{\rho}$ and $\Omega_z$.
Nevertheless, we  present both
the precession and the oscillation frequencies of the analytic
metric in comparison to those of the numerical metric, together with
the relative difference between the analytic and the numerical
oscillation frequencies, in order to get a clearer picture of the
comparison. All these plots, again for the model $\#10$  of the AU
EOS, the characteristics of which are
presented in Table II of the supplement of
\cite{PappasMoments}, are shown in Figure \ref{freqfig}.

Generally, the relative differences in $\kappa_{\rho}$ between the
numerical and the analytic metrics is small and in some cases they
could climb up to 10 per cent near the ISCO. This is due to  the
fact that the radial oscillation frequency tends to zero as the
ISCO is approached, causing an increase of the relative difference.
In contrast, for $\kappa_z$, the relative differences are always below 1 per cent
at the ISCO. Now the picture is inverted for the precession
frequencies, which is related to the fact that in this case the
small quantities are the $\Omega_z$'s. The overall picture we obtain
is that the analytic frequencies capture quite well the behavior
of the numerical frequencies both qualitatively and
quantitatively. This is especially evident in the case of the
$\kappa_z$ frequency where in some cases it becomes greater than
the orbital frequency, an effect more prominent in the models of
EOS L, which the two-soliton metric captures quite well. The
importance of this effect and its relevance to QPOs is further
discussed by \cite{PappasQPOs}. 

\begin{figure*}
\centering
\includegraphics[width=0.45\textwidth]{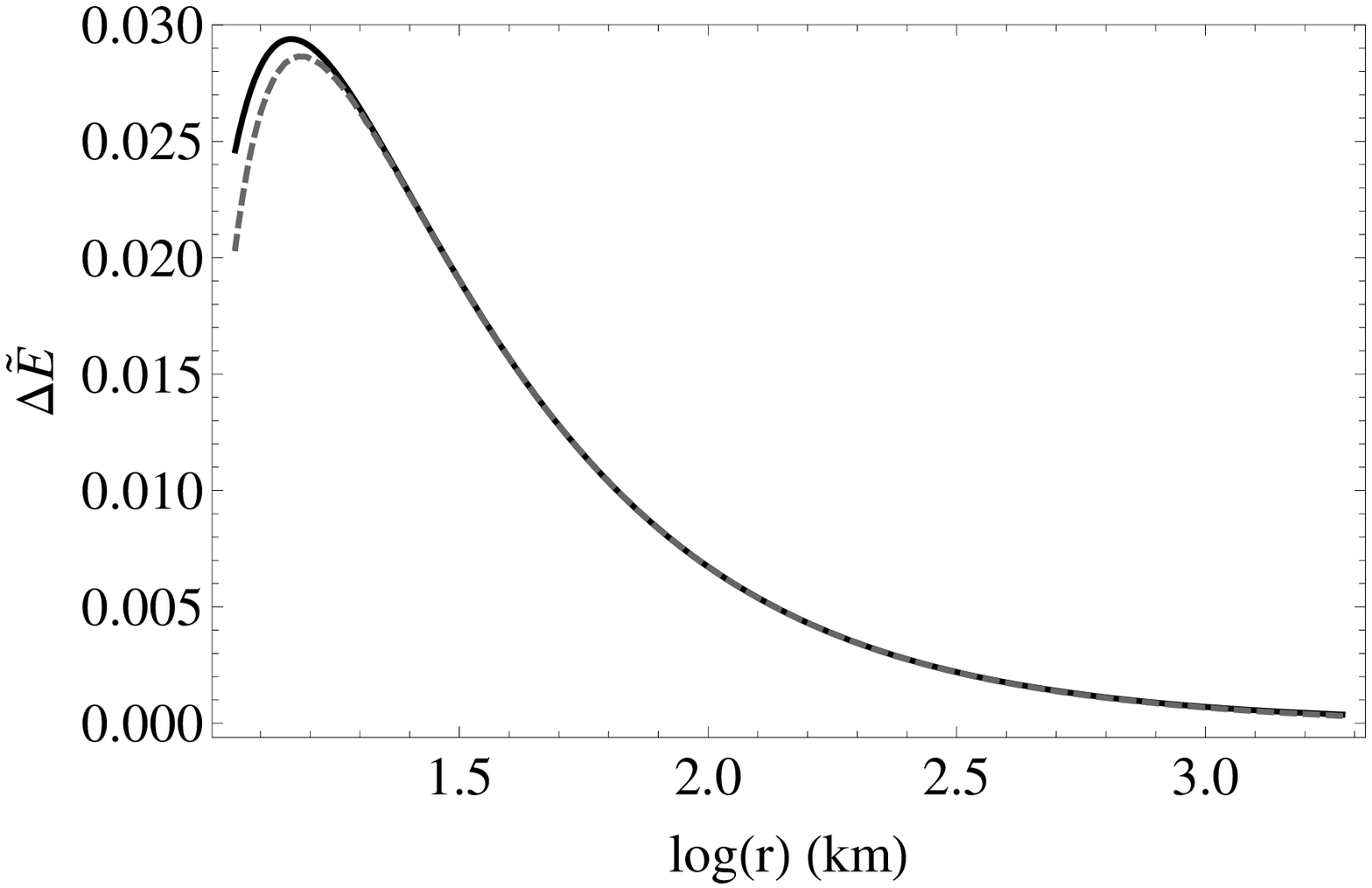}
\includegraphics[width=0.45\textwidth]{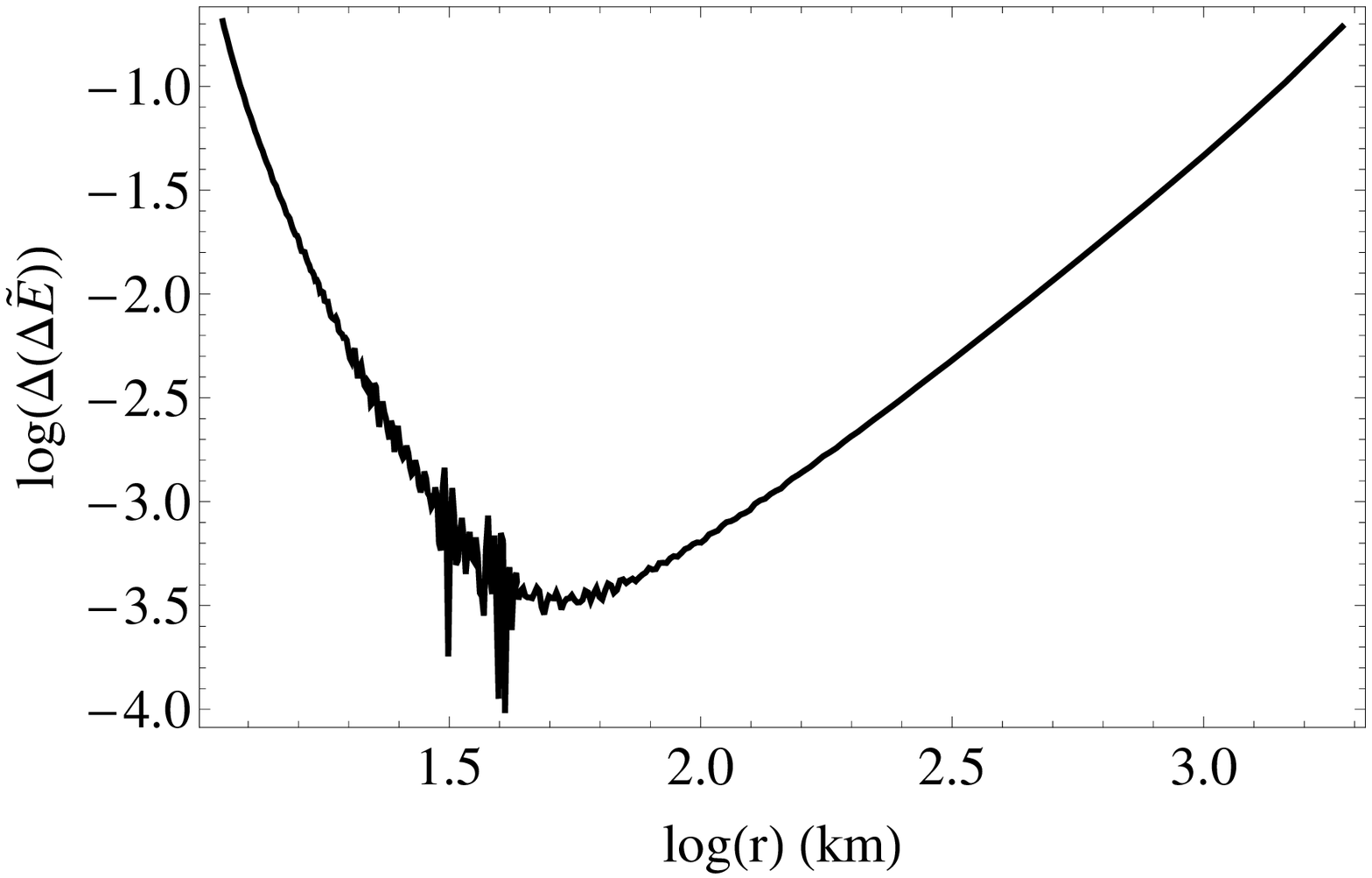}
\caption{\label{DEfig} Plots showing the analytic (solid) and
numerical (dotted) $\Delta\tilde{E}$ (left) as well as the logarithm of their
relative difference (right) for the model $\#10$ of AU EOS presented in
Pappas~\&~Apostolatos (2012).}

\end{figure*}

The final comparison criterion is the quantity $\Delta\tilde{E}$.
The numerical computation of this quantity has similar difficulties
with the precession frequencies; these issues could be fixed by performing the
same tricks to avoid numerical oscillations.
In Figure \ref{DEfig} we show for the same model of a
rotating neutron star as in the previous cases, the quantity $\Delta\tilde{E}$ computed from the
numerical and the analytic metric on the left, and their relative
difference on the right. Again we see that the two-soliton metric
 describes with high accuracy the $\Delta\tilde{E}$ which we obtain
 numerically from the numerical models.
 We remind here that this quantity is relevant for the emitted
spectrum of a thin accretion disc and its temperature profile, as
well as for the efficiency of the disc, i.e., the amount of
kinetic energy transformed to radiation.

We close this section with Table \ref{modelAUcomp}, where we
present for all  numerical models of EOS AU (the multipole
moments of which are given in the Tables of \cite{PappasMoments}) the
parameters and the type of the two-soliton metric along with a
few quantities of astrophysical interest, and specifically their
comparison between the ones calculated using the analytic and the
numerical space-times. These quantities are, the circumferential
radius at the ISCO $R_{\rm{ISCO}}$, the efficiency
$\eta=1-\tilde{E}_{\rm{ISCO}}$ of a thin accretion disc (if there was one
around the particular neutron star), the orbital frequency
at the position of the ISCO $\Omega_{\rm{ISCO}}$ (this is a
frequency expected to show up in QPOs if the latter are related to
the orbital motion), and finally the vertical oscillation frequency at the
 ISCO $(\kappa_z)_{\rm{ISCO}}$ (this could also be related to QPOs). The Table shows that the relative
differences between the numerical and the analytic quantities is of the order
of 1 per cent or lower.

\begin{table*}
 \centering
  \caption{For each neutron star model, that was constructed by the RNS code using the AU EOS, we have computed a
  number of parameters that are related to the particular analytic two-soliton metric which better approximates
  the numerical metric. These parameters are (i) the type (Case) of the
two-soliton space-time that each particular model corresponds to, (ii) the parameters $k$ and $b$ of the
two-soliton as well as the spin parameter $j$ and, (iii) the relative
difference between the following analytic and numerical quantities: $R_{\rm{ISCO}}$, $\eta$,
  $\Omega_{\rm{ISCO}}$ and $(\kappa_z)_{\rm{ISCO}}$, for the models that the ISCO lies outside the surface
  of the star. The  models shown here are the same models presented in Pappas~\&~Apostolatos (2012). The rest of the
  physical parameters of these models, such as the mass and all the other multipole moments of
  the models, can be found there. All the relative differences are given as a percentage.}
\begin{tabular}{|c|c|c|c|c|c|c|c|c|}\hline
model& Case & $j$ & $b$  &  $k$ & $\Delta R_{\rm{ISCO}}$ & $\Delta\eta$  & $\Delta\Omega_{\rm{ISCO}}$  & $(\Delta\kappa_{z})_{\rm{ISCO}}$  \\
       &       &     &($km$)&($km^2$)&  (per cent) & (per cent) & (per cent) & (per cent) \\ \hline
 1 & III & 0. & -3. & 0 & 0.003 & 0.032 & 0.006 & - \\
 2 & Ia & 0.2015 & -0.0784 & -0.5271 & 0.013 & 0.019 & 0.003 & 0.024 \\
 3 & IIa & 0.3126 & -0.1305 & -1.2503 & 0.283 & 0.142 & 0.027 & 0.141 \\
 4 & IIa & 0.414 & -0.1858 & -2.1664 & - & - & - & - \\
 5 & IIa & 0.4749 & -0.224 & -2.8362 & - & - & - & - \\
 6 & IIa & 0.5297 & -0.2626 & -3.517 & - & - & - & - \\
 7 & IIa & 0.5789 & -0.3014 & -4.1972 & - & - & - & - \\
 8 & IIa & 0.617 & -0.3351 & -4.7738 & - & - & - & - \\
 9 & IIa & 0.651 & -0.368 & -5.3277 & - & - & - & - \\
10 & IIa & 0.6618 & -0.3793 & -5.5114 & - & - & - & -
 \\ \hline
11 & III & 0. & -3. & 0 & 0.003 & -0.021 & -0.004 & - \\
12 & III & 0.194524 & -0.0915 & -0.1528 & -0.021 & 0.022 & 0.002 & -0.034 \\
13 & III & 0.309849 & -0.1669 & -0.4288 & -0.018 & 0.025 & 0.003 & -0.071 \\
14 & III & 0.406932 & -0.2417 & -0.8163 & 0.019 & 0.03 & 0.005 & 0.113 \\
15 & III & 0.485572 & -0.3158 & -1.2698 & 0.148 & 0.114 & 0.019 & 0.04 \\
16 & III & 0.550214 & -0.3897 & -1.7658 & 0.391 & 0.204 & 0.035 & 0.217 \\
17 & III & 0.603381 & -0.4628 & -2.279 & 0.778 & 0.463 & 0.077 & 0.477 \\
18 & III & 0.645447 & -0.5317 & -2.7705 & 1.274 & 0.83 & 0.137 & 0.815 \\
19 & III & 0.676639 & -0.5916 & -3.197 & 1.798 & 1.205 & 0.197 & 1.15 \\
20 & III & 0.706299 & -0.6585 & -3.6602 & 2.453 & 1.831 & 0.297 &
1.638
 \\ \hline
21 & III & 0.510282 & -0.3726 & -0.7958 & 0.043 & 0.026 & 0.005 & 0.15 \\
22 & III & 0.510617 & -0.3717 & -0.8179 & 0.054 & 0.046 & 0.008 & 0.058 \\
23 & III & 0.514032 & -0.3731 & -0.8721 & 0.063 & 0.05 & 0.009 & 0.031 \\
24 & III & 0.520506 & -0.3789 & -0.9409 & 0.083 & 0.016 & 0.004 & 0.065 \\
25 & III & 0.547452 & -0.4083 & -1.1827 & 0.164 & 0.072 & 0.013 & 0.153 \\
26 & III & 0.587439 & -0.4607 & -1.5504 & 0.346 & 0.206 & 0.034 & 0.164 \\
27 & III & 0.626593 & -0.5214 & -1.9574 & 0.63 & 0.391 & 0.064 & 0.362 \\
28 & III & 0.659098 & -0.5806 & -2.3502 & 0.991 & 0.645 & 0.104 & 0.64 \\
29 & III & 0.694585 & -0.6577 & -2.8456 & 1.551 & 1.018 & 0.162 & 0.948 \\
30 & III & 0.713165 & -0.7054 & -3.1406 & 1.95 & 1.349 & 0.214 &
1.221
 \\ \hline
\end{tabular}
\protect\label{modelAUcomp}
\end{table*}

\section{Conclusions}

In this work we have tested whether the four-parameter two-soliton analytic metric,
which was derived by \cite{twosoliton}, can be used as a trustworthy approximation for the space-time around
all kind of neutron stars.
%
%
To match the particular analytic metric to a specific neutron star model,
which was produced through the numerical code RNS of \cite{Sterg}, we have
used as matching conditions the first four non-zero multipole
moments. Our choice was justified, apart from theoretical reasoning, by comparing the numerical metrics with
different analytic metrics produced by slightly varying two of their moments (quadrupole and spin-octupole).
The comparison showed clearly that the best matching comes from imposing the
condition that the parameters of the analytic metric should be such
that the analytic space-time acquires the first four non-vanishing moments of the numerical metric.

Having demonstrated the appropriateness of the matching conditions, we proceeded to compare the various numerical
neutron star space-times with the corresponding analytic space-times. To perform the comparison, we have assumed several
criteria having in mind that they should correspond to geometric and physical properties of the space-time,
with a special interest in physical quantities that could be associated to astrophysical
processes that are usually observed from the vicinity of neutron stars.

The result of these comparisons is that the two-soliton space-time
can reproduce the properties of the space-time around realistic
neutron stars, and in particular it can reproduce all
astrophysically interesting  properties. Probably the most
important fact is that the analytic metric can capture properties
of the neutron star space-time that a corresponding Kerr space-time
could not, such as the behavior of the precession frequencies of
almost circular and almost equatorial orbits. A typical
example is shown in Figure 8, where we present the analytic and
numerical frequencies of the precession of the orbital plane for a
model constructed using the L EOS. The possible importance and
implication of this, i.e,. the capability of the two-soliton to
capture this particular behavior in contrast to the Kerr geometry, was
further discussed in \cite{PappasQPOs}.

Generally the two-soliton metric can be a very useful tool for studying
phenomena that happen around all kind of neutron stars and are quite sensitive
to more realistic and accurate geometries than the ones used
so far. Relying on a single analytic metric for all neutron stars is
practically more favorable than using numerical space-times, or more than one analytic
metrics depending on the type of the neutron star. Thus,
the two-soliton metric can be further used for more elaborate applications
such as the ones explored by \cite{psaltis,Psaltis1,Psaltis2}.

\begin{figure}
\centering
\includegraphics[width=0.45\textwidth]{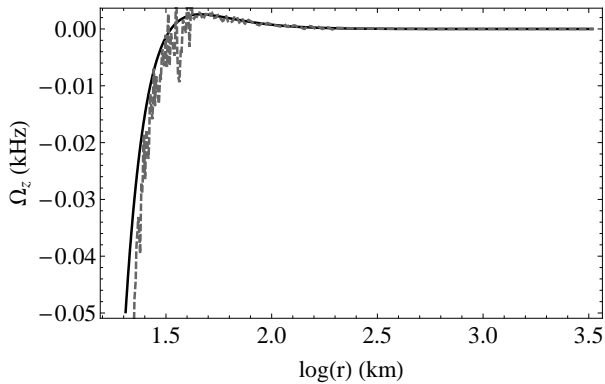}
\caption{Plot showing the numerical (dotted) and analytic (solid)
precession frequencies $\Omega_z$ for the model $\#10$ of the L
EOS. One can see how the frequencies that are calculated from the analytic
metric capture the qualitative behavior of the corresponding numerical
frequencies. The parameters (i.e., the multipole moments) for this
model can be found in Pappas~\&~Apostolatos (2012).}
\end{figure}

\section*{Acknowledgments}

We would  like to thank Kostas Kokkotas and Kostas Glampedakis
for many useful discussions, and Nikos Stergioulas for
providing us access to his RNS numerical code as well as solid advice.
This work has been supported by the I.K.Y. (IKYDA 2010).
G.P. would also like to acknowledge DAAD scholarship number A/12/71258.

\appendix
%
%
%
%

\section[]{Neutron star models}

In order to construct the analytic space-time exterior to a compact
object, one has to choose the appropriate multipole moments. For
neutron stars, these moments could be computed from numerical models that are
constructed with realistic equations of state. There are several
schemes developed for numerically integrating stellar models (see
\cite{Sterg}, and for an extended list of numerical schemes see
\cite{Lrr}). We have used Stergioulas's RNS code for the
construction of the models.

In order to cover more space on the ``neutron
star parameter space'', we constructed numerical
neutron star models with equations of state of varying
stiffness. For that purpose we have chosen AU as a typical soft
EOS, FPS as a representative moderate stiff EOS, and L to describe
stiff EOS. Having the numerical models ready, we proceeded in
evaluating their multipole moments according to the algorithm described in
\cite{PappasMoments}. The parameters then used to construct the analytic space-time models,
i.e., $M,\;a,\;k$, and $b$, are evaluated from the first four
multipole moments ($M,\;J,\;M_2,\;S_3$) of each model by inverting  equations
(\ref{moments2}).

For the specifics of the various models chosen here, we have followed
\cite{BertSter}. We have constructed the same constant rest-mass
sequences as the ones presented in \cite{BertSter} for the
corresponding equations of state. For every equation of state, 3
sequences of 10 models were constructed, that corresponded to:
\begin{enumerate}
\item a sequence corresponding to a neutron star of $1.4
M_{\odot}$ in the non-rotating limit,
\item a sequence terminating at the maximum-mass neutron star in
the non-rotating limit,
\item a supermassive sequence that does not terminate at a
non-rotating model at its lower rotation limit.
\end{enumerate}

All the sequences end at the mass-shedding limit on the side of
fast rotation, i.e., at the limit were the angular velocity of a
particle at the equator is equal to the Keplerian velocity at that
radius. These sequences are the so called {\it evolutionary
sequences}.

All the parameters for the computed models can be found in
\cite{PappasMoments}.

\section[]{The functions of the two-soliton}

In this Appendix we present the full expressions for writing the
metric functions of the two-soliton. The determinants that are
given in Sec.~2 and appear in the formulas for the metric
functions, can be substituted with the following expressions,
starting with $E_{\pm}=A\mp B$, where the functions $A,\,B$ are
given as:

\bea A\!\!\!\!&=&\!\!\!\!-16 d k \left(r_- r_++R_- R_+\right) M^2\nn\\
\!\!\!\!&-&\!\!\!\!\left[A_1^- \left(R_- r_+ + r_- R_+\right)- A_2^-
\left(R_- r_+-r_- R_+\right) \right] \kappa _+^2 \nn\\
\!\!\!\!&+&\!\!\!\! \left[A_1^+ \left(r_- R_-+r_+ R_+\right)-
A_2^+ \left(r_- R_--r_+ R_+\right) \right]\kappa _-^2 ,\eea
\bea
A_1^{\pm}\!\!\!\!&=&\!\!\!\!\left(4(a-b)^2 (a b\pm d-k)-4 ((a-2 b) b\pm d)M^2\right),\\
A_2^{\pm}\!\!\!\!&=&\!\!\!\!4 i \left((a-b) (a b\pm d-k)-b
M^2\right)\kappa _{\pm}, \eea
\bea B\!\!\!\!&=&\!\!\!\!2\kappa _- \kappa _+ M\left[
\left(R_+-R_-\right)
 B_1^- + \left(r_--r_+\right) B_1^+\right.\nn\\
\!\!\!\!&+&\!\!\!\!\left.\left(R_-+R_+\right)B_2^--
\left(r_-+r_+\right)B_2^+\right],\eea
\bea B_1^{\pm}\!\!\!\!&=&\!\!\!\! i\left(2 k
(a-b)+b\left(M^2-a^2+b^2\pm \kappa _-
\kappa_+\right)\right)\nn\\
\!\!\!\!&\times&\!\!\!\!\left(\kappa _+ \pm\kappa _-\right),\\
B_2^{\pm}\!\!\!\!&=&\!\!\!\! 2d\left(M^2-a^2+b^2 \pm\kappa _-
\kappa_+\right).\eea
The determinant $H$ can be substituted as $H=-L$, where $L$ is
given by the expression,

\bea L\!\!\!\!&=&\!\!\!\! \left(r_- r_++R_- R_+\right)L_1 + \left(r_-+R_-+r_++R_+\right)L_2\nn\\
   \!\!\!\!&-&\!\!\!\! \left(r_--R_-+r_+-R_+\right)L_3+ \left(r_- R_--R_+ r_+\right)L_5^-\nn\\
    \!\!\!\!&+&\!\!\!\!(r_--r_++R_--R_+)L_4^- +\left(r_- R_+-R_- r_+\right)L_5^+\nn\\
    \!\!\!\!&+&\!\!\!\!(r_--r_+-R_-+R_+)L_4^+ +\left(R_- r_++r_- R_+\right)L_6^+\nn\\
    \!\!\!\!&+&\!\!\!\!\left(R_- r_-+r_+ R_+\right)L_6^-,\eea 
%
%
\bea   L_1\!\!\!\!&=&\!\!\!\!-16 d k (i a-i b+M-z)M^2,\\
   L_2\!\!\!\!&=&\!\!\!\!i 4dM(a+i M) \left[M^4-2 \left(a^2-b^2-2 k\right)M^2\right.\nn\\
       \!\!\!\!&+&\!\!\!\!\left.(a-b)^2 \left((a+b)^2-4 k\right)\right],\\
   L_3\!\!\!\!&=&\!\!\!\! 4dM\left[i a^3-M a^2-i \left(b^2+M^2+2 k\right) a+M^3\right.\nn\\
      \!\!\!\!&+&\!\!\!\! \left.2 i b k+b^2 M\right] \kappa _- \kappa _+,\eea
\bea   L_4^{\pm}\!\!\!\!&=&\!\!\!\! 4\kappa _{\mp}M\left\{i b M^5+a b M^4\right.\nn\\
              \!\!\!\!&-&\!\!\!\! i \left(2 b a^2-k a-b \left(b^2\pm d+2 k\right)\right) M^3\nn\\
   \!\!\!\!&-&\!\!\!\!\left(2 b a^3-k a^2-b\left(b^2 \pm d+3 k\right) a\right.\nn\\
   \!\!\!\!&+&\!\!\!\!\left.k \left(2 b^2 \mp d+k\right)\right) M^2\nn\\
   \!\!\!\!&+&\!\!\!\!\left(2 k^2-\left(a^2+3 b a \mp 2 d\right) k+b (a+b) \left(a^2 \mp d\right)\right)\nn\\
   \!\!\!\!&\times&\!\!\!\!i (a-b) M \nn\\
   \!\!\!\!&+&\!\!\!\!\left. (a-b) (a b-k)\left((a+b) \left(a^2 \mp d\right)+(b-3 a)
k\right)\right\}, \eea
\bea   L_5^{\pm}\!\!\!\!&=&\!\!\!\! +4 i \left[b (M-z) a^4+\left(b^2 \mp d-k\right) (M-z) a^3\right.\nn\\
   \!\!\!\!&+&\!\!\!\!\left((z-M) b^3+i M^2 b^2\right.\nn\\
   \!\!\!\!&+&\!\!\!\!\left.\left(2M^2 \pm d+5 k\right) (z-M) b \mp i d M^2\right) a^2\nn\\
   \!\!\!\!&+&\!\!\!\!\left((z-M) b^4+\left(M^2 \mp d-5 k\right) (z-M) b^2\right.\nn\\
   \!\!\!\!&-&\!\!\!\!\left.4 i k M^2 b+(k \pm d) \left(M^2+4 k\right) (M-z)\right) a \nn\\
   \!\!\!\!&-&\!\!\!\! i b^4 M^2-i b^2 M^2 \left(M^2 \mp d-2 k\right)\nn\\
   \!\!\!\!&+&\!\!\!\!i M^2 \left(2 k (k \pm d)\pm d M^2\right)+ b\left(b^2 \left(M^2 \pm d+k\right)\right.\nn\\
   \!\!\!\!&+&\!\!\!\!\left.\left.\left(M^4+(k \pm d) M^2-4 k (k \pm d)\right)\right) (M-z)\right] \kappa _{\mp},\eea
\bea   L_6^{\pm}\!\!\!\!&=&\!\!\!\!\pm 4 \left[b (M-z) a^5+((k \pm d) (z-M) +i b M^2) a^4\right.\nn\\
   \!\!\!\!&+&\!\!\!\!\left(2 (z-M) b^3+2 \left(M^2+2 k\right) (z-M) b\right.\nn\\
   \!\!\!\!&-&\!\!\!\!\left.i k M^2\right) a^3\nn\\
   \!\!\!\!&+&\!\!\!\!\left(-2 i M^2 b^3+2 (5k \pm d) (M-z) b^2\right.\nn\\
   \!\!\!\!&-&\!\!\!\!i \left(2 M^4+3 k M^2\right) b \nn\\
   \!\!\!\!&+&\!\!\!\!\left.\left((k \pm 2 d) M^2+4 k (k \pm d)\right) (M-z)\right) a^2\nn\\
   \!\!\!\!&+&\!\!\!\!\left\{(M-z) b^5-2 \left(2 k-M^2\right) (M-z) b^3\right.\nn\\
   \!\!\!\!&+&\!\!\!\!7 i kM^2 b^2+\left(M^4+4 k M^2-8 k (k \pm d)\right) (M-z) b\nn\\
   \!\!\!\!&+&\!\!\!\!\left.i k M^2 \left(M^2+2 (k \pm d)\right)\right\} a\nn\\
   \!\!\!\!&+&\!\!\!\!i b^5 M^2-i b^3 \left(3 k-2 M^2\right)M^2\nn\\
   \!\!\!\!&+&\!\!\!\!i b M^2 \left(M^4+3 k M^2-2 k (k \pm d)\right)\nn\\
   \!\!\!\!&+&\!\!\!\!\left(M^2 \left(2 k (k \pm d)\pm d M^2\right)+b^4 (k \pm d)\right) (z-M)\nn\\
   \!\!\!\!&+&\!\!\!\!\left.b^2 \left(4 k (k \pm d)-(3k \pm 2 d) M^2\right) (M-z)\right]. \eea
The determinant $G$ can be expressed as $G=-E$, where $E$ is given
by the expressions,
\bea  E\!\!\!\!&=&\!\!\!\!E_1\left(r_- r_++R_- R_+\right)+E_2\left(r_-+R_-+r_++R_+\right)\nn\\
       \!\!\!\!&+&\!\!\!\!E_3\left(r_--R_-+r_+-R_+\right) +E_4^+(r_-R_+-R_-r_+)\nn\\
       \!\!\!\!&+&\!\!\!\!E_4^-(r_-R_--r_+R_+)+ E_5^+(R_-r_++r_-R_+)\nn\\
       \!\!\!\!&+&\!\!\!\!E_5^-(r_-R_-+r_+R_+)+ E_6^+(r_--r_+-R_-+R_+)\nn\\
       \!\!\!\!&+&\!\!\!\!E_6^-(r_--r_++R_--R_+),\eea
\bea  E_1\!\!\!\!&=&\!\!\!\! 16 d k (-i a+i b+M)M^2,\\
      E_2\!\!\!\!&=&\!\!\!\! 4 d \kappa _-^2\kappa _+^2(i a+M-z)M,\\
      E_3\!\!\!\!&=&\!\!\!\! 4 d \left[2i k (a-b)+\left(b^2+M^2-a^2\right)(M- z+ia)\right] \nn\\
         \!\!\!\!&\times&\!\!\!\! \kappa _-  \kappa _+ M,\eea
\bea  E_4^{\pm}\!\!\!\!&=&\!\!\!\!-4 i \left[b M^4+i \left(b^2 \mp d\right) M^3\right.\nn\\
              \!\!\!\!&-&\!\!\!\!(a+b) \left(2 a b-b^2 \mp d-k\right) M^2\nn\\
      \!\!\!\!&-&\!\!\!\! i \left(2 k^2+2 \left(b^2-2 a b \pm d\right) k \right.\nn\\
      \!\!\!\!&+&\!\!\!\!\left.(a^2-b^2) \left(b^2 \mp d\right)\right) M\nn\\
      \!\!\!\!&+&\!\!\!\!\left.(a-b) \left((a+b)^2-4 k\right) (a b \mp d-k)\right] \kappa _{\mp} M,\eea
\bea   E_5^{\pm}\!\!\!\!&=&\!\!\!\!\pm4M \left[i b M^5-(a b \mp d) M^4\right.\nn\\
                \!\!\!\!&-&\!\!\!\!i \left(-2b^3+2 a^2 b-3 k b-a k\right) M^3\nn\\
               \!\!\!\!&+&\!\!\!\!\left(2 k^2-\left(a^2+4 b a-3 b^2 \mp 2 d\right) k\right.\nn\\
               \!\!\!\!&+&\!\!\!\!\left.2 (a^2-b^2)(a b \mp d)\right) M^2\nn\\
               \!\!\!\!&+&\!\!\!\!i (a-b) \left((a-b) b (a+b)^2+2k^2\right.\nn\\
               \!\!\!\!&-&\!\!\!\!\left.\left(a^2+4 b a-3 b^2 \mp 2 d\right) k\right) M\nn\\
               \!\!\!\!&-&\!\!\!\!\left.(a-b)^2 \left((a+b)^2-4 k\right) (a b \mp d-k)\right],\eea
\bea  E_6^{\pm}\!\!\!\!&=&\!\!\!\!4 \kappa _{\mp} M \left\{b a^5-(k+i b (M-z)) a^4\right.\nn\\
               \!\!\!\!&-&\!\!\!\!\left(b^3+\left(2 M^2 \pm d+3 k\right) b-i k(M-z)\right) a^3\nn\\
               \!\!\!\!&+&\!\!\!\!\left(i (M-z) b^3+5 k b^2+i \left(\pm d+2 \left(M^2+k\right)\right) \right.\nn\\
              \!\!\!\!&\times&\!\!\!\!\left.(M-z) b+k \left(M^2 \pm d+3 k\right)\right) a^2\nn\\
              \!\!\!\!&+&\!\!\!\!\left(\left(M^2 \pm d-k\right) b^3-3 i k(M-z) b^2\right.\nn\\
              \!\!\!\!&+&\!\!\!\!\left(M^4+(3 k \pm d) M^2-4 k^2\right) b\nn\\
              \!\!\!\!&-&\!\!\!\!\left.i k \left(M^2+2 (k \pm d)\right) (M-z)\right) a\nn\\
              \!\!\!\!&-&\!\!\!\! (k \mp d) k M^2+b^2 k \left(-2 M^2 \mp d+k\right)\nn\\
              \!\!\!\!&-&\!\!\!\!i b^3 \left(M^2 \pm d\right)(M-z)\nn\\
              \!\!\!\!&-&\!\!\!\!\left. i b \left(M^4+(2k \pm d) M^2-2 k (k \pm d)\right) (M-z)\right\},\eea
and finally the determinant $K_0$ can be expressed as
\be
K_0=-16d\kappa_-^2\kappa_+^2. \ee
In order to get the metric functions one should substitute these
expressions to the equations (\ref{metricfunc1}-\ref{metricfunc2}).
We should note here that the above expressions are not
equal to the corresponding determinants since a common factor to
all the determinants, i.e., the quantity $\prod_{k=1}^n e_k e^*_k$,
has been simplified out of the expressions.

\section[]{Transformation to Weyl-Papapetrou coordinates for a general metric}

As we have mentioned, a stationary and axially symmetric
space-time $g_{\mu\nu}(x^{\mu})$ can be cast in the form of the
Papapetrou line element (\ref{Pap})

\[ ds^2=-f(dt-\omega
d\phi)^2+f^{-1}\left(e^{2\gamma}(d\rho^2+dz^2)+\rho^2d\phi^2\right),
\]
by an appropriate coordinate transformation, where the metric functions
are functions of the Weyl-Papapetrou coordinates $(\rho,z)$.
These coordinates as expressed as functions of the previous coordinates $x^{\mu}$ are harmonic conjugate
functions. That is, the coordinate $\rho(x^{\mu})$ is defined as
$\rho^2=(g_{t\phi})^2-g_{tt}g_{\phi\phi}$ and satisfies the
Laplace equation while $z(x^{\mu})$ is it's harmonic conjugate.
 Thus the integrability conditions for $z$ are the Cauchy-Riemann conditions
which can be used to calculate that coordinate. In an earlier work
(see \cite{Pappas}) we have shown the correct way to integrate these
conditions when one has to transform a metric given in quasi-isotropic
coordinates

\be\label{CSTmetr} ds^2=-e^{2\nu}dt^2+e^{2\psi}\left(d\phi-\omega
dt\right)^2+e^{2\mu}\left(dr^2+r^2d\theta^2\right),\ee
which are commonly used in numerical integrations of the
Einstein field equations, to the Papapetrou form. In that case,
the quasi-isotropic coordinates, $r,\theta$ enter the metric in a
way that can be easily cast in a cartesian form and thus the
Cauchy-Riemann conditions are given from the usual expressions

\[ \frac{\p z}{\p \varpi}=-\frac{\p\rho}{\p\zeta},\]
\[ \frac{\p z}{\p \zeta}=\frac{\p\rho}{\p\varpi},
\]
where we have defined the cartesian coordinates
$\varpi=r\sin\theta=r\sqrt{1-\mu^2},\;\zeta=r\cos\theta=r\mu$.

In the case of a metric given in a general form though,
like the Hartle-Thorne metric, the Cauchy-Riemann conditions can
not be used as given in the previous equations. So, one has to
calculate the general form of these conditions. The general
expressions for the Cauchy-Riemann conditions can be evaluated
from the orthogonality condition that the functions
$\rho(r,\theta)$ and $z(r,\theta)$ must satisfy. Thus, for a
metric given in the form

\be
ds^2=g_{tt}dt^2+2g_{t\phi}dtd\phi+g_{rr}dr^2+g_{\theta\theta}d\theta^2+g_{\phi\phi}d\phi^2,
\ee
the orthogonality condition between $\rho$ and $z$ is
$\nabla\rho\cdot\nabla z=0$, which gives the expressions for the
Cauchy-Riemann conditions
\be \frac{\p z}{\p
r}=\sqrt{\frac{g_{rr}}{g_{\theta\theta}}}\frac{\p \rho}{\p
\theta}, \ee
 \be     \frac{\p z}{\p \theta}=-\sqrt{\frac{g_{\theta\theta}}{g_{rr}}}\frac{\p \rho}{\p r},
\ee
where we have expressed the conditions in the $r,\theta$
coordinates and the function $\rho(r,\theta)$ is defined as
\be \rho(r,\theta)=\sqrt{(g_{t\phi})^2-g_{tt}g_{\phi\phi}}. \ee
These general Cauchy-Riemann conditions are used in the same way
as prescribed in \cite{Pappas} in order to evaluate the $z$
coordinate for the Hartle-Thorne metric and then compare the latter
metric with the numerical metric and the two-soliton metric on the axis of
symmetry. The metric functions given by the metric in the general
form can be directly associated to the metric functions in the
Papapetrou line element (\ref{Pap}) by comparing the corresponding
$g_{tt},~g_{t\phi}$ and $g_{\phi\phi}$ values. The only component that
needs to be evaluated then is the $g_{\rho\rho}=g_{zz}$ and it is given
by the equation
\be
g_{\rho\rho}=g_{zz}=f^{-1}e^{2\gamma}=\left(\frac{1}{g_{rr}}(\rho_{,r})^2
+\frac{1}{g_{\theta\theta}}(\rho_{,\theta})^2\right)^{-1}.
\ee

\section[]{The Hartle-Thorne metric.}

This is a metric produced by \cite{HT} as an expansion up to order
$O(\varepsilon^3)$, where $\varepsilon=\Omega/\Omega_*$ is a
parameter that characterizes the rotation of a star, and
corresponds to the exterior space-time of slowly rotating
relativistic stars ($\Omega_*=\sqrt{M/R^3}$ is the Kepler limit).
The components of the metric can be found in \cite{Bertietal} and
are given by the expressions,

\bea
g_{tt}&=&-\left(1-\frac{2M}{r}\right)\left(1+j^2F_1-qF_2\right),\\
g_{rr}&=&\left(1-\frac{2M}{r}\right)^{-1}\left(1+j^2G_1+qF_2\right),\\
g_{\theta\theta}&=&r^2\left(1+j^2H_1-qH_2\right),\\
g_{\phi\phi}&=&\sin^2\theta g_{\theta\theta},\\
g_{t\phi}&=&\left(\frac{2jM^2}{r}\right)sin^2\theta , \eea
where
\bea L&=&80 M^6+8 r^2 M^4+10 r^3 M^3+20 r^4 M^2\nn\\
      &-&45 r^5 M+15r^6,\nn\\
     p&=&\frac{1}{8M r^4 (r-2 M)},\nn\\
     W&=&\left(48 M^6-8 r M^5-24 r^2 M^4-30 r^3 M^3\right.\nn\\
      &-&\left.60 r^4 M^2+135 r^5M-45 r^6\right) u^2\nn\\
      &+&\left(16 M^5+8 r M^4-10 r^3 M^2-30 r^4M+15 r^5\right)\nn\\
      &\times&(r-M)  ,\nn\\
   A_1&=&\frac{15 r (r-2 M) \left(1-3 u^2\right) \ln\left(\frac{r}{r-2 M}\right)}{16 M^2},\nn\\
   A_2&=&\frac{15 \left(r^2-2 M^2\right) \left(3 u^2-1\right) \ln\left(\frac{r}{r-2 M}\right)}{16 M^2},\nn\\
   F_1&=&-pW+A_1,\nn\\
   F_2&=&5 p r^3 (r-M) \left(2 M^2+6 r M-3 r^2\right) \left(3u^2-1\right)-A_1,\nn\\
   G_1&=&p \left(-72 r M^5-3 \left(L-56 M^5 r\right)u^2+L\right)-A_1,\nn\\
   H_1&=& \frac{\left(16 M^5+8 r M^4-10 r^3 M^2+15 r^4 M+15r^5\right) }{8 M r^4}\nn\\
      &\times&\left(1-3 u^2\right)+A_2,\nn\\
   H_2&=&\frac{5 \left(2 M^2-3 r M-3 r^2\right) \left(1-3u^2\right)}{8 M r}-A_2,\nn\eea
with $u=\cos\theta$. We will now try to evaluate the first five
multipole moments of the Hartle-Thorne space-time, following
\cite{Ryan}. From the metric functions one can calculate the
rotation frequency of circular equatorial orbits, which is given
by  equation (\ref{Omega4}). $\Omega$ can be expressed with respect to a
new parameter $\upsilon=(M\Omega)^{1/3}$. This new parameter is a function of
$r$, which can in turn be expressed as a function of another new parameter,
$x=(M/r)^{1/2}$. Then the parameter $\upsilon$ can be expanded as a series in $x$.
This series can then be inverted and thus the parameter $x$ can be expressed as a
power series in $\upsilon$. Similarly, we can calculate the energy per mass, $\tilde{E}$, of a test particle
in circular orbit (\ref{Etilde4}) as a function of the parameter $x$.
This quantity can then be expressed as an expansion on $x$ which by substituting the
previously obtained power series can then be expressed as a series expansion in $\upsilon$. From that, one can
calculate the invariant quantity
\be
\Delta\tilde{E}=-\Omega\frac{d\tilde{E}}{d\Omega}=-\frac{\upsilon}{3}\frac{d\tilde{E}}{d\upsilon},
\ee
which is related to the multipole moments of the space-time as they are
given by \cite{Ryan}. After following all the previously mentioned calculations,
the expansion of the above expression is,
\bea
\Delta\tilde{E}&=&\frac{\upsilon^2}{3}-\frac{\upsilon^4}{2}-\frac{20
j \upsilon^5}{9}+\left(q-\frac{27}{8}\right) \upsilon^6-\frac{28 j
\upsilon^7}{3}\nn\\
 &+&\left(\frac{80 j^2}{27}+\frac{70
   q}{9}-\frac{225}{16}\right) \upsilon^8+\left(-6 q j-\frac{81 j}{2}\right) \upsilon^9\nn\\
   &&+\left(\frac{115
j^2}{18}+\frac{35 q^2}{12}+\frac{935
   q}{24}-\frac{6615}{128}\right) \upsilon^{10}\nn\\
   &+&\left(-\frac{1408
j^3}{243}-\frac{572 q j}{27}-165 j\right)
\upsilon^{11}+O(\upsilon^{12}).
\eea
From comparing Ryan's expressions to this expansion, we see
that the Hartle-Thorne metric, as given by Berti, has a rotation
parameter $j$ defined as $j=-J/M^2$, a quadrupole parameter
$q=Q/M^3$ and the following moments $S_3,\;M_4$ are equal to zero. The
above expansion seems to be consistent with the aforementioned
multipole moments up to $O(\upsilon^{11})$.

\bsp \label{lastpage}


\begin{thebibliography}{99}
\bibitem[\protect\citeauthoryear{Berti~\&~Stergioulas}{2004}]{BertSter} Berti E., Stergioulas N., 2004, MNRAS, 350
1416
\bibitem[\protect\citeauthoryear{Berti et al.}{2004}]{Bertietal} Berti E., White F., Maniopoulou A., Bruni M., 2005,
MNRAS, 358, 923
\bibitem[\protect\citeauthoryear{Boutloukos et al.}{2006}]{boutloukos} Boutloukos S., van der Klis M., Altamirano D.,
Klein-Wolt M., Wijnands R., Jonker P.G., Fender R.P., 2006, ApJ, 653, 1435
\bibitem[\protect\citeauthoryear{Baub{\"o}ck,~Psaltis,~{\"O}zel~\&~Johannsen}{2012}]{Psaltis1}
Baub{\"o}ck M., Psaltis D., {\"O}zel F., Johannsen T., 2012, ApJ, 753, 175
\bibitem[\protect\citeauthoryear{Baub{\"o}ck,~Psaltis~\&~{\"O}zel}{2012}]{Psaltis2}
Baub{\"o}ck M., Psaltis D., {\"O}zel F.,  preprint (arXiv:1209.0768[astro-ph])
\bibitem[\protect\citeauthoryear{Ernst}{1968a}]{ernst1} Ernst F.J., 1968, Phys. Rev., 167, 1175
\bibitem[\protect\citeauthoryear{Ernst}{1968b}]{ernst2} Ernst F.J., 1968, Phys. Rev., 168, 1415
\bibitem[\protect\citeauthoryear{Fodor,~Honselaers~\&~Perjes}{1989}]{fodor} Fodor G., Honselaers C., Perjes Z.,
1989, J. Math. Phys., 30, 2252
\bibitem[\protect\citeauthoryear{Geroch}{1970}]{geroch} Geroch R., 1970, J. Math. Phys., 11, 2580
\bibitem[\protect\citeauthoryear{Hansen}{1974}]{hansen} Hansen R.O., 1974, J. Math. Phys., 15, 46
\bibitem[\protect\citeauthoryear{Hartle~\&~Thorne}{1968}]{HT} Hartle J. B., Thorne K. S., 1968, ApJ, 153, 807
\bibitem[\protect\citeauthoryear{Hauser~\&~Ernst}{1981}]{HauserV} Hauser I.,Ernst F. J., 1981, J. Math. Phys., 22, 1051
\bibitem[\protect\citeauthoryear{Krolik}{1999}]{krolik} Krolik J.H., 1999, Active galactic nuclei: from
the central black hole to the galactic environment. Princeton Univ. Press, Princeton, NJ
\bibitem[\protect\citeauthoryear{Lamb}{2007}]{lamb}Lamb F.K., 2007, preprint
(arXiv:0705.0030[astro-ph])
\bibitem[\protect\citeauthoryear{Lukes-Gerakopoulos}{2012}]{Lukes} Lukes-Gerakopoulos G., 2012 Phys. Rev. D, 86, 044013
\bibitem[\protect\citeauthoryear{Manko~\&~Sibgatulin}{1993}]{SibManko} Manko V.S., Sibgatullin N.R., 1993,
Class. Quantum Grav., 10, 1383
\bibitem[\protect\citeauthoryear{Manko,~Martin~\&~Ruiz}{1995}]{twosoliton} Manko V.S., Martin J., Ruiz J.E., 1995,
J. Math. Phys., 36, 3063
\bibitem[\protect\citeauthoryear{Manko,~Mielke~\&~Sanabria-G{\'o}mez}{2000}]{Mankoetal} Manko V. S., Mielke E. W.,
Sanabria-G{\'o}mez J. D., 2000, Phys. Rev. D, 61, 081501
\bibitem[\protect\citeauthoryear{Pach{\'o}n,~Rueda~\&~Sanabria-G{\'o}mez}{2006}]{Pachon} Pach{\'o}n L.A.,
Rueda J.A., Sanabria-G{\'o}mez J.D., 2006, Phys. Rev. D, 73, 104038 
\bibitem[\protect\citeauthoryear{Pach\'{o}n,~Rueda~\&~Valenzuela-Toledo}{2012}]{PachonNew} Pach\'{o}n L.A.,
Rueda J.A., Valenzuela-Toledo C.A., 2012, ApJ, 756, 82
\bibitem[\protect\citeauthoryear{Papapetrou}{1953}]{Papapetrou} Papapetrou A., 1953, Ann. Phys., 12, 309
\bibitem[\protect\citeauthoryear{Pappas}{2009}]{Pappas2} Pappas G., 2009, Journal of Physics: Conference Series, 189, 012028
\bibitem[\protect\citeauthoryear{Pappas}{2012}]{PappasQPOs} Pappas G., 2012, MNRAS, 422, 2581
\bibitem[\protect\citeauthoryear{Pappas~\&~Apostolatos}{2008}]{Pappas} Pappas G., Apostolatos T.A., 2008,
Class. Quantum Grav., 25, 228002
\bibitem[\protect\citeauthoryear{Pappas~\&~Apostolatos}{2012}]{PappasMoments} Pappas G., Apostolatos T.A., 2012,
Phys. Rev. Lett., 108, 231104
\bibitem[\protect\citeauthoryear{Psaltis}{2008}]{psaltis} Psaltis D., 2008, Living Reviews in Relativity, 11, 9
\bibitem[\protect\citeauthoryear{Ruiz,~Manko~\&~Martin}{1995}]{manko2} Ruiz E., Manko V.S., Martin J., 1995,
Phys. Rev. D, 51, 4192
\bibitem[\protect\citeauthoryear{Ryan}{1995}]{Ryan} Ryan F.D., 1995, Phys. Rev. D, 52, 5707
\bibitem[\protect\citeauthoryear{Shafee et al.}{2006}]{Narayan} Shafee R., McClintock J. E., Narayan R., Davis S. W.,
Li L.-X., Remillard R. A., ApJ, 636, L113
\bibitem[\protect\citeauthoryear{Sibgatulin}{1984}]{sib1} Sibgatullin N.R., 1991,
Oscilations and Waves in Strong Gravitational and Electromagnetic
Fields. Springer, Berlin [original in Russian, 1984, Nauka,
Moscow]
\bibitem[\protect\citeauthoryear{Sotiriou~\&~Pappas}{2005}]{Sotiriou}Sotiriou T.P., Pappas G., 2005, Journal of Physics:
Conference Series, 8, 23
\bibitem[\protect\citeauthoryear{Stella}{2000}]{stella} Stella L., 2000, preprint
(astro-ph/0011395)
\bibitem[\protect\citeauthoryear{Stergioulas}{2003}]{Lrr} Stergioulas N., 2003,
Living Reviews in Relativity, 6,
http://relativity.livingreviews.org/lrr-2003-3
\bibitem[\protect\citeauthoryear{Stergioulas~\&~Friedman}{1995}]{Sterg} Stergioulas N., Friedman J.L., 1995, ApJ, 444, 306
\bibitem[\protect\citeauthoryear{Stute~\&~Camenzind}{2002}]{Stuart} Stute M., Camenzind M., 2002, MNRAS, 336, 831
\bibitem[\protect\citeauthoryear{Teichm{\"u}ller,~Fr{\" o}b~\&~Maucher}{2011}]{Teich}
Teichm{\" u}ller C., Fr{\"o}b M.B., Maucher F., 2011, Class. Quantum Grav., 28 155015
\bibitem[\protect\citeauthoryear{van der Klis}{2006}]{derKlis} van der Klis M., 2006,
in Lewis W.H.G., van der Klis M., eds, Compact Stellar X-Ray
Sources. Cambridge Univ. Press, p. 39
\bibitem[\protect\citeauthoryear{Xanthopoulos}{1979}]{XanthMoments} Xanthopoulos, B. C., 1979,
J. Phys. A: Math. Gen., 12, 1025
\bibitem[\protect\citeauthoryear{Xanthopoulos}{1981}]{Xanth32} Xanthopoulos, B. C., 1981,
J. Math. Phys., 22, 1254

\end{thebibliography}
\end{document}